\documentclass[acmsmall, manuscript, nonacm]{acmart}
\usepackage[most]{tcolorbox}
\usepackage{caption}
\usepackage{array}
\usepackage{diagbox}
\usepackage{hyperref}
\usepackage{supertabular}
\usepackage{booktabs}
\usepackage{float}
\usepackage{enumitem}
\usepackage{graphicx}
\usepackage{comment}
\usepackage{multirow}
\usepackage{tabularx}
\usepackage{colortbl}
\usepackage{longtable}
\usepackage[resetlabels,labeled]{multibib}
\usepackage{subcaption}
\usepackage{makecell}
\usepackage{url}
\usepackage{soul}
\usepackage[title]{appendix}

\usepackage{wasysym} 

\newcites{PS}{Appendix - Selected Primary Studies}
\newcommand{\customcitePS}[1]{[\citenum{#1}]}

\newcounter{PScounter}
\AtBeginDocument{%
  \providecommand\BibTeX{{%
    \normalfont B\kern-0.5em{\scshape i\kern-0.25em b}\kern-0.8em\TeX}}}

\acmJournal{CSUR}
\acmVolume{1}
\acmNumber{1}
\acmArticle{1}
\acmMonth{1}

\begin{document}

\title{A Multivocal Review of MLOps Practices, Challenges and Open Issues}

\author{Beyza Eken}
\authornote{Both authors contributed equally to this research.}
\email{beken@sakarya.edu.tr}
\orcid{0000-0002-6824-2765}
\affiliation{
   \institution{Faculty of Computer and Information Sciences, Sakarya University}
   \country{Turkey}}
\author{Samodha Pallewatta}
\authornotemark[1]
\email{skpallewatta92@gmail.com}
\affiliation{
 \institution{CREST - The Centre for Research on Engineering Software Technology and The University of Adelaide}
  \country{Australia}}

\author{Nguyen Khoi Tran}
\email{nguyen.tran@adelaide.edu.au}
\orcid{}
\affiliation{
 \institution{CREST - The Centre for Research on Engineering Software Technology and The University of Adelaide}
  \country{Australia}}

\author{Ayse Tosun}
\email{tosunay@itu.edu.tr}
\orcid{0000-0003-1859-7872}
\affiliation{
  \institution{Faculty of Computer and Informatics Engineering, Istanbul Technical University}
  \country{Turkey}}

\author{Muhammad Ali Babar}
\orcid{}
\email{ali.babar@adelaide.edu.au}
\affiliation{
 \institution{CREST - The Centre for Research on Engineering Software Technology and The University of Adelaide}
  \country{Australia}}

\begin{abstract}

MLOps has emerged as a key solution to address many socio-technical challenges of bringing ML models to production, such as integrating ML models with non-ML software, continuous monitoring, maintenance, and retraining of deployed models. Despite the utility of MLOps, an integrated body of knowledge regarding MLOps remains elusive because of its extensive scope due to the diversity of ML productionalization challenges it addresses. Whilst the existing literature reviews provide valuable snapshots of specific practices, tools, and research prototypes related to MLOps at various times, they focus on particular facets of MLOps, thus fail to offer a comprehensive and invariant framework that can weave these perspectives into a unified understanding of MLOps. This paper presents a Multivocal Literature Review that systematically analyzes a corpus of 150 peer-reviewed and 48 grey literature to synthesize a unified conceptualization of MLOps and develop a snapshot of its best practices, adoption challenges, and solutions.

\end{abstract}



\maketitle

\section{Introduction} \label{introduction}
There is a huge trend of incorporating artificial intelligence (AI) and machine learning (ML) components into software systems to construct ML-enabled systems and take advantage of their well-established inferential and predictive capabilities. IBM Global AI Adoption Index for 2022 reports that AI/ML adoption is increasing rapidly, with 44\% of the organizations working on embedding AI/ML in their products or processes in 2022 \cite{ibmindex2022}. As of January 2023, 34\% of the experimental models are registered as candidates for production, which is a significant increase compared to 20\% in 2022 \cite{DataBricksMLOps}.

Despite the interest in ML-enabled systems, most organizations that build ML pilots fail to bring their ML models to production \cite{practitionersKhalid} due to novel operationalizing challenges of ML-enabled software systems compared to traditional software. \textit{Productionalization of ML models} refers to the process of transitioning ML models from a laboratory setting to a production environment, where they must handle live traffic from real users and provide inference services reliably at scale while ensuring safety. Successful productionalization of ML models requires organizations to address multiple challenges, both technical and social in nature. For instance, one prominent issue is the integration of ML models into the surrounding software systems \customcitePS{S66}. To enable this integration, software engineers must be able to package the trained model \customcitePS{S66}, \customcitePS{S660}, \customcitePS{I41}, \customcitePS{I66}, \customcitePS{I19}, \customcitePS{S243} and implement the necessary data transformation logic to feed data from the surrounding software components to the ML model with correct syntax and semantics \customcitePS{S510}. Failure to do so could result in erroneous predictions, such as when feeding metric measurements to a model that expects imperial measurements. The details regarding syntax, semantics, and data transformation used by ML models might not be apparent or readily documented. Consequently, resolving this technical challenge requires software engineers to collaborate closely with data scientists and ML engineers despite these disciplines' differing languages and cultures \customcitePS{I109}.

Another common problem in the productionalization of ML models lies in the monitoring and maintaining ML-enabled software in production \customcitePS{AwsMLOps2021}. The IT operations team must implement additional telemetry to monitor the behaviour of ML models, allowing them to detect when models suffer from concept drifts or violate safety constraints \customcitePS{AmazonAI}. Again, this task requires a skill set different from that typically found among IT operations teams. It necessitates collaboration between IT staff and data scientists with distinct disciplinary cultures and lexicon \customcitePS{I109}.

Additionally, retraining and updating models when problems such as drifts are detected in production represents another challenge in operating ML-enabled systems. The retraining process necessitates new data, likely sourced from the production environment. This data may contain sensitive personal information and must undergo appropriate preprocessing \customcitePS{AmazonAI}. Developing this data pipeline requires collaboration between software engineers, IT operations, data engineers, and data scientists. During the model training phase, it’s critical to maintain provenance records of experiments for analysis and compliance purposes \customcitePS{G3}, \customcitePS{IBM2023}. After a new model is created, software engineers and IT operators face the challenge of rolling out the new models quickly and safely and need to rely on strategies like A/B testing, canary deployments, and blue/green deployment \customcitePS{AwsLens}, \customcitePS{S454}. Velocity and safety ultimately drive the success of ML model productionalization.

The concept of MLOps emerged as a key solution to address these challenges of ML model productionalization. MLOps was first proposed by the research community in 2015 to expedite ML lifecycle management while achieving the high scalability required by business applications \cite{IbmMLOps, sculley2015hidden}. Research indicates that the effective operationalization of ML-enabled systems is contingent on employing robust MLOps practices, processes, and technologies \cite{practitionersKhalid}. A survey conducted by Databricks in 2023 indicates that, although the percentage of ML models reaching production remains low, there has been a significant increase with the rapid advancements in MLOps solutions (e.g., MLFlow, Hugging Face, SageMaker, Google Vertex AI).

Despite the widespread prevalence of MLOps, an evidence-based body of knowledge regarding MLOps practices and tools grounded in a unified conceptualization of the MLOps concept remains elusive for several reasons. First, the sheer scope of MLOps encompasses a broad array of practices, techniques, and corresponding technologies, complicating the effort to systematize the knowledge. For example, MLOps can be viewed from the perspective of ML “operations” (i.e., bringing ML models from the lab into deployed software systems), leading to research related to the packaging of ML models, interoperability with surrounding software systems, and testing of ML-enabled systems (e.g., \customcitePS{S510}, \customcitePS{S163}, \customcitePS{S243}, \customcitePS{I43}, \customcitePS{S454}). MLOps is also often conceptualised as "DevOps for ML", leading to the development of practices and tools for defining and managing automated pipelines of ML lifecycle activities (e.g., \customcitePS{S57}, \customcitePS{S163}). Additionally, MLOps is also viewed from the perspective of systematic management of ML experiments and related assets (e.g., datasets, feature sets, trained models), leading to practices such as applying version control to ML assets and recording experiments to support both analysis and provenance for compliance and security (e.g.,\cite{S565,S204}). On-demand and continuous monitoring of model performance and safety in a production environment is yet another critical practice associated with MLOps (e.g.,\cite{A36}). Finally, given that ML model productionalization involves collaboration among multiple teams, many MLOps best practices emphasize the need for organizational and cultural changes to facilitate necessary cross-team collaboration in handling ML productionalization (e.g., establishing project teams \customcitePS{G3}, \customcitePS{AwsLens}, building specialized ML engineering teams).

The diversity of information sources on MLOps compounds the issue. Particularly, gray literature such as training materials and best practices from various platform providers \customcitePS{G3},  \customcitePS{IBM2023}, \customcitePS{NVIDIA2023}, \customcitePS{AwsSecure}, \customcitePS{AwsMLOps2021} play a significant role in MLOps since it is inherently driven by practice. This gray literature must be compared and contrasted with peer-reviewed literature to synthesise a comprehensive and unbiased understanding of the field, which requires substantial time and effort. This challenge is exacerbated by the previously mentioned diversity in what constitutes “MLOps”. There is considerable confusion surrounding MLOps due to its complexity, diversity, and broad scope \cite{NVIDIA2023}, creating obstacles for both practitioners and researchers navigating the current MLOps landscape, hindering effective adoption and subsequent improvements in MLOps practices and solutions. 

In this paper, our objective is to build an evidence-based knowledge base for MLOps based on a unified conceptualization of the MLOps concept by integrating input from the gray literature and peer-reviewed literature to address the challenges that practitioners and researchers face when working with MLOps. The unified conceptualization of MLOps in this context denotes a coherent and comprehensive model that accounts for the various points of view on MLOps with respect to its definition, activities carried out within its scope, and the roles and responsibilities involved. Such a conceptualization creates a framework for identifying and classifying the diverse practices, processes, and technologies under the MLOps umbrella. “Evidence-based” in this context signifies that the knowledge base and conceptualization will be synthesized inductively from the literature. This knowledge base's benefit is providing practitioners with a framework for understanding, assessing, and adopting MLOps practices while offering researchers a roadmap for open research opportunities. We applied the Systematic Multivocal Literature Review (MLR) and thematic analysis methodologies to address the research goal. The MLR methodology enables us to systematically identify, select, and extract information from peer-reviewed and gray literature on MLOps published up to September 2023. The thematic analysis allows us to systematically analyze the extracted data and synthesize the conceptualization of MLOps, best practices, adoption challenges, and solutions to build a robust body of knowledge for researchers and practitioners interested in MLOps.

The multifaceted nature of MLOps and the abundance of primary studies classified under this umbrella have led to many literature reviews exploring various aspects of MLOps. For example, Steidl et al. \cite{steidl2024past} conducted a tertiary analysis that identified 33 reviews encompassing 1,397 primary studies on MLOps and found minimal overlap in the coverage of these reviews. Whilst the existing literature reviews provide valuable snapshots of specific practices, tools, and research prototypes related to MLOps at various times, they focus on particular facets of MLOps (e.g., the DevOps angle, the experiment management angle, the continuous monitoring angle), thus fail to offer a \textbf{comprehensive and invariant} framework that can weave these perspectives into a unified understanding of MLOps. Such a framework is crucial because it aligns current and future practices and tools with their challenges. Although snapshots of MLOps technologies and practices can quickly become outdated as tools evolve, a robust framework allows for analyzing and categorizing present and future developments, paving the way for creating new snapshots. Thus, the systematic synthesis of a conceptual framework for MLOps from both peer-reviewed and gray literature represents our study's core focus and contribution (Section \ref{RQ1}). Building upon this foundation, we identify and categorize the state-of-the-art practices and techniques for implementing MLOps, forming a "snapshot" of the literature (Section \ref{RQ2}). Finally, we highlight practitioners' challenges when adopting MLOps and identify future research directions (Section \ref{RQ3}).

\section{Methodology}\label{methodology}

A Multivocal Literature Review study (MLR) allows researchers to collect and present information from the industry when the formal/academic literature is not sufficient \cite{garousi2019guidelines}. 
Since MLOps is an emerging and practitioner-driven topic, only considering academic publications is not sufficient to cover all the knowledge related to this topic. Industry practitioners have been implementing and getting experience in MLOps to integrate ML-enabled systems into their software development pipelines, their knowledge is a crucial element in understanding MLOps. Thus, gray literature such as industrial white papers, blog posts, and company websites, are the key sources for expanding the knowledge obtained from academic literature. 
Therefore, we believe an MLR would be the most suitable methodology. 

\begin{figure}[ht]
   \centering
   \includegraphics[width=0.8\linewidth]{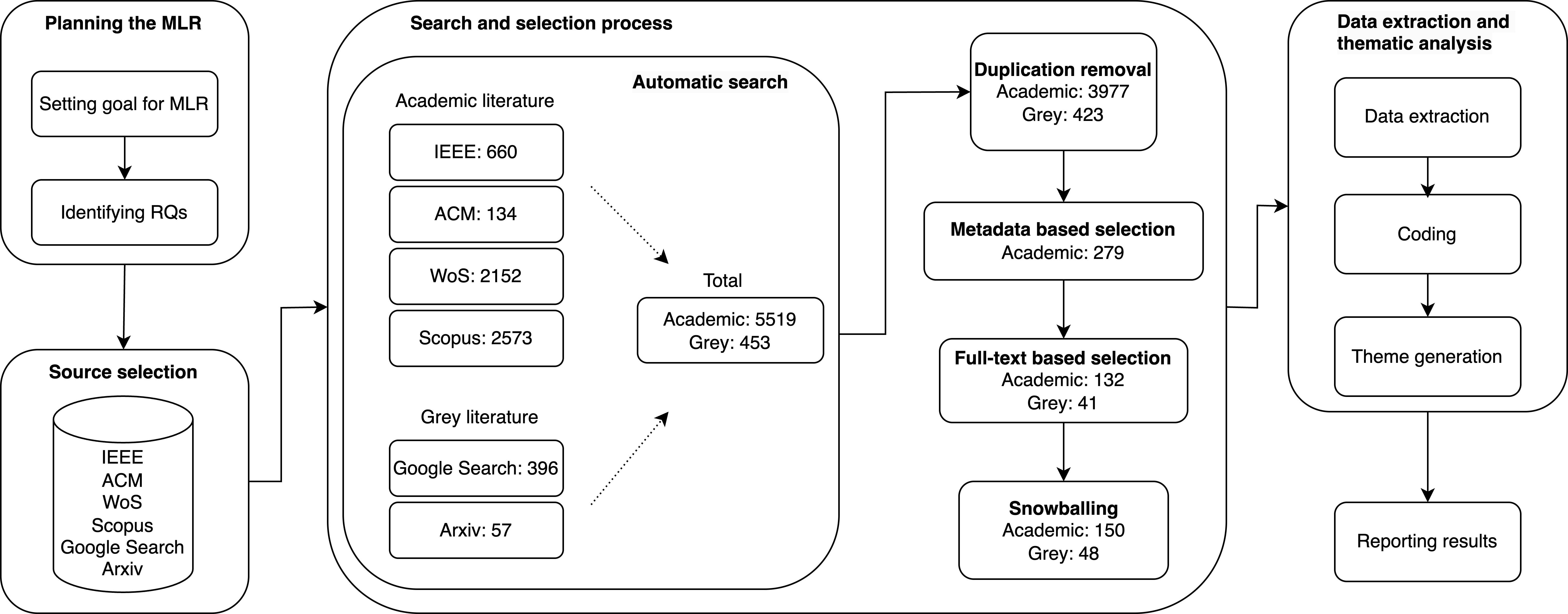}
   \caption{Research Methodology}
   \label{fig:ResearchMethodology} 
\end{figure}

We performed our MLR following the guidelines reported by Garousi et al. \cite{garousi2019guidelines}. Consequently, our MLR consists of three main stages: (1) \textit{planning the review}, (2) \textit{conducting the review}, and (3) \textit{reporting the review}. In the planning stage, we defined the research questions and developed the protocol for identifying and selecting literature. In the conducting phase, we followed the specified protocol to identify and choose literature, extract data from the literature, and synthesize the extracted data to answer the research questions. We employed the Thematic Analysis methodology based on the guidelines proposed by Braun et al. \cite{thematicAnalysisBraun} to carry out data synthesis. In the final stage, we write up the synthesis results. Figure \ref{fig:ResearchMethodology} represents our MLR methodology.

\subsection{Planning the Review}

We designed three main research questions (RQs) to systematically explore MLOps through academic and gray literature to address our research goals: (i) establish a foundational understanding of how MLOps is defined and perceived, (ii) identify and map current practices, and (iii) identify challenges and solutions within the field. Table \ref{tab:rqs} outlines the RQs in detail and the motivations behind each.

\begin{table}[htbp]
    \caption{Research Questions addressed in this MLR}
    \label{tab:rqs}
    \centering
    \resizebox{\textwidth}{!}{%
    \begin{tabular}{p{6cm}|p{14cm}}   
         \textbf{Research Questions} & \textbf{Motivation} \\
         \hline
         \textbf{RQ1} - Conceptualization of MLOps & Exploring foundational aspects of MLOps is important for creating a common and standardized understanding of MLOps, optimizing workflows, enabling efficient adoption of MLOps practices across diverse contexts, and fostering collaboration.

        \\
         \textit{\textbf{RQ1.1} - How is MLOps defined in the literature?} & RQ1.1 aims to provide insights on common and varying definitions of MLOps to facilitate better understanding of its foundational aspects to contribute towards creating a standardized and consistent usage within the community. 

         \\
         \textit{\textbf{RQ1.2} - What are the primary activities and tasks that define an MLOps workflow?} &
         RQ1.2 identifies the core components of MLOps, focusing on the essential activities and tasks within an MLOps workflow. 
         Outlining these key elements, provides a foundation for process optimization and standardization, while offering practitioners actionable insights for efficient adoption and clarifying the operational impact of MLOps in their specific contexts.
         
         \\
         \textit{\textbf{RQ1.3} - What are the practitioner roles and responsibilities in MLOps?} & RQ1.3 highlights the responsibilities of people who take part in MLOps settings. Establishing a common understanding of roles and their responsibilities leads to more efficient workflows and productive collaboration practices within the complex MLOps team structure.
         
         \\
         \hline
         \textbf{RQ2} - What are the state-of-the-art best practices and techniques for implementing MLOps? & 
         A complete view of how MLOps activities and tasks are performed in practice covering, frameworks, processes, tools and best practices to implement them is required for companies to plan for successful implementation of MLOps by minimizing the technical debt and improving the maintainability of the ML systems as they scale and grow in complexity over time. To this end, RQ2 aims to provide state-of-the-art MLOps practices and techniques covering social and technical aspects across the entire life cycle of ML models.

         \\
         \hline
        \textbf{RQ3} - What are the challenges practitioners face when adopting MLOps, and what solutions have been proposed to address them? & Gaining a deeper understanding of the challenges practitioners face when implementing MLOps along with proposed solutions  in the current literature, is crucial to see the bottlenecks in MLOps adoption. 
        RQ3 addresses the state-of-the-art challenges that would be beneficial to practitioners to overcome those when they adopt MLOps for productionalizing their ML-enabled software systems. Further, existing challenges often lead to new and highly impactful research directions and innovative solutions. 
        
        \\
         \hline
    \end{tabular}
    }
\end{table}

\subsection{Study Search and Selection Protocol} 
We based our protocol on the guidelines provided by Garousi et al. \cite{garousi2019guidelines} for conducting MLR. Our protocol consists of three main steps: automatic search method using inclusive search strings, primary study selection by applying quality assessment and snowballing to identify further studies \cite{wohlin2014guidelines}. 

\subsubsection{Automatic Search}  

For academic literature, we searched IEEEXplore Digital Library, ACM Digital Library, Web of Science, and Scopus using complex Boolean queries incorporating all related keywords into the search string to maximize the retrieval of relevant content from these databases. For gray literature, we used simpler, flexible search string as complex queries would limit the results of the search engine. 

We developed and evolved search strings through an exploratory search process, incorporating relevant keywords (e.g., continuous test, mlsecops) while avoiding restricting keywords (e.g., technologies, challenges, and tools) to capture relevant content. Based on this process, we reached two different search strings for peer-reviewed and gray literature (Table \ref{tbl:search_strings}).

\begin{table}[ht]
    \caption{Search Strings}
    \label{tbl:search_strings}
    \centering
    \resizebox{\textwidth}{!}{%
    \begin{tabular}{p{3cm}|p{15cm}}   
         Academic literature &  mlops OR "machine learning operations" OR mlsecops OR mlflow OR kubeflow OR cd4ml OR ( ( devops OR "continuous software engineering" OR cse OR "CI/CD" OR "continuous integration" OR "continuous deployment" OR "continuous delivery" OR "continuous test*" OR "continuous verification" OR "continuous security" OR "continuous compliance" OR "continuous evolution" OR "continuous use" OR "continuous trust" OR "continuous monitoring" OR "continuous release" OR "continuous build" OR "continuous systems engineering" ) AND ( "machine learning" OR "artificial intelligence" OR "deep learning" OR ml OR ai OR dl ) ) \\
         \hline
         Gray literature & mlops OR "machine learning operations" OR (devops AND "machine learning") OR ("continuous software engineering" AND "machine learning") \\
    \end{tabular}
    }
    \label{tab:search_strings}
\end{table}

The search process was independently executed by two authors. Individual results were combined in a shared file, and non-overlapping items—potentially due to location-based differences—were reviewed and discussed for inclusion.

\subsubsection{Primary Study Selection from Academic Literature}
We removed those duplicated items from our pool of studies, which reduced the number of academic studies from 5519 to 3977. Afterwards, we applied a two-step study selection process to the academic literature: (1) metadata-based selection and (2) full text-based selection. The metadata-based selection is conducted by considering only the paper title, paper keywords, and abstract of the paper, whereas the full-text selection is carried out by reviewing the full text of studies. 

During the metadata-based selection, we followed the inclusion and exclusion criteria listed in our online appendix \cite{onlineAppendix}. 
On the other hand, during the full text-based selection, we assessed the quality of each study considering the research aim, research design, methodology, results, limitations of the study, and the value added to research or practice by the study. A more detailed list of quality assessment criteria can be found in our online appendix \cite{onlineAppendix}. 
These criteria are adopted from \cite{kitchenhamSLR2007} and \cite{garousi2019guidelines}, discussed and agreed by all authors. 

At the beginning of both metadata and full-text-based selection steps, a pilot selection is carried out by all authors on a sample of the study pool. This pilot study is crucial to increase the inter-rater agreement and establish a common terminology among authors who later continue with study selection individually on a subset of relevant studies. Pilot study samples were picked randomly to avoid the bias that might occur during database search since the most relevant studies are naturally listed at the top of the search. We picked six studies for metadata-based selection and eight for full-text selection. These studies were reviewed by all the authors and the results were discussed. 

After completing the pilot study, the study pool was split into three by considering the bias of study order, and the selection processes were carried out separately by three authors. 
When a decision cannot be made on a study by a single author, a second and if necessary a third reviewer assessed the same study, and the final decision was made accordingly. If authors were unable to reach a consensus, the study was added to the final list of primary studies in order to avoid the exclusion of any relevant content. We ended up having 279 studies after the metadata-based selection process. After reading the full text and assessing the quality of the studies based on our criteria, there were 132 academic studies left. 

As the final step in academic literature selection, we applied forward and backward snowballing based on the guidelines by Wohlin et al. \cite{wohlin2014guidelines}. We received 39 more studies as a result of snowballing on the selected academic studies. Following the same inclusion and exclusion, and quality criteria we used in the prior steps, we included 18 out of the 39 studies found by snowballing, creating a final set of \textbf{150 peer-reviewed literature}.

\subsubsection{Primary Study Selection from Gray Literature}
First, we collected a list of studies by applying our search string on Google Search and Arxiv. We stopped our search on Google when the results reached theoretical saturation \cite{garousi2019guidelines}, i.e., meaning that no new content related to MLOps was found. The gathered results roughly covered 50\% of all search results. At the end of the study collection, we had a total number of 453 studies including search engine results and pre-prints. 
Second, we applied the inclusion and exclusion criteria \cite{onlineAppendix} to select primary studies. This involved removing video contents, non-free materials, and duplicates already identified through academic search. We ended up having 423 studies. 
Third, we applied our quality assessment criteria, reported in our online appendix \cite{onlineAppendix}, which are adopted from \cite{garousi2019guidelines}, on the remaining studies. We iteratively simplified the quality criteria provided by Garousi et al. \cite{garousi2019guidelines} to ensure we captured all the relevant content. For example, we excluded many of the methodology related criteria as MLOps is still an emerging field and many are still trying to define it and its associated technologies and concepts. Hence studies tend to skip reporting experimental data or methodology. 
After the quality assessment, we selected studies that met at least seven out of nine criteria listed in the table. As the last step, a snowballing process was performed, leading to a final set of \textbf{48 gray literature}. 

\subsection{Data Extraction and Thematic Analysis} 
Thematic analysis is a qualitative research methodology widely used in systematic literature reviews to understand patterns within the extracted data by identifying, analyzing, and reporting themes \cite{dissanayake2022, croft2022}. The thematic analysis begins with researchers familiarizing themselves with the data to gain a comprehensive understanding. Based on this understanding, they systematically create and assign \textit{codes} to the data to signify interesting features and potential themes. The codes are then sorted into \textit{themes}, which capture significant patterns in the data. Finally, themes are reviewed, accompanied by clear definitions and names for reporting. We followed the guidelines for thematic analysis established by Braun et al. \cite{thematicAnalysisBraun} to carry out the analysis.

\textit{Understanding:} The process commenced with all three authors engaging in data extraction to facilitate an in-depth understanding of the source material. We performed a pilot data extraction on six papers. To enhance inter-rater agreement, we discussed the results to refine the data extraction process and the predefined data-extraction form \cite{garousi2019guidelines, kitchenhamSLR2007}. Subsequently, data extraction was carried out separately, with the primary studies divided among the authors.

\textit{Coding:} The coding process began after the completion of data extraction. We utilized a shared spreadsheet to maintain an up-to-date Code Book. Additionally, we maintained an Obsidian \footnote{https://obsidian.md/} vault to supplement the details in the Code Book within the spreadsheet. This vault contains a collection of notes, each providing a detailed description of a code along with related data extracted from the primary studies. These artefacts were shared and version-controlled, enabling authors to coordinate, validate the generated codes, and reuse codes when applicable. Each of the three authors focused on coding for one specific research question, with biweekly catch-up meetings to synchronize our efforts.

\textit{Theme generation:} We analyzed the generated codes and combined them to create a hierarchy of candidate themes (including themes, sub-themes, and sub-sub-themes). Each author independently generated themes corresponding to their coded research question. The other two authors then reviewed the generated themes for validation and refinement. Finally, we defined the final names and definitions of the refined themes before reporting the results.

\subsection{Threats to Validity} \label{threatsToValidity}
Potential threats to the validity of our study primarily concern the processes of study search, study selection, data extraction, and theme analysis. 
We collected studies from well-known academic databases and general search engines to ensure a comprehensive selection of the studies. Overview of selected studies and their distributions over time is available at our online appendix \cite{onlineAppendix}.
While going through search results, we excluded non-free resources, which is a limitation of our study.
Additionally, we applied backward and forward snowballing techniques to identify potentially missed studies.
We designed search strings to ensure broad coverage while avoiding restrictive keywords. Complex queries including many keywords were used in academic databases, while simpler queries were applied for gray literature to accommodate search engine limitations. This approach minimized the risk of omitting critical studies while maintaining relevance to MLOps.
To ensure consistency in applying inclusion/exclusion criteria and quality assessment, we conducted pilot evaluations. All authors independently assessed randomly selected studies and discussed results to align on terminology and evaluation standards. In cases of uncertainty, decisions were made through consensus, ensuring reliability in the quality assessment process.
We used a standardized template to extract key study details (e.g., objectives, methodologies, findings) \cite{onlineAppendix}. A pilot extraction phase helped align understanding, with discrepancies resolved through discussion.
For theme analysis, we employed an iterative coding process where three authors independently coded the data, resolving disagreements through discussion.

\section{RQ1: Conceptualization of MLOps}  \label{RQ1}

\subsection{RQ1.1 - How is MLOps Defined in the Literature?}

Eight types of definitions for MLOps have been identified in the existing literature. These definitions can be considered the conceptual foundation for MLOps work, as they encompass what authors of the work perceive to be the primary challenges in ML productionalization, thereby delineating the scope of MLOps solutions (e.g., best practices, tools) to be proposed.

\begin{figure}[t] 
   \centering
   \includegraphics[width=0.8\linewidth]{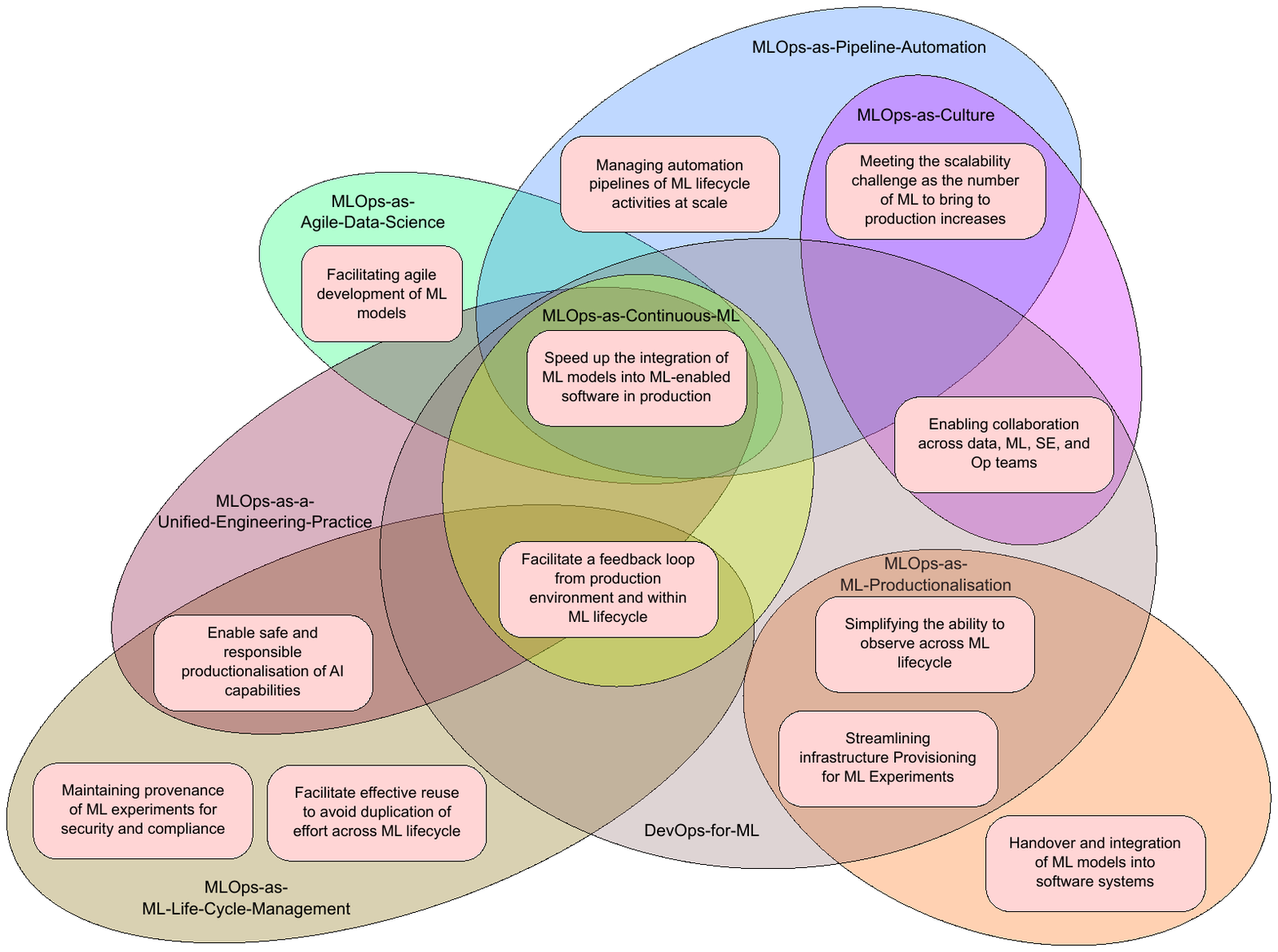}
   \caption{MLOps definitions and the ML productionalization problems they address}
   \label{fig:RQ11} 
\end{figure}

The following subsections introduce the identified types of MLOps definitions, ranging from the most common to the least common. For example, 34 studies define MLOps as ``continuous ML,'' 32 refer to it as ``DevOps for ML,'' and 18 studies describe it as encompassing cultural and organizational practices for ML. We provide the complete statistics in our online appendix \cite{onlineAppendix}. Figure \ref{fig:RQ11} visualizes the relationship between eight definition types based on the types of ML productionalization problems on which they focus.

\subsubsection{\textbf{MLOps-as-Continuous-ML}} 
This definition considers \textbf{MLOps as the integration of continuous engineering practices (e.g., Continuous Integration (CI) and Continuous Delivery (CD)) into the lifecycle of ML-enabled software systems}. The goal of MLOps-as-Continuous-ML is to enable rapid development and operations through continuous practices, allowing for the swift adaptation of existing solutions to changing conditions (e.g., business needs, concept shifts) \customcitePS{MLOpsMicrosoft}, \customcitePS{MLOpsAWS}, \customcitePS{GoogleCloud}. Moreover, continuous practices facilitate feedback loops between different phases critical in ML projects due to the experimental nature of ML engineering. For example, ML models are often updated after the initial monitoring phase, and feature engineering alongside model serving tasks are repeated to accommodate varying model needs \customcitePS{FIRST-10}.

The concept of MLOps-as-Continuous-ML is inspired by the well-known software engineering practice of CI/CD (continuous integration and delivery), which aims to integrate software to deliver high-quality products frequently (see Appendix \ref{Background-DevOps}), particularly emphasizing the parallels between delivering software to production and delivering ML to production. Continuous training (CT) \customcitePS{FIRST-GL-15} is often introduced as the ML-specific element. MLOps-as-Continuous-ML integrates continuous practices into many phases of the ML lifecycle \customcitePS{FIRST-WA-5}, including model design \customcitePS{FIRST-5}, model training \customcitePS{FIRST-1}, serving \customcitePS{FIRST-GL-1}, validating \customcitePS{FIRST-GL-26}, monitoring \customcitePS{FIRST-5}, model delivery \customcitePS{FIRST-28}, and the integration of pipelines \customcitePS{GoogleCloud}.

\subsubsection{\textbf{DevOps-for-ML}} 
This definition considers MLOps as \textbf{an extension of DevOps practices for ML}. It should be noted that while continuous engineering practices are a key and perhaps the most well-known element of DevOps, the scope of DevOps itself extends beyond this to encompass practices such as streamlining infrastructure provisioning via Infrastructure-as-Code, facilitating data-driven decision-making by providing developers with on-demand observability, as well as fostering the cultural shifts necessary to enable these cross-silo activities. These cultural shifts might include promoting collaboration between development and operations teams, encouraging a mindset of continuous improvement, and embracing automation.

MLOps originates from the DevOps concept \customcitePS{FIRST-1}, \customcitePS{FIRST-29}, \customcitePS{FIRST-32}, \customcitePS{I41} and targets the development (dev) and operationalization (ops) of ML systems \customcitePS{FIRST-30}, \customcitePS{FIRST-GL-28}, \customcitePS{GoogleCloud}. MLOps adopt practices and cultural aspects associated with DevOps \customcitePS{FIRST-32}, \customcitePS{MLOpsAWS}, \customcitePS{S243}, aiming to standardize and streamline the development, deployment, operation, and management of ML systems \customcitePS{AwsMLOps2022}, \customcitePS{IbmMLOps}. MLOps unifies the development and operational cycles carried out by different roles, ensuring the automated, frequent, and continuous delivery of high-performing ML systems \customcitePS{FIRST-GL-1}.

\subsubsection{\textbf{MLOps-as-Pipeline-Automation}} 
This definition considers MLOps as \textbf{the practice of automating the repetitive activities in the ML lifecycle by constructing and managing automated pipelines}. Details regarding ML lifecycle activites would be discussed in section \ref{RQ1.2-Activities}. Some benefits of automation include more rapid development and deployment \customcitePS{FIRST-28} and easier management of the infrastructure to which ML models are deployed \customcitePS{FIRST-GL-7}. MLOps advocates for automation \customcitePS{GoogleCloud} at all stages of ML system development and operations, including activities such as data collection, model building and deployment, and monitoring \customcitePS{FIRST-1}, \customcitePS{FIRST-17}, \customcitePS{FIRST-30}. Automation simplifies deployment by providing a more straightforward release and environment configuration process, allowing researchers to focus more on model optimization \customcitePS{FIRST-44}.

\subsubsection{\textbf{MLOps-as-Culture}} 
Developing and operationalising an ML-enabled system necessitates collaboration between many different roles, e.g., business analysts, ML engineers, and software engineers. As the number of ML applications and their complexity grows in the community, the need for establishing a standard for new roles and responsibilities and organisational collaboration practices becomes more prevalent. This definition considers MLOps as \textbf{the set of tools and managerial practices that lead to more sustainable development and operationalisation of ML-enabled systems} \customcitePS{FIRST-GL-5} through enabling collaboration across diverse teams, e.g., data science team and IT professionals \customcitePS{FIRST-45}. 

\subsubsection{\textbf{MLOps-as-Agile-Data-Science}} 
As emphasized by several studies \customcitePS{FIRST-GL-16},  \customcitePS{MLOpsMicrosoft}, \customcitePS{FIRST-GL-26},  \customcitePS{FIRST-WA-6}, MLOps shares a similar philosophy with DevOps, but has unique aspects related to data and model. Further, MLOps is closely linked with agility in data science \customcitePS{FIRST-30}, \customcitePS{FIRST-32}, \customcitePS{FIRST-36}, \customcitePS{FIRST-GL-19}. Specifically, producing and releasing ML models iteratively is the application of agile principles \cite{fowler2001agile} in ML projects. Hence, this definition describes MLOps as \textbf{the intersection between data engineering, data science, machine learning, and operation to facilitate an agile end-to-end practice}.


\subsubsection{\textbf{MLOps-as-ML-Productionalisation}} 
This definition considers MLOps as \textbf{the process of operating ML models, which involves bringing them into a production environment} \customcitePS{FIRST-7}. The purpose of MLOps from this perspective is to minimize the gap between experimental development and the production environment \customcitePS{I45}, as well as to facilitate the transfer of developed ML models to production at scale by ensuring reproducibility, reliability, and efficiency \customcitePS{FIRST-GL-6}, \customcitePS{FIRST-23}. Research based on this definition of MLOps focuses on the processes, standards, and tools for packaging and integrating ML models into software systems, as well as on streamlined workflows from development to operationalization \customcitePS{FIRST-22}, \customcitePS{FIRST-41}, \customcitePS{FIRST-GL-29}, \customcitePS{FIRST-WA-6}.

\subsubsection{\textbf{MLOps-as-ML-Life-Cycle-Management}} 
This definition considers MLOps as \textbf{the set of practices and tools to manage lifecycle activities of MLOps and their corresponding artefacts to ensure reproducibility, stability, optimization, and scalability}. One focus of MLOps research from this perspective is the handling of triggers, loops, and common artefacts that link various MLOps lifecycle activities \customcitePS{FIRST-SB-13}, such as managing multiple versions of experiments, datasets, models, and metadata related to models \customcitePS{I55}. Another focus is the management, standardization and optimization of workflows across lifecycle activities \customcitePS{FIRST-GL-33}, \customcitePS{FIRST-WA-4}, \customcitePS{FIRST-WA-5}. Another focus is permeating the lifecycle with responsible practices such as fairness of predictions \customcitePS{AmazonAI}.

\subsubsection{\textbf{MLOps-as-a-Unified-Engineering-Practice}} Several studies define MLOps as \textbf{unifying ML-enabled systems' development, deployment, and operations}. This definition refers to supporting practices for developing \customcitePS{FIRST-GL-29}, deploying \customcitePS{GoogleCloud}, monitoring \customcitePS{FIRST-25}, and scaling \customcitePS{FIRST-GL-26} ML models quickly and efficiently, and managing their infrastructure configuration complexity \customcitePS{FIRST-GL-23} to ensure ML-enabled systems operate optimally and securely \customcitePS{FIRST-GL-23} within the predefined business thresholds \customcitePS{FIRST-GL-26}.

\subsection{RQ1.2 - What are the Primary Activities and Tasks of MLOps?}\label{RQ1.2-Activities}
\begin{figure*}[h] 
   \centering
   \includegraphics[width=1\linewidth]{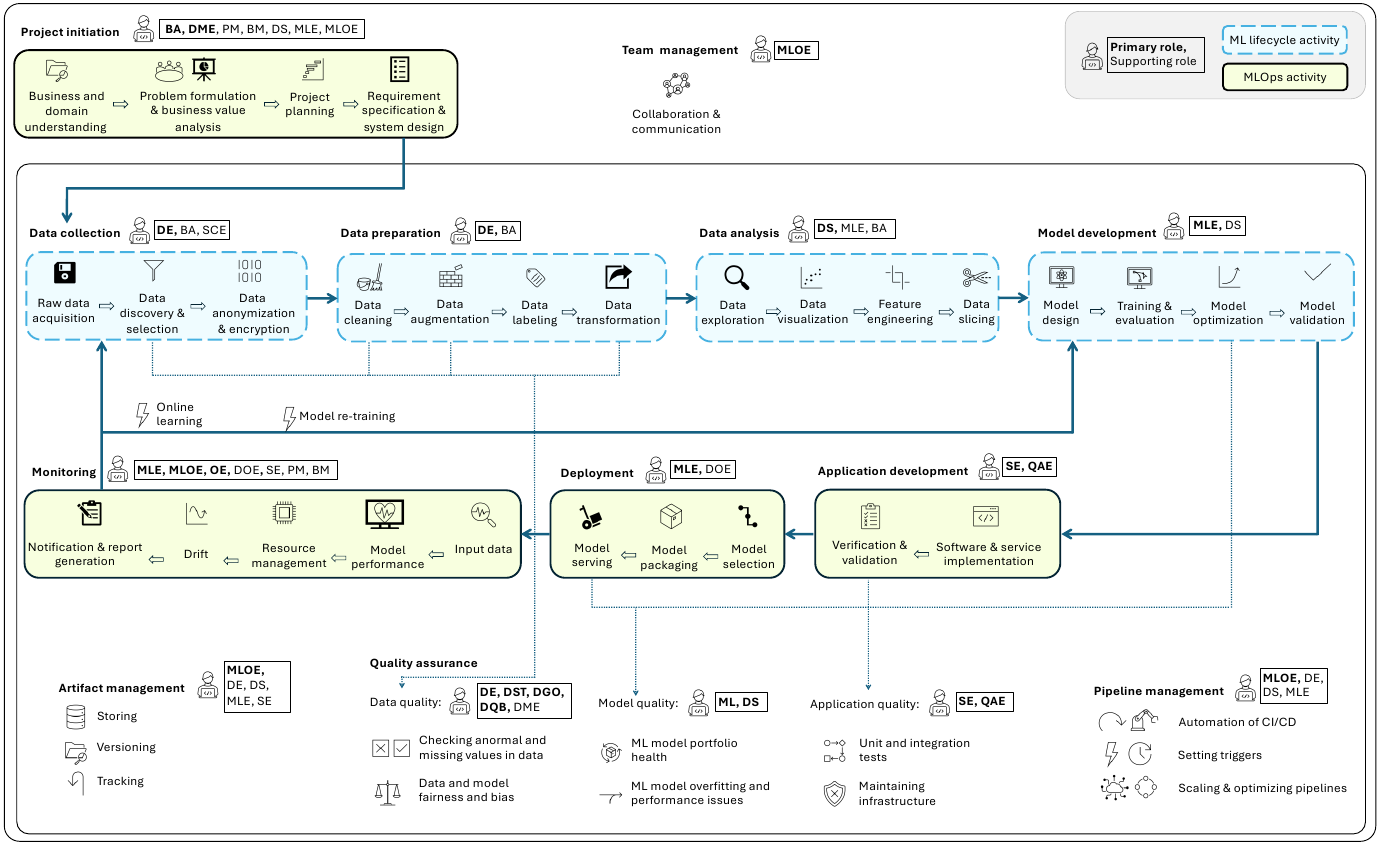}
   \caption{A pipeline for ML development and operations, and associated roles: Domain Experts (DME); Business Analysts (BA); Data Engineers (DE); Data Scientists (DS); ML Engineers (MLE); MLOps Engineers (MLOE); DevOps Engineers (DOE); Software Engineers (SE); Quality Assurance Engineers (QAE); Operational Engineers (OPE); Technical and Solution Architects (ARC); Security Experts (SE); Business Owners/Managers (BM); Product Owners/Managers/Experts (PM); Data Governance Officer (DGO), Data Quality Board (DQB), Data Steward (DST)}
   \label{fig:pipeline} 
\end{figure*}

Figure \ref{fig:pipeline} shows a comprehensive MLOps pipeline that includes core MLOps activities and roles associated with them. We elaborate more on the activities in this section and roles in section \ref{RQ1.3-RolesAndResponsibilities}.

\subsubsection{\textbf{ML project initiation}}
ML project initiation mainly includes activities such as business analysis, domain understanding, project planning, specifying system requirements, and ML system design. Many stakeholders take part in these activities, i.e., business analysts, domain experts, data engineers and scientists, and ML and MLOps engineers. 
Specification of business and system requirements is crucial for planning and execution of an ML project, and successful deployment and operation of the ML system \customcitePS{I8}, \customcitePS{AwsLens}.
The process begins with formulating the business problem to establish a clear objective \customcitePS{FIRST-41}, helping assess business value \customcitePS{AwsLens}, determine the feasibility of engineering an ML system \customcitePS{G3}.
Further steps include obtaining security objectives of the company, e.g., confidentiality and privacy requirements of the data used through ML system \customcitePS{I8}, and regulatory rules \customcitePS{FIRST-WA-15}.
Moreover, ML system planning incorporates software project management knowledge areas \cite{washizaki2024guide} such as defining project scope \customcitePS{FIRST-GL-23}, specifying ML use cases \customcitePS{FIRST-GL-5}, designing model architecture, pipeline, and infrastructure \customcitePS{FIRST-GL-23}, \customcitePS{FIRST-GL-5}, identifying performance metrics for model evaluation \customcitePS{AwsLens}, and establishing monitoring strategies with feedback mechanisms for business owners \customcitePS{FIRST-GL-5}.

\subsubsection{\textbf{ML lifecycle activities}}

\textbf{Data collection} is the first step in the ML pipeline, involving the \textbf{acquisition of raw data} that will be processed in later steps to train a ML model. Often, raw data is gathered from relevant data sources, e.g., sensors \customcitePS{FIRST-24}, GitHub code repositories \customcitePS{FIRST-27}, or cameras \customcitePS{FIRST-34}. 
Some studies have more exploratory data collection steps. For example, data collection is combined with \textbf{data discovery} and \textbf{data selection}, i.e., collecting data for a drug discovery task requires performing experiments to produce data, followed by selection of data by domain experts to be used in later phases \customcitePS{FIRST-37}. 
Alternatively, some cases follow a data management plan prepared by data engineers to collect data to be fed into the ML pipeline \customcitePS{I40}.
For security-critical systems, data collection process is applied with \textbf{anonymization and encryption} of the collected information to ensure data privacy \customcitePS{FIRST-31}.

\textbf{Data preparation} aims to process raw data for ML model building through cleaning the raw data, augmentation, labeling data points, transformation, and feature engineering. 
\textbf{Data cleaning} involves detecting and fixing errors in raw data, such as removing duplicates \customcitePS{FIRST-24}, filtering invalid and irrelevant data points \customcitePS{FIRST-27}, imputing missing values in data \customcitePS{FIRST-37}. This step is also known as \textbf{ data filtering} in the literature \customcitePS{FIRST-27}.
\textbf{Data augmentation} increases sample size to address limited data availability, with techniques like oversampling and simulation \customcitePS{FIRST-5}.
\textbf{Data labeling} marks data points according to observations in the problem domain, involves data engineers, data scientists, and domain experts \customcitePS{I40}. While data labeling is a manual task in many cases, tools also support automated labeling  \customcitePS{AwsLens}.
\textbf{Data transforming} adjusts the data format according to the needs of the ML system to be built, for example, removing characters for compatibility with other file systems \customcitePS{FIRST-27}.
It also includes data representation changes, like distributional transformations on numerical data and data discretization for smaller units \customcitePS{FIRST-31}.
           
\textbf{Data analysis} includes \textbf{data exploration} based on data statistics, e.g., the mean and standard deviation, and quantiles\customcitePS{FIRST-30}, \customcitePS{FIRST-SB-14} to better understand the data before model development. It helps to understand and assess which features \customcitePS{AwsApn} and data slices \customcitePS{FIRST-30} best meet ML model requirements.
Data analysis also includes \textbf{data visualization} providing dashboards to facilitate the team's evaluations of data \customcitePS{FIRST-25}, \customcitePS{FIRST-42}, \customcitePS{DataBricksMLOps}, \customcitePS{Canonical}.
Moreover, data analysis supports data validation, which assesses the health of data by identifying anomalies or unexpected properties that could impact model quality \customcitePS{FIRST-SB-14}.
\textbf{Feature engineering} processes features to enhance model performance and accuracy \customcitePS{FIRST-SB-16}, \customcitePS{S262}. It includes sub-steps such as feature transformation, feature selection, feature extraction, feature construction, and feature storage \customcitePS{I89}, \customcitePS{DataBricksMLOps}, \customcitePS{IbmMLOps}, \customcitePS{GoogleCloud}. Feature engineering is mostly seen as the responsibility of data engineers and scientists but also uses the expertise of other roles such as security \customcitePS{I8} and product experts \customcitePS{FIRST-31}.
Another mentioned activity is \textbf{data slicing} \customcitePS{I171}, which is the process of finding the subset of data that performs according to the use case. It overlaps with model building and evaluation steps since it explores the interest in data through evaluating the ML model's prediction performance on different data slices. 

\textbf{Model development} is a core step in ML life-cycle, focused on constructing ML models. It includes model design, training, evaluation, selection, validation, and re-training. Model building partially overlaps with data analysis as it often involves exploring various training approaches, i.e., ML algorithm and hyperparameter configurations \customcitePS{DataBricksMLOps}.
\textbf{Model design} refers to designing and selecting the appropriate architecture and tools for ML model construction to ensure successful model training and prediction performance. Examples include designing neural network architectures \customcitePS{FIRST-39}, choosing the right algorithm for the problem \customcitePS{FIRST-24}, and optimizing ensemble models \customcitePS{FIRST-31}.
\textbf{Model training} builds the ML model for prediction, using a learning algorithm to train it on the prepared data \customcitePS{AwsLens}, \customcitePS{I9}. 
\textbf{Model evaluation} assesses predictive performance of the ML model, by testing it with a held-out test set, which includes a separate set of data points not used in training \customcitePS{FIRST-29}.
Prediction performance is measured using relevant metrics for the use case \customcitePS{I18}.
\textbf{Model optimization} improves the model's prediction performance. Hyperparameter tuning is a common method applied \customcitePS{FIRST-24}, \customcitePS{GoogleCloud}.
\textbf{Model validation} evaluates the model based on business goals \customcitePS{FIRST-GL-5}, quality standards \customcitePS{FIRST-SB-14}, and regulatory requirements \customcitePS{DataBricksAuto}.
The common method for validating the quality of an ML model is comparing its predictive performance against a baseline model or a fixed threshold value \customcitePS{FIRST-SB-14}, \customcitePS{S33}, \customcitePS{GoogleCloud}. 

\subsubsection{\textbf{Application development}} step addresses the implementation of software and/or service that will be interfaced with the deployed ML solution \customcitePS{AwsMLOps2022}, \customcitePS{S57}. 
Additionally, this step involves verification, validation, and quality assurance activities of the developed application such as unit testing, integration testing, refactoring, maintenance to ensure reliability, functionality, and seamless interaction with the ML model \customcitePS{I40}, \customcitePS{I53}.

\subsubsection{\textbf{Deployment}} 
The deployment phase integrates trained ML models into production environments, making model predictions accessible to users. Depending on the use case, deployment is achieved by integrating the ML model into a software system \customcitePS{I40}, serving it as a service \customcitePS{FIRST-14}, or creating containerized images of ML models \customcitePS{I41}. 
Deployment phase differs from routine ML activities \customcitePS{S66}, as it involves transferring the model to another team or infrastructure \customcitePS{DataBricksAuto}.
As Neghawi et al. \customcitePS{I40} note, "the deployment stage is a pivotal moment in the MLOps process," due to the transition from an experimental (ML development) to an actualized environment (ML production).
Key deployment activities include \textbf{model selection} for choosing the most appropriate model among various models built for the same ML system, i.e., the best performing one \customcitePS{I5}, or based on client's requirements \customcitePS{I64}. 
Deployment of a model is decided based on model validation results, which often involve supervision of a manager \customcitePS{FIRST-34}, and collaboration of data scientists, business stakeholders, and ML engineers \customcitePS{DataBricksMLOps}. 
Another key activity is \textbf{model packaging} for reproducibility and sharing \customcitePS{FIRST-24}, creating containerized model images with dependencies for inference \customcitePS{FIRST-GL-5} 
, and selecting deployment methodologies based on the target ML system \customcitePS{FIRST-24}, \customcitePS{I16}. 
\textbf{Model serving} refers to making the ML model accessible for generating new predictions. A common approach is embedding the model within an application for user queries \customcitePS{FIRST-29}. Many studies favor the \textit{Model-as-a-Service} paradigm, where models are exposed through REST API endpoints for scalability and ease of access \customcitePS{FIRST-29}, \customcitePS{I41}. 
Deployment requires collaboration from ML and MLOps engineers, as well as software engineers, to ensure smooth integration into a production environment.

\subsubsection{\textbf{Monitoring}} 
Monitoring is a key MLOps activity applied by many studies \customcitePS{FIRST-26}, \customcitePS{FIRST-32} and supported by many tools \customcitePS{AmazonAI},\customcitePS{DataBricksMLOps}, \customcitePS{Canonical}, \customcitePS{RedHatOpenShiftAI}. It helps improve system health by providing feedback, often triggering the repetition of relevant MLOps activities. 
Ensuring operational stability and quality of an ML system in production is done by continuously observing data and model changes, model performance, system resources, application and its dependencies to capture and report the issues in a timely manner \customcitePS{S46}, \customcitePS{I40}. 
Input data changes, data and model biases, and concept drifts can be detected setting pre-defined thresholds and rules on data \customcitePS{AwsLens}, or monitoring the \textbf{model performance} \customcitePS{RedHatOpenShiftAI}.
When a change, bias, or drift is captured through monitoring, it often results in adjusting the dataset for model training and/or model optimization \customcitePS{S261}.
Captured model performance degrades invokes model re-training \customcitePS{DataBricksMLOps} or relevant ML processes \customcitePS{I41}. 
Other common monitored subjects are \textbf{resource usage}, e.g., memory usage and number of API calls \customcitePS{FIRST-24}, non-functional requirements like \textbf{security} \customcitePS{FIRST-32}.
Monitoring changes around system dependencies helps providing notifications regarding specific modifications \customcitePS{Canonical}.
Information collected during monitoring is stored and used to produce \textbf{notifications and reports} of important situations which often leads to an invocation of earlier tasks in the pipeline \customcitePS{AmazonAI}.
Dashboards \customcitePS{DataBricksMLOps} and reports \customcitePS{GoogleCloud} are common methods for presenting this information. 
Moreover, monitoring enhances users' and business stakeholders' trust by enabling system observability \customcitePS{I170}, 
e.g., providing insights on how pipeline tasks are executed \customcitePS{FIRST-GL-33}.
While MLOps engineers are primarily responsible for monitoring \customcitePS{I171}, \customcitePS{AwsLens}, other roles, such as product performance experts, site reliability engineers, and solution engineers, also utilize monitoring data to analyze issues \cite{FIRST-31}. In more mature systems and organizations, such as those in highly regulated industries like financial services \customcitePS{FIRST-WA-15}, monitoring also includes explainability features to enhance understanding of model predictions, including decision visualizations \customcitePS{FIRST-17} and rule-based root-cause analysis \customcitePS{FIRST-23}.

\subsubsection{\textbf{Pipeline management}}
An MLOps pipeline consists of interconnected ML life-cycle tasks and operations, which can become easily complex due to their dependencies. Pipeline management addresses pipeline design, orchestration, and optimization to execute MLOps tasks efficiently and reliably.
\textbf{Pipeline automation} minimizes human intervention by executing MLOps tasks automatically, simplifying repetition of complex ML operations \customcitePS{FIRST-42}. Studies have explored automation for different use cases, such as pipelines for data downloading, processing, model training, and testing \customcitePS{FIRST-3}, or for model configuration, serving, and monitoring \customcitePS{FIRST-44}.
Automation relies on \textbf{pipeline triggers} to enable task execution. Triggers can be sequential, i.e., executing a task upon the previous task's completion \customcitePS{FIRST-42}, periodical e.g., executing according to a pre-scheduled period \customcitePS{FIRST-25}, event-driven, e.g., deployment can be triggered when a new ML model is available \customcitePS{RedHatOpenShiftAI}, e.g., training can be triggered when new data is available or the model performance is reduced \customcitePS{I41}\customcitePS{I41}, or a concept drift is detected \customcitePS{GoogleCloud}. 
Pipeline automation is often tailored to specific use cases. For instance, one study automates data downloading, processing, model training, and testing \customcitePS{FIRST-3}, while another focuses on automating model configuration, serving, and monitoring \customcitePS{FIRST-44}.
Moreover, \textbf{scaling pipelines} is essential in today's infrastructures such as diverse environments including edge devices and multi-tenant platforms. For example, data pipelines must adapt to manage the evolving customer requirements \customcitePS{FIRST-42}, tools like Airflow support scalable, dependency-aware pipelines \customcitePS{Airflow}.

\subsubsection{\textbf{Artifact management}} 
Versioning artifacts is crucial in MLOps projects just as in traditional software engineering. Common MLOps artifacts are code, data, feature, models, experiments, metadata, and project documents.
Managing artifacts involves keeping a record of history and different versions of artifacts. It facilitates collaboration among team members \customcitePS{FIRST-33}, enables debugging and system backups \customcitePS{S33}, supports artifact reuse \customcitePS{FIRST-32}, ensures auditability for quality and risk assessments \customcitePS{FIRST-24}, \customcitePS{S262}, and enhances project transparency \customcitePS{FIRST-41}.
Studies focus on versioning various MLOps components, including data science experiments \customcitePS{I22}, codes \customcitePS{FIRST-24}, ML models \customcitePS{FIRST-SB-17}, training metadata such as dataset versions \customcitePS{S262}, parameter weights used for model training \customcitePS{S33}, tracking software versions running with the ML model \customcitePS{FIRST-41}, logs of pipeline executions \customcitePS{S33}, logs of monitoring events \customcitePS{FIRST-33}, and logs of container data \customcitePS{FIRST-GL-23}.
Further, data \customcitePS{FIRST-24} and model \customcitePS{S261} provenance are stored which are the records that keep the origin and lineage of data and model in ML enabled systems \customcitePS{FIRST-SB-18}.

\subsubsection{\textbf{Quality assurance}}
Quality assurance covers ensuring data quality, model quality, and ML application's quality.
Quality is managed across various MLOps phases, but data quality is closely associated with data cleaning, analyzing data, model training, model selection, and monitoring phases \customcitePS{FIRST-24}. Model quality is managed through data collection, model training and evaluation, and monitoring phases \customcitePS{AwsLens}. Moreover, ML application's quality is ensured through traditional software testing strategies \customcitePS{FIRST-WA-17}. 
Data quality involves assessing the scientific quality of data with domain experts, e.g., whether data includes realistic values, correcting data anomalies, checking data completeness, e.g., addressing problem-related missing values, checking consistency in data format and structure, data heterogeneity, and fairness of collected data \customcitePS{FIRST-SB-14}, \customcitePS{FIRST-24}, \customcitePS{AmazonAI}. 
Model quality involves assessing the model's predictive performance through performance metrics and test sets \customcitePS{FIRST-21}, validating model bias and fairness \customcitePS{FIRST-WA-17}, maintaining ML portfolio health \customcitePS{FIRST-GL-23}, monitoring model to detect overfitting issues \customcitePS{FIRST-SB-11}, dimensionality reduction \customcitePS{FIRST-24}.
Application quality involves unit testing, acceptance testing \customcitePS{FIRST-GL-29}, maintaining containerized images and identifying vulnerabilities \customcitePS{FIRST-GL-23}.

\subsubsection{\textbf{Team management}}
Various people from different backgrounds, such as ML engineers and software engineers, need to work together in MLOps settings \customcitePS{I109}. 
Facilitating communication among these individuals is crucial to maintaining collaboration and ensuring smooth operation of ML system to achieve business goals \customcitePS{I8}. Activities associated with team management include clear identification of requirements, planning the project, ensuring transparency in project documentation and model evaluation, involving individuals with relevant expertise in meetings and negotiations, and investing in education of team members \customcitePS{I109}.

\subsection{RQ1.3 - What are the Practitioner Roles and Responsibilities in MLOps?}\label{RQ1.3-RolesAndResponsibilities}

\textbf{Domain experts} possess deep knowledge of the problem domain, such as clinical neurologists for an ML-enabled psychological monitoring system \customcitePS{I68} or urban planners for a smart city system \customcitePS{FIRST-30}.
They are described as ``irreplaceable actors within ML projects'' \customcitePS{FIRST-24} due to their crucial role in domain discovery \customcitePS{FIRST-41}, engineering domain specific requirements \customcitePS{FIRST-24}, and fulfilling business goals \customcitePS{FIRST-41}, \customcitePS{I40}. 
Moreover, they contribute to data collection, cleaning, labeling, hypothesis development, model evaluation, and result interpretation \customcitePS{FIRST-32}, \customcitePS{I109}, \customcitePS{AwsLens}. 
They either provide domain knowledge and processed data to the MLOps team in the early stages of a project and/or collaborate directly with data engineers, data scientists and ML engineers during the ML phases to validate data and model \customcitePS{FIRST-24}, \customcitePS{I55}, \customcitePS{I68}, \customcitePS{I168}.

\textbf{Business analysts} define the problem, set clear objectives, and assess whether an ML-enabled system can provide a solution to the problem at hand \customcitePS{FIRST-GL-26}, \customcitePS{I28}, \customcitePS{FIRST-WA-9}.
MLOps team collaborates with business analysts to gain a better understanding of the problem and convert it to a successful ML solution that aligns with the business goals.
Their role may overlap with domain experts, as both contribute to understanding the problem domain. Business analysts also define business requirements, key performance indicators (KPIs), and model quality metrics, leveraging query processing on structured datasets to analyze business problems \customcitePS{FIRST-GL-26}, \customcitePS{NVIDIA2023}.

\textbf{Data engineers} are mainly responsible for fundamental data activities, enabling data operations in a high-quality, secure, and scalable way \customcitePS{FIRST-1}, 
\customcitePS{MLOpsAWS}. Data engineers' typical set of tasks include identifying suitable data sources, verifying data legality, collecting data from various sources, cleaning them, and preparing labeled data \customcitePS{FIRST-5}, \customcitePS{FIRST-SB-3}, \customcitePS{FIRST-WA-3}. 
This data engineering process is characterized as "extract-transfer-load" which involves transferring data into consumable formats for data scientists and ML engineers \customcitePS{AwsApn}, \customcitePS{AwsSecure}. 
Beyond data preparation, data engineers set up and manage data storage infrastructures, build automated data pipelines \customcitePS{FIRST-1}, and enable big data storage solutions such as centralized data lakes \customcitePS{FIRST-GL-4} and streaming platforms \customcitePS{I66}. Their role extends to security, privacy, quality, and governance of data \customcitePS{FIRST-GL-26}, \customcitePS{AwsSecure}. 
Some organizations define specialized roles, such as \textbf{data curator}, responsible for custom data acquisition \customcitePS{FIRST-19}, and \textbf{data expert}, applying noise reduction techniques \customcitePS{FIRST-5}.

\textbf{Data scientists} are mainly responsible for exploring and analyzing the data made available by data engineers, training, tuning, and evaluating models to derive insights \customcitePS{I22}, \customcitePS{MLOpsAWS}.
Particularly, their initial task is to understand the business problem and its domain, evaluate the feasibility of an ML-enabled solution \customcitePS{FIRST-24}, \customcitePS{DataBricksMLOps}. At this step, data scientists work with domain experts and business analysts to discuss objectives and key results (OKRs) \customcitePS{FIRST-GL-26}.
Data scientists also conduct feature engineering to identify the best combinations of features for training effective ML models \customcitePS{S58}. Tasks such as preparing training, test, and validation sets, choosing suitable ML algorithms and training approaches, and selecting appropriate evaluation metrics (e.g., accuracy) are also carried out by data scientists \customcitePS{AwsApn}. 
Moreover, as data engineers build the initial part of the ML pipeline, data scientists are responsible for designing and building scalable and high quality data analytics pipelines for experimentation and model training \customcitePS{DataBricksMLOps}. Additionally, they ensure the trackability of multiple versions of experiments, datasets, and models \customcitePS{I22}. 
Also, they monitor the early results after the deployment, collaborating with business analysts and ML engineers, whether the model meets technical, business, and regulatory requirements \customcitePS{DataBricksMLOps}. If necessary, data scientists iterate on the model to improve its performance \customcitePS{FIRST-WA-3}.

\textbf{ML engineer} and data scientist roles often overlap, with some studies suggesting they are interchangeable \customcitePS{I36}, \customcitePS{I41}. However, many studies distinguish the role of the ML engineers with their primary tasks being model optimization, deployment, and monitoring \customcitePS{DataBricksMLOps}, \customcitePS{NVIDIA2023}. Whereas, experimenting with datasets and models, and optimizing the model and its parameters are tasks shared with data scientists \customcitePS{I48}, \customcitePS{luley2023concept}.
In particular, ML engineers contribute to monitoring and production activities, collaborating with data scientists and business stakeholders to ensure the deployed model meets performance and business requirements \customcitePS{DataBricksMLOps}, \customcitePS{AwsMLOps2021}. 
They are also responsible for executing appropriate deployments, such as online serving, and monitoring the model's performance in the production environment \customcitePS{DataBricksMLOps}, \customcitePS{AwsMLOps2021}.

\textbf{MLOps engineers} are responsible for building automated pipelines, maintaining operational stability of the entire ML system, and facilitating collaboration across roles \customcitePS{AwsSecure}, \customcitePS{I40}.
While data engineers, data scientists, and ML engineers design and implement the pipelines for data collection, analysis, model development, and deployment, MLOps engineers focus on building, optimizing, and integrating these pipelines, including services like cloud and containers \customcitePS{AwsSecure}. They manage dependencies, versioning, tracking mechanisms, and platform governance. MLOps engineers also monitor data, models, and applications along with ML engineers \customcitePS{I40}, and step in when human intervention is needed—such as when re-training is required due to data shifts or resolving dependencies between data and models, \customcitePS{I171}. Effective communication and collaboration management across roles during these tasks is also the responsibility of MLOps engineers \customcitePS{I8}.

\textbf{DevOps engineers} also play a role in ensuring the seamless deployment, integration, and stability of machine learning models within production systems. 
Their responsibilities often overlap with those of MLOps engineers, operational engineers, and software engineers \customcitePS{FIRST-SB-7}, \customcitePS{FIRST-GL-4}.
While some studies indicate DevOps engineers in traditional software projects are replaced with MLOps engineers for ML projects \customcitePS{Canonical}, a few mention DevOps and MLOps engineers work within the same team.
For example, Neghawi et al. \customcitePS{I40} report that while software engineers develop and validate applications, DevOps engineers construct, test, and deploy working systems, and MLOps engineers continuously monitor the entire system to ensure operational stability.  

\textbf{Software engineers} develop the final applications and services that support ML production. Application developer is another common reference to the same role \customcitePS{FIRST-1}.
They often collaborate with data scientists and ML engineers in data processing and analysis, and model building \customcitePS{IbmMLOps}, \customcitePS{I109}. Besides, deployment, packing, releasing, configuring, and monitoring of ML systems are also reported as the role of software engineers \customcitePS{S43}, \customcitePS{I55}, \customcitePS{NVIDIA2023}. 
Additionally, \textbf{quality assurance engineers} are responsible for testing the software application and services to ensure product quality \customcitePS{FIRST-36}.

\textbf{Operational engineers} or operational team are not consistently referenced across all studies. But some highlight their role as taking over after data scientists and/or ML engineers hand over the ML model \customcitePS{FIRST-24}. They are responsible for deployment operations including packaging, releasing, and configuring ML systems \customcitePS{I55} and continuous monitoring of the model's performance \customcitePS{S58}. 
They serve as a bridge between data science and software engineering teams, facilitating iterative improvements by reporting issues, providing feedback, and ensuring the model remains efficient and reliable in real-world applications \customcitePS{FIRST-GL-1}. In the absence of explicit mention of operational engineers, these tasks are carried out by ML \customcitePS{DataBricksMLOps}, and sometimes software \customcitePS{I55}, and MLOps engineers \customcitePS{AwsSecure}.

\textbf{Technical and solution architects} are responsible for designing and building ML systems and services that serve as the backbone of ML operations in production \customcitePS{FIRST-GL-5}. Their role includes secure design, selecting appropriate technologies and design paradigms, tracking and monitoring models in production, and integrating cloud-based applications and web services \customcitePS{I8}, \customcitePS{I96}. Also, referred as IT team or IT architects in the literature \customcitePS{I8}.

\textbf{Security experts} ensure the integrity, confidentiality, and availability of ML assets by defining security objectives and requirements at the project's initiation \customcitePS{I8}. 
Some studies refer to them as a dedicated security team \customcitePS{AwsMLOps2022} or security engineers \customcitePS{FIRST-29}, while others define them as data and software engineers with security expertise \customcitePS{I8}. Their role involves securing the overall ML system, addressing potential attacks during data ingestion and feature engineering, and enforcing secure coding practices \customcitePS{I8}.

\textbf{Compliance teams} focus on mitigating unfairness and bias in ML systems throughout the model training, deployment, and monitoring phases \customcitePS{AwsMLOps2022}, \customcitePS{AmazonAI}.
Recently, \textbf{risk professionals} have become involved in MLOps projects to manage data and model-related risks \customcitePS{I96}. Specifically, \textbf{model risk manager} is responsible for ML model inventory management, model explainability, and pipeline management \customcitePS{AwsLens}, \customcitePS{AwsSecure}, while other studies only mention risk professionals for managing any data and technology related risks \customcitePS{I96}.
\textbf{Data steward} and \textbf{data quality board} specifically focus on data quality management \customcitePS{I92}, and \textbf{data governance officer} focus on data governance and privacy \customcitePS{DataBricksMLOps}.

\textbf{Business owners}, \textbf{business managers}, and business stakeholders are not clearly distinguished in the literature yet. However, they are often associated with organization level decision making \customcitePS{MLOpsAWS}, supervision of other roles in development and operations \customcitePS{I8}, and evaluating the production of ML systems from an operational and regulatory perspective \customcitePS{FIRST-GL-4}. 
\textbf{Product experts}, \textbf{owners}, and \textbf{managers} are associated with defining and optimizing product features to enhance performance and usability. For example, product experts identify feature set candidates based on specific performance issues and apply feature engineering techniques alongside to improve model outcomes \customcitePS{FIRST-31}. Product managers make data-driven decisions by reviewing relevant collected and visualized information \customcitePS{FIRST-34}.

\textbf{Collaborative activities among roles:} In a typical MLOps pipeline (Figure \ref{fig:pipeline}), various roles collaborate to fulfill different responsibilities. Often, one role is the primary owner of an activity, while other roles provide support as collaborators \customcitePS{FIRST-GL-26}.
Data engineers, data scientists, and ML engineers often collaborate with domain experts and business analysts during data collection and preparation, data analysis, and model development and deployment stages \customcitePS{AwsSecure}. 
Particularly, data engineers work together with data scientists and ML engineers to determine the appropriate data format for subsequent steps \customcitePS{AwsLens}.
Additionally, data engineers, scientists, and ML engineers often collaborate with business and domain experts to interpret both the data and the model \customcitePS{AwsLens}. 
ML engineers and occasionally data scientists accompany the MLOps engineers at the initial monitoring to ensure the model's stability and step in when the model or features need optimization \customcitePS{I40}.
Meanwhile, MLOps engineers interact with every role to ensure the optimization of the automated pipeline, the operational stability of the ML system, and the smooth collaboration across various roles \customcitePS{I8},\customcitePS{MLOpsMicrosoft}.

\begin{tcolorbox}
MLOps is perceived as a multidisciplinary field integrating continuous engineering practices (CI/CD), DevOps principles, data science, machine learning, and software engineering to systematically manage the lifecycle, development, deployment, and operation of ML-enabled software systems.
It encompasses traditional ML activities with core MLOps tasks such as continuous deployment and monitoring, while placing increasing emphasis on \textit{quality assurance and compliance} and \textit{efficient management of pipelines and artifacts}, and cross-role collaborations. 
Emerging roles such as \textit{MLOps engineers}, \textit{data governance officers}, and \textit{compliance teams} illustrate increasing specialization within MLOps, emphasizing its multidisciplinary character. Findings of RQ1 reflect the evolution of MLOps into a more structured, mature, and systematically governed approach for operationalizing ML systems.
\end{tcolorbox}

\section{RQ2: What are the state-of-the-art practices and techniques proposed for implementing MLOps?}\label{RQ2}
Here, we report the state-of-the-art work practices, processes, techniques, and supporting technologies that aid organizations in implementing MLOps. These MLOps practices 
cover a broad spectrum of social and technical aspects across the life cycle of ML models. We have grouped and reported them under seven categories below. 

\begin{figure}[t] 
   \centering
   \includegraphics[width=1\linewidth]{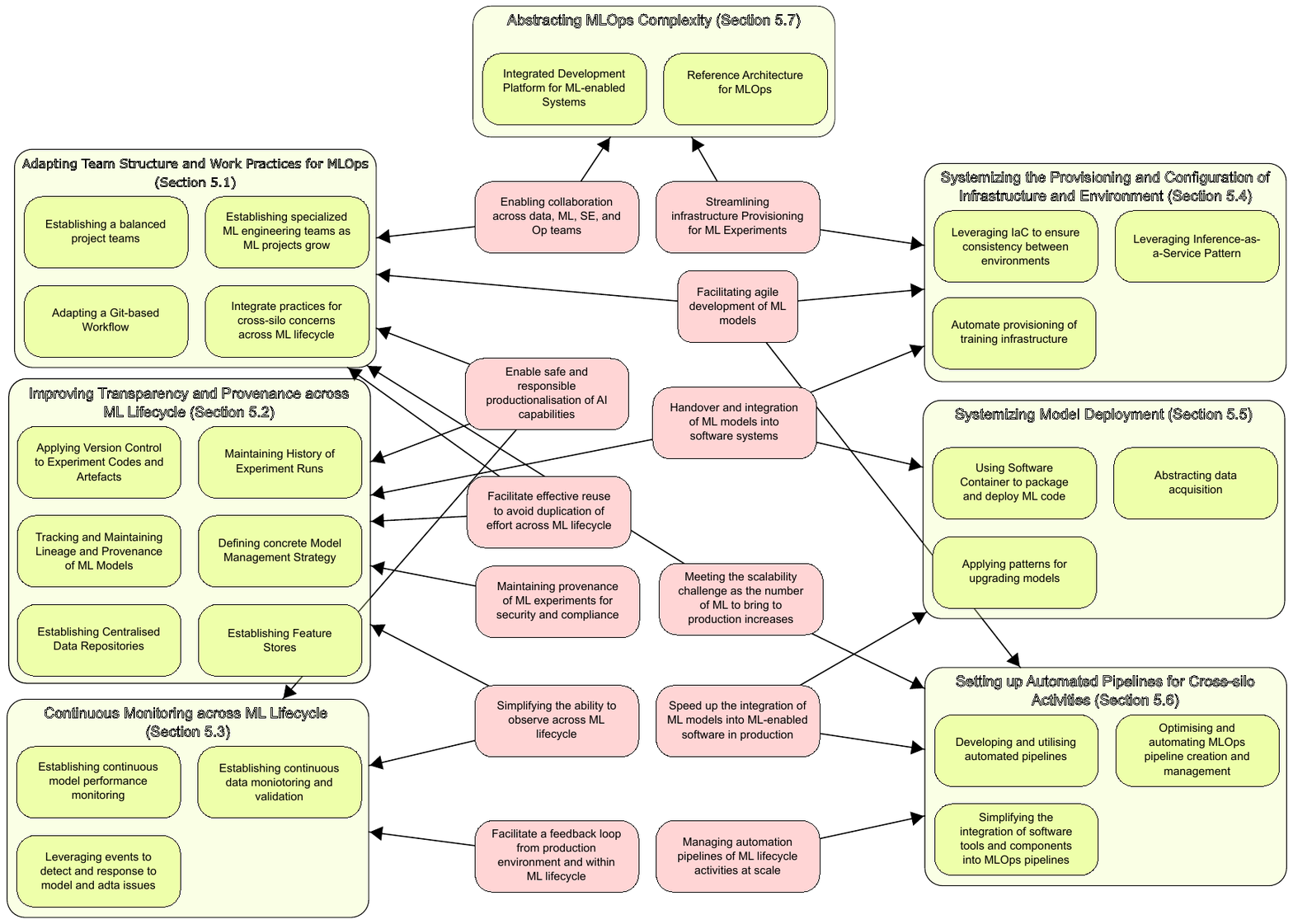}
   \caption{MLOps solutions and the corresponding ML productionalization challenges}
   \label{fig:RQ2} 
\end{figure}

\subsection{Adapting Team Structure and Work Practices for MLOps}\label{solution:work_practices}
The productionalization of ML models is a multifaceted challenge that requires the expertise of stakeholders with diverse skill sets, from domain knowledge to data science, software engineering, system operation, and cyber security. Following best practices are introduced to enable the successful implementation of MLOps by fostering effective collaboration among such diverse stakeholders.

\vspace{1.5mm} \noindent \textbf{Establishing balanced project teams:} The AWS Well-Architected Machine Learning framework \customcitePS{AwsLens} and Azure Machine Learning best practices \customcitePS{G3} recommend establishing \textit{project teams}, which comprise of domain expert, data engineers, data scientists, ML engineers, MLOps engineers, IT auditor, and cloud security engineers to work with ML-based software features and systems, rather than building functional silos for each capability, e.g. ML engineering team, security team. 
A project team should be configured with balanced capabilities such that it can carry out all phases of the life cycle of an ML model. 

\vspace{1.5mm} \noindent \textbf{Establishing specialized ML engineering teams as ML projects grow:} As organizations grow in ML utilization, the number of ML projects maintained by a project team could rapidly outgrow the ability of the team to scale. At this state, a helpful practice is establishing a specialized ML engineering team that focuses on the deployment, operations, and maintenance of ML models. These capabilities are common across ML projects and uncoupled to domain specific ML problems. By centralizing these capabilities into a specialized team, an organization can develop, manage, and utilize these capabilities more efficiently. Specialized ML engineering team along with standardized processes and reusable patterns and artifacts lay a foundation for establishing ``AI Factory'', scaling ML productionalization to ML \textit{industrialization} \customcitePS{G3}.

\vspace{1.5mm} \noindent \textbf{Adopting a Git-based workflow:} 
ML project teams can leverage Git-based workflows to coordinate and carry out the day-to-day activities of ML projects. 
Git allows team members to work on isolated branches without disrupting the stable version of the project as well as enabling the automation of CI/CD activities.
Gunny et al., \customcitePS{S627} proposes a branch structure consisting of two stable branches (i.e., \texttt{dev} and \texttt{main}) and several short-lived feature branches to integrate model training, testing, and deployment activities into pull requests under two main stages. 
Pull requests from feature branches to \texttt{dev} branch would trigger model training and testing within a staging environment, thus ensuring the model's performance.
Pull requests from \texttt{dev} to \texttt{main} branch would trigger the packaging of the model as software containers and deployment of the packaged model into a production environment. 


\vspace{1.5mm} \noindent \textbf{Integrate practices for cross-silo concerns across ML life cycle:} Cross-silo concerns include AI fairness, AI explainability, and security. Instead of addressing these concerns at any particular phase, they should be a cross-silo effort, which each part of the life cycle implementing different aspects. Amazon AI Fairness and Explainability Whitepaper \customcitePS{AmazonAI} recommends integrating bias measurement before-, during-, and after-training, as well as in production. The detected bias then can be addressed using pre-processing, training tuning, and post-processing. 
Security is another cross-silo concern of ML life cycle. Zhang and Jaskolka \customcitePS{I8} introduce the concept of Secure Machine Learning Operations (SecMLOps) based on the People, Processes, Technology, Governance, and Compliance (PPTGC) framework. SecMLOps embeds security controls across the whole life cycle. In the requirement phase, security objectives and requirements of ML models and data are defined, and security models are established. These models drive the development of security policies regarding data, model, physical security, personnel security, administrative security, and network security across the ML life cycle and all involved parties. These models also drive the development and integration of security monitoring and incident response components that monitor ML models in production.

\subsection{Improve Transparency and Provenance Across ML Life Cycle}
The goal of improving the transparency across the ML life cycle is twofold. Firstly, improving transparency helps project teams avoid or reduce duplicated work, such as building and running data pre-processing and feature engineering pipelines. A closely related advantage is avoiding unexpected dependencies between projects, such as via a shared feature set. Secondly, maintaining the provenance of data, models, and features facilitates project teams to identify the root causes of production issues and roll back ML systems to the previous working state when production issues happen. The improvement of transparency and provenance can happen along three axes: models and experiments, datasets, and features.

\vspace{1.5mm} \noindent \textbf{Applying version control to experiment codes and artifacts:} 
Existing guidelines \customcitePS{G3}, \customcitePS{AwsLens} advocated using version control systems such as Git to manage all inputs that go into experiments, such as software codes for model training and testing, datasets, environment configurations, and other experiment parameters. Reproducibility is a key benefit of version control. However, the benefit of this practice extends beyond reproducibility. It provides project teams with the ability to roll back the system to the last known functional version of an ML system, thus facilitating recoverability and improving the overall fault tolerance of these systems.

\vspace{1.5mm} \noindent \textbf{Maintaining history of experiment runs:} The probabilistic nature of many ML models makes it possible for a set of version controlled source codes and artifacts to produce ML models with varying performance characteristics across experiment runs. Therefore, it is beneficial to maintain the historical data of experiment runs of the model \customcitePS{G3}, in addition to the version controlled source codes and artifacts.  Experiment versioning plays an important role in ensuring the repeatability of experiments and improving their conclusiveness \customcitePS{S627}. The historical data of experiment runs could provide additional transparency into the decision to upgrade models. This data would also be invaluable for planning the resource provisioning and optimization for the model training process. Experiment management tools such as MLFlow \customcitePS{S204} and IBM AI Factsheets \customcitePS{IBM2023} enable experiment tracking along with model and data versioning \customcitePS{NVIDIA2023}. By integrating tools like AutoML with Experiment management tools, practitioners can automate the generation of the best suited models \customcitePS{NVIDIA2023}. 

\vspace{1.5mm} \noindent \textbf{Tracking and Maintaining Lineage and Provenance of ML Models:} Lineage and dependencies among models and datasets are other aspects that project teams need to capture to develop a comprehensive picture of ML model life cycle. Closely related to lineage is the model's provenance, containing its machine-readable metadata and historical events within its life cycle, such as how and by whom it was trained and tested. The IBM MLOps handbook \customcitePS{IBM2023} highlights two main needs for capturing metadata and recording historical events regarding the life cycle of ML models: improving the productivity of ML model productionization by making ML model lifecycle related metadata visible to all stakeholders, and facilitating the regulatory compliance checking and explainability of the models. Academic literature proposes multiple provenance extraction tools such as Vamsa \customcitePS{S565}, Geyser \customcitePS{S565} and MLflow2PROV \customcitePS{S204}. Vamsa is a knowledge-base driven \textit{static} provenance extraction tool. Geyser extends Vamsa for dynamic provenance extraction, provenance storage and querying. MLflow2PROV was designed for extracting provenance from version control system, experiment management systems and store in a central database for querying and visualization. IBM AI Factsheets \customcitePS{IBM2023} capture model and process metadata such as model details, training information, metrics, etc., and store in a central meta-store. 

\vspace{1.5mm} \noindent \textbf{Defining concrete model management strategy:} A model management strategy defines how models are identified, named, and labeled. An organization might also include requirements on the type and format of metadata, provenance, and lineage information to be maintained by project teams. A critical element of a model management strategy is the process and criteria for replacing models. For example, the Azure Machine Learning best practices \customcitePS{G3} recommended labeling a baseline model as ``champion'' and new models as ``challengers.'' Strategies can be defined to leverage experiment results captured in the historical data of experiment runs to decide when and how to replace the ``champion'' model. 

\vspace{1.5mm} \noindent \textbf{Establishing centralized data repositories:} Centralized repositories provide project teams with visibility into the available data and its utilization to avoid redundant effort and potential dependency problems \cite{I96}. Centralization of data also helps organizations to enforce a consistent access control policy and allows them to implement best practices regarding data privacy and bias that are relevant across projects. Centralizing data also facilitates the implementation of techniques such as real-time monitoring and alerting of data drifts \customcitePS{AwsSecure} and automated detection of data dependencies. For instance, Boue et al. \customcitePS{I170} propose a dependency mapping framework to analyze and notify MLOps engineers of changes in data sources and their effects. Another advantage of centralized data repositories is resource utilization efficiency because they allow organizations to employ specialized and scalable data technologies, such as transactional databases supporting structured or semi-structured data \customcitePS{S262}, data lakes \customcitePS{DataBricksAuto}, and data warehouses \customcitePS{NVIDIA2023} where version controlled datasets are stored to maintain links between data, code, and models.

\vspace{1.5mm} \noindent \textbf{Establishing feature stores:} Features can become a point of friction between the data science and operation sides of an ML model, especially if the feature construction code is not well documented and lacks scalability. In these cases, the feature engineering code might need to be reconstructed. However, this process might introduce subtle changes to the features that make model's performance deviate from the experiment results. At the same time, reusing features and feature engineering pipelines might lead to undocumented coupling between projects and unexpected consequences of changes. Establishing centralized feature stores lays a foundation to address these challenges, because they provide visibility into existing features and feature engineering pipelines to avoid redundant effort. Moreover, feature stores facilitate feature governance by making features visible and reusable among different stakeholders in a secure manner \customcitePS{AwsApn}, \customcitePS{G2}, \customcitePS{NVIDIA2023}. Some existing feature store platforms and technologies include Amazon SageMaker Feature Store\footnote{\url{https://aws.amazon.com/sagemaker/feature-store/}} and Databricks Feature Store \footnote{\url{https://www.databricks.com/resources/demos/tutorials/data-science-and-ai/feature-store-and-online-inference}}.

\subsection{Continuous Monitoring Across ML life cycle}

\vspace{1.5mm} \noindent \textbf{Establishing continuous model performance monitoring:} 
The concept of ``model performance'' has multiple facets, such as prediction accuracy, robustness, fairness, explainability, and safety. 
Once the required performance metrics have been established during project initiation, they can be implemented as automated monitoring jobs. 
Various tools exist for monitoring production ML models. 
For instance, AWS Sagemaker Model Monitor \customcitePS{AwsMLOps2021}, \customcitePS{S625} enables detecting data drift, concept drift, bias drift, and feature attribution drift of deployed ML models. 
Sagemaker Clarify \customcitePS{AmazonAI} can detect bias to ensure fairness of models deployed on Amazon Sagemaker. 
Similarly IBM OpenScale \customcitePS{IBM2023} monitors models for model accuracy, fairness and data drift.
Databricks Lakehouse Monitoring \customcitePS{DataBricksMLOps} maintains inference tables with request-response data to detect data and model drift. 
Jayalath and Ramaswamy \customcitePS{S46} highlight the importance of user feedback for the success of deployed ML models and propose a monitoring framework to that utilizes user feedback to investigate data deficiencies.
Visual analytic tools such as Slice Teller \customcitePS{I171} would be invaluable in aiding human experts in such cases when human inspection of model's \textit{local} performance on some particular segments of data is necessary. 



\vspace{1.5mm} \noindent \textbf{Establishing continuous data monitoring and validation:} 
As organizations have more data over time, they often face issues of data drift, bias, data quality and security (e.g., poisoning). Continuous monitoring and validation of datasets are practices to address these challenges. 
Data validation tools such as Tensorflow DataValidation \footnote{https://www.tensorflow.org/tfx/tutorials/data\_validation/tfdv\_basic} and Amazon Deequ \footnote{https://aws.amazon.com/blogs/big-data/test-data-quality-at-scale-with-deequ/} are used to write declarative data quality constraints to catch data quality issues as they occur. 
Further, Tu et al. \customcitePS{A36} extend this concept to auto-program data quality constraints by leveraging statistical information from past executions.

\vspace{1.5mm} \noindent \textbf{Leveraging events to detect and response to model and data issues:} 
The continuous monitoring of models and datasets can produce the signals to facilitate further analysis and response. Event-driven programming approach is used to take advantage of these signals \customcitePS{G3}. In particular, monitoring platforms are considered as event emitters. These events can be published on a centralized event bus that is available to all teams. Authorized team members can subscribe to events from the bus to be notified for further analysis and response.

\subsection{Systemizing the Provisioning and Configuration of Infrastructure and Environment}
The performance of ML models, particularly deep learning variants, depends not only on the input data and model parameters but also on the software environment surrounding them. Therefore, it is critical to maintain the consistency of the software dependencies and configurations across training, testing, staging, and production environments. Practices such as infrastructure-as-code (IaC), automated infrastructure provisioning, and Inference-as-a-Service provide the necessary systematization to address the environment consistency challenge.

\vspace{1.5mm} \noindent \textbf{Leveraging IaC to ensure consistency between environments:} The IaC practice ensures that software environments are constructed and configured consistently according to predefined specifications. These specifications can be captured as software artifacts, stored in source code repositories, and managed by version control systems. Treating environment specifications as code artifacts enables project teams to collaborate on the development, maintenance, and improvement of these configurations. Existing IaC tools such as Ansible and Terraform can be readily adapted to ML projects. Existing literature has also proposed ML-specific tools, such as the Pinto utility \customcitePS{S627} that automatically creates software environment based on the specification of an ML project. 

\vspace{1.5mm} \noindent \textbf{Leveraging Inference-as-a-Service:} Inference-as-a-Service is an approach to centralize the execution of ML models to a service provider. Instead of setting up local environment to run a model, entities and software systems needing model's inference would request the inference results from the service provider via a predefined service API instead. By removing the need to run models locally, Inference-as-a-Service can reduce the number of production environments to be setup, thus mitigating the risk of organizations using inconsistent environments. Another advantage of this approach is the maximization of resource utilization: the centralization of AI inference can replace multiple under-utilized infrastructure across entities that run the model locally with a fully-saturated infrastructure at the centralized service provider \customcitePS{S627}. NVIDIA Triton Inference Server is an open-source software that could be leveraged to implement AI inference services. 

\vspace{1.5mm} \noindent \textbf{Automate provisioning of training infrastructure:} An advantage of the systematization of environment creation is the possibility to provide data scientists with the ability to provision training infrastructures such as virtual machines by themselves. For instance, AWS \customcitePS{AwsLens} provides managed training infrastructures for optimum automated resource provisioning during iterative training process through Amazon Sagemaker. Sagemaker monitors the resource utilization of GPUs, CPUs and network bandwidth to make auto scaling of training infrastructure, thus abstracting the infrastructure management activities from the users and reducing training costs through use of on-demand computing resources. Raffin et al., \customcitePS{S262} propose the use of Infrastructure As Code (IaC) and declarative configurations (i.e., Yaml manifests of Kubernetes defining the desired state of operation) to automate provisioning of underlying infrastructure in a reliable, reproducible and scalable manner.

A related problem is the monitoring of system metrics (GPU, CPU and I/O utilization) to provide insights into overall performance of the training infrastructure. Rauschmayr et al., \customcitePS{S615} introduces a profiling tool for Deep Learning model training clusters, which can be integrated to Amazon Sagemaker as an add-on. Profile monitors the system utilization metrics related to training jobs to identify performance bottlenecks (GPU, CPU and I/O  bottlenecks) and provide insights for optimizing resource usage during model training phase.

\subsection{Systemizing Model Deployment}
The process of translating data science codes into production codes and maintaining them brings about a variety of challenges. Some patterns for model deployment and maintenance have emerged from the literature and best practices. 

\vspace{1.5mm} \noindent \textbf{Using software container to package and deploy ML code:} 
A common practice is packaging ML code with all of its dependencies within software containers, and then applying optimizations such as parallelization and load balancing upon the containerized workloads \customcitePS{S510}, \customcitePS{S163}. Containerized workloads can also be operated at scale by software container orchestrate platforms such as Kubernetes \customcitePS{S243}. By encapsulating ML models as docker containers practitioners can reap the benefits of container management platforms such as AWS Fargate which provides serverless options for ML model deployment \customcitePS{AwsLens}. Kubernetes enables automated deployments, monitoring, and scaling of ML models to ensure efficient operation of deployed ML models, thus overcoming limitations of manual model deployments \customcitePS{I43}. Syrigos et al. \customcitePS{I43} propose an extension for Kubernetes Scheduler where container scheduling decisions are made in a latency-sensitive manner in addition to CPU and RAM of the computing resources to deploy ML models.

\vspace{1.5mm} \noindent \textbf{Abstracting data acquisition:} Deployed ML systems do not exist in vacuum. Instead, they need to pull data from various sources and perform pre-processing tasks such as merging datasets and cleaning. Therefore, containerizing ML codes by itself is inadequate for addressing productionalization challenges. An approach to abstracting the data acquisition activities as services can help attain scalability of data processing with increasing volumes of data. \customcitePS{S510} proposes a pattern to abstract data sources and introduce a software agent that pulls and feeds data to containerized ML code. This design introduces the necessary decoupling to parallelize the containerized ML code. 

\vspace{1.5mm} \noindent \textbf{Applying patterns for upgrading models:} 
To prevent potential failures after updates to ML models, various deployment strategies for a safe upgrade of ML models in production are proposed. AWS \customcitePS{AwsLens} highlights the importance of conducting a trade-off analysis to compare multi-version deployment strategies like A/B testing, canary deployments, and blue/green deployment to select the best approach. Such strategies facilitate rollback or roll-forward decisions to be made after evaluation of the performance of the updated model without affecting the availability of the deployed service. To this end, Guissouma et al. \customcitePS{S454} propose a multi-version execution platform that facilitates the execution of shadow versions alongside the current active version, measures safety metrics of the updates through monitoring, and safe activation of the updated version if it outperforms the current active version.


\subsection{Setting up Automated Pipelines for Cross-silo Activities} \label{sec:sol_pipelines}

\vspace{1.5mm} \noindent \textbf{Developing and utilizing automated pipelines:} 
might be able to save time by having their well-defined model validation pipelines approved, saving time on having the models themselves validated and approved. 
MLOps pipelines can be implemented with various tools. At a basic level, software scripts (e.g., Bash or Python) for submitting training or testing jobs can form a rudimentary version of automated pipelines. Moving up a level, organizations can adapt the existing tools and platform for CI/CD to MLOps workflows. GitHub Actions, CircleCI, ArgoCD, and Jenkins were among the tools suggested by the existing literature and industry best practices for adaptation to MLOps. For instance, Stieler and Bauer \customcitePS{S57} proposed a Git workflow for an ``active learning life cycle'', which contains four interconnected iterations of data management, model training, software development, and system operations. In the proposed workflow, a CI/CD tool hosted on a resource-rich machine with GPUs executes ML pipelines for training, testing, and integrating ML models based on the code committed by developers to a source control management system. Alternative, generic workflow engines such as Apache Airflow\footnote{\url{https://airflow.apache.org/use-cases/mlops/}} can be adopted to MLOps pipelines. For instance, the FLScalize \customcitePS{S163}, an MLOps platform for federated learning, utilizes Airflow to orchestrate the interaction sequences among federated learning servers and clients as well as other monitoring workflows. Finally, there have been more ML-specific workflow orchestration tools (e.g., SageMaker Pipelines, IBM Watson Pipelines, Google Vertex AI Pipelines, Netflix Metaflow, ZenML) for defining and running MLOps pipelines. 

\vspace{1.5mm} \noindent \textbf{Optimising and automating MLOps pipeline creation and management:} 
By specifying pipelines explicitly, it is possible to automate, optimize, and manage the execution of MLOps pipelines. The automation helps overcome the uncertainties that ML workflows face such as changing requirements, changing system context due to other workflows competing for resources, workflow orchestrator's failure, or hardware crashes \cite{I86}. For instance, Scanflow-K8s \customcitePS{I86} was proposed as a framework for autonomic management and supervision of ML workflows in Kubernetes clusters. 
Moreover, automating the creation of MLOps workflows contributed to reducing the setup overhead, particularly during the initial states of ML development. Carqueja et al. \customcitePS{I92} proposed a lightweight Quick Machine Learning framework, QML, to address this problem. Automated creation also facilitates optimized deployment of data analytic pipelines against the heterogeneous infrastructure available for training or operating the models. Díaz-de-Arcaya et al. \customcitePS{S632} proposed Orfeon, an AIOps framework for optimizing the deployment of analytical pipelines, taking as input the static pipeline definition by data scientists and the dynamic metrics from the target infrastructure.


\vspace{1.5mm} \noindent \textbf{Simplifying the integration of software tools and components into MLOps pipelines.} 
Software-centric integration approach \customcitePS{S646} is proposed in Looper where predefined API end points are provided for interconnecting steps in MLOps pipelines. Modularisation through microservice-based design is recommended in \customcitePS{AwsLens}, \customcitePS{I18}, \customcitePS{I16} to enhance the integration of pipeline components.

\subsection{Abstracting MLOps Complexity}\label{sec:sol_complexity}
\vspace{1.5mm} \noindent \textbf{Integrated development platform for ML Systems.} 
Integrated development platforms provide a collaborative environment for whole-life-cycle-management for ML models and abstract the complexity from end users (e.g., data scientists, researchers, and other ML practitioners) \customcitePS{S646}. They provide interactive control planes accessible by ML practitioners to manage end-to-end ML life cycle \customcitePS{NVIDIA2023}. Managed cloud-based platforms such as Amazon SageMaker \customcitePS{AwsMLOps2022} and IBM Watson \customcitePS{IbmMLOps} provide support for data preparation, model training and experiment tracking, automated deployment on managed infrastructure and monitoring throughout the life cycle. Sagemaker \customcitePS{AwsMLOps2022} is developed with a modular and customizable architecture to support the integration of open-source tools to meet practitioner requirements. MLflow, Kubeflow, and Metaflow are open-source end-to-end platforms to simplify the ML lifecycle \customcitePS{Canonical}. Databricks Lakehouse Platform \customcitePS{DataBricksAuto} provides a data-native platform which together with MLflow provides a single platform for managing data, conducting data analysis, model training, and model serving. Peng et al. \customcitePS{W115} proposes a platform for deploying models within Kubernetes and Istio to enable cloud-native environments by providing Source to Image (S2I) and model deployment support.

\vspace{1.5mm} \noindent \textbf{Reference Architecture of an MLOps system.} Functional decomposition and reference architecture of MLOps systems guide the composition or development of MLOps pipelines and platforms depending on domain-specific requirements. To this end, academic literature presents domain specific architectures for various domains such as manufacturing domain \customcitePS{S262}, Industry 4.0 \customcitePS{I16}, transnational machine learning in health domain \customcitePS{W127}, and Gravitational Wave Physics \customcitePS{S627}.

\begin{tcolorbox}
State-of-the-art MLOps practices and techniques prioritize automation and optimization of MLOps lifecycle, service-oriented deployment, artifact tracking, and collaboration.
Recent practices increasingly focus on domain-specific reference architectures tailored for particular contexts, 
systematic artifact management emphasizing provenance and lineage tracking, 
and frameworks and practices that support addressing compliance, fairness, and security. 
These developments illustrate a clear evolution toward more structured, scalable, and comprehensive MLOps implementations, capable of addressing emerging complexities in ML operationalization across diverse application domains.
\end{tcolorbox}

\section{RQ3: What are the Challenges Practitioners Face when Adopting MLOps, and What Solutions have been Proposed to Address them?}  \label{RQ3}
In \textit{RQ2} we discuss work practices, techniques and technologies that organizations need to implement to enable MLOps. However, there are challenges that hinder teams from efficiently adopting MLOps into their organization's ML-based software system productionalization process. Thus, we identify such challenges and highlight reported solutions from the literature, and summarize them. 

\subsection{Socio-technical Challenges}
MLOps is a continuously evolving, technology-intensive domain which requires a considerable cultural shift in organization for its successful adoption (as discussed in section \ref{solution:work_practices}). Our thematic analysis of literature resulted in three major challenges arising due to interactions between social and technical aspects when adopting MLOps.

\subsubsection{Cultural Changes} The introduction of MLOps practices disrupts the processes followed during the development and management of traditional software systems. Thus, the successful adoption of MLOps practices depends on the openness and willingness of the teams to change and embrace the cultural changes introduced by MLOps \customcitePS{I9}.  Works such as \customcitePS{I9}, \customcitePS{S58} and \customcitePS{AwsMLOps2022} discuss the inter-disciplinary collaboration requirements of the MLOps practices as one of the major cultural changes. Successful adoption of MLOps requires continuous collaboration among interdisciplinary teams that previously used to work in silos, which may result in conflicts.  As a solution to this, it is important to have MLOps best practices in place to enable smooth cultural transitions. One way to achieve this is the introduction of a clear and well-structured communication plan, and having a clear definition of the business objectives/goals and metrics as possible solutions to help teams understand novel MLOps practices introduced to organization culture \customcitePS{I9}. Furthermore, methods need to be implemented for swiftly identifying and resolving conflicts arising related to responsibility boundaries and different goals/ expectations of multi-disciplinary stakeholders \customcitePS{S58}, \customcitePS{I109}. \customcitePS{S58} proposes a i* Strategic Actor Rationale to model the actor relationships to identify and resolve conflicts during collaboration. However, addressing the challenges associated with cultural changes is unique to each organization. For example, in the financial services domain, due to the complex regulatory environment, teams have to operate in silos \customcitePS{AwsMLOps2022}. Thus, financial service institutions benefit from tools and platforms (e.g., ValidMind \footnote{https://validmind.com/}) that enable continuous collaboration between data scientists and compliance teams to achieve efficient risk management throughout the productionalization lifecycle. Hence, necessary cultural changes must be implemented (through MLOps best practices and tools) after careful analysis of the existing organizational structure and strategically implement the changes to maximize their positive impact \customcitePS{I9}. To this end, Nvidia \customcitePS{NVIDIA2023} recommends a human-centric approach to guide the selection and adoption of MLOps tools, techniques, and practices.

\subsubsection{MLOps Fragmentation} \label{sec:Fragmentation} Rapid evolution and growth in MLOps practices, solutions and tools create industry-wide fragmentation of MLOps due to the increase in \textbf{\textit{complexity}} and \textbf{\textit{diversity}} of MLOps. Being a broad domain consisting of practices, processes, solutions and tools to productionalize ML systems, MLOps is inherently complex \customcitePS{NVIDIA2023}.  Moreover, the popularity of MLOps and the use of ML in various domains that lack expertise in ML (e.g., physical sciences \customcitePS{S627}) result in the introduction of diverse approaches and tools particular to different domains, use cases and organizations \customcitePS{NVIDIA2023}. This results in accidental complexities, and it is challenging to separate accidental complexity from essential complexity \customcitePS{NVIDIA2023}. Standardizing MLOps process, tooling, etc. and generation of domain-specific reference architectures to capture domain-specific ML productionalization requirements contribute to solving these issues. For instance, \customcitePS{S262} extends generic MLOps to meet the requirements of the Manufacturing Domain where ML-enabled software is deployed within distributed Edge-Cloud environments. However, as production ML systems are novel and the rapid evolution of MLOps, standardization needs a huge effort and should be done by bringing practitioners, researchers and industry MLOps solution providers together.

\subsubsection{Responsible MLOps} Achieving Responsible MLOps is a challenging multidisciplinary effort that brings people, technology and processes together throughout the MLOps life cycle \customcitePS{I8}. This includes various aspects such as compliance with regulations and standards, AI fairness assurance through bias detection/mitigation, transparency and accountability through explainable AI and security. However, the MLOps process lacks explicit consideration for responsible AI practices and provides only limited guidance concerning these aspects, which makes their implementation challenging for practitioners \customcitePS{I8} \customcitePS{I24}. Being a fast-evolving field, there's a shortage of experts who can navigate these challenging aspects successfully to implement responsible MLOps \customcitePS{I24}. Moreover, varying regulatory requirements across different domains demand different levels of collaboration with compliance teams and quality assurance teams throughout the MLOps lifecycle. For instance, highly regulated domains such as financial services require compliance by design as an MLOps practice. Developing comprehensive MLOps frameworks by integrating responsible AI practices can help practitioners overcome this. As an initial step, academic literature presents conceptual frameworks that integrate security (SecMLOps framework presented in \customcitePS{I8}) and explainable AI practices (Explainable AI framework presented in \customcitePS{I40}) into the MLOps process. 

To facilitate Responsible MLOp, practitioners require tools that support end-to-end assurance of these aspects, which includes provenance support tools during developments and experiment activities \customcitePS{DataBricksMLOps},  tools for bias mitigation during different stages of the ML life cycle and explain models and their predictions \customcitePS{AmazonAI}, and tools for automated ML compliance checking \customcitePS{I8}. Moreover, detailed documentation regarding responsible MLOps support of the tools and MLOps platforms can enable users to make informed decisions \customcitePS{I8} on which tools to use. As an example, \customcitePS{AwsSecure} provides the architectural components of a secure MLOps platform using AWS services, which can be used by practitioners to decide if AWS services meet their responsible AI requirements.

\subsection{MLOps Pipeline Related Challenges} 

MLOps best practices recommend using automated pipelines for managing the life cycle of the ML-enabled software systems (as discussed in section \ref{sec:sol_pipelines}). This section details challenges faced by practitioners during the implementation of MLOps pipelines.

\subsubsection{Complexity} 
While there are many powerful tools such as Kubeflow, FedML, Polyaxon for creating MLOps pipelines,  high infrastructure cost, high expertise required for pipelines operation, high setup and maintenance overhead of the pipelines slows down their adoptions \customcitePS{I92}. This is prevalent especially during the initial stages of ML projects or during ML projects driven by small companies. In order to avoid these complexities, practitioners tend to choose manual workflows over automated pipelines. IBM’s
2022 Global AI adoption Index report \cite{ibmindex2022} also highlights that the success rate of MLOps adoption of small companies is considerably lower compared to large corporations due to existing complexity. Design and development of MLOps pipeline creation frameworks with low setup, maintenance overhead and low infrastructure cost can overcome these challenges. DagsHub \footnote{https://dagshub.com/docs/} and Quick Machine Learning framework (QML) \customcitePS{I92} are lightweight frameworks designed to act as an abstraction layer between ML practitioners and their chosen pipelines and enable creation of lightweight pipelines to meet ML workflows. 

\subsubsection{Optimized deployment} IoT domains require MLOps pipelines to operate within distributed computing environments which include resource constrained and geographically distributed edge devices and centralized cloud data centers \customcitePS{I16}. This creates the challenge of optimally placing modularized pipeline components across these resources, especially to meet the increasing resource requirements during model training and model serving phases \customcitePS{I43}. This challenge is further amplified by the availability of continuous dynamic data streams. Thus, manual provisioning of pipeline components is error prone and provides sub-optimal results. Strategic modularization of pipeline components to suit the business domain characteristics and designing dynamic resources provisioning frameworks and algorithms can help overcome these issues \customcitePS{I16}, \customcitePS{I43}. Furthermore, training pipelines need to adjust dynamically (i.e., resource optimized model training, model splitting, etc. \customcitePS{I47}) to enable parallel or distributed model training to ensure efficient operation of the MLOps pipelines. 

\subsubsection{Managing large scale pipelines}
Large-scale enterprises that adopt MLOps practices have to operate thousands of recurring and interdependent production pipelines. Such pipelines are prone to silent failures caused by data quality issues in the data pipelines such as schema-drift, increasing nulls, change of units, change of value standards, and change of data volume \customcitePS{A36}, \customcitePS{S521}. Such issues generate cascading failures in pipelines. Thus, practitioners face the challenge of maintaining large-scale pipelines without failures. Moreover, due to the sheer scale of pipelines and their complexities, manual monitoring of data pipelines and manual root cause analysis for model mispredictions incur extra human cost \customcitePS{A36} \customcitePS{I32}. To overcome these challenges, practitioners should automate pipeline operations monitoring as much as possible. This includes automation of data quality issue detection \customcitePS{A36} and automated root causes analysis for detected errors \customcitePS{I32}. 

\subsubsection{Continual learning} 
With the popularity of IoT, streaming big data is available for training powerful ML models to derive insights from data \customcitePS{I66} \customcitePS{I53}. However, high volume and velocity of streaming big data, make batch training infeasible due to cost of storage and cascading lag created by storing and retraining stages on recurring pipelines. While continual learning pipelines have emerged as a solution to this, they need to be scalable enough to withstand high-velocity data streams and make swift model updates accordingly. To enable  executing model updates without pausing the inference pipelines, practitioners need to create and use elastically scalable online learning pipeline architectures with dynamic approaches to scale pipelines components vertically and horizontally (e.g., StreamMLOps \customcitePS{I66}).

\subsubsection{Automation decision dilemma}\label{sec:automationDecision}
Although the practitioners tend to automate wherever possible when implementing MLOps, ML lifecycle consists of activities where human feedback is crucial and full automation can create negative impacts, especially considering the responsible AI practices \customcitePS{DataBricksAuto}. For example, model validation processes require communication and approval from regulatory agencies, fairness and bias identification and analysis, root cause analysis from monitoring alerts, etc. Moreover, companies may have to prioritize other tasks over allocating time and effort on automating some of the tasks. Thus, caution must be taken during automation such that the tedious tasks are automated and human insights are incorporated wherever it is needed throughout the lifecycle \customcitePS{DataBricksMLOps},  \customcitePS{DataBricksAuto}, \customcitePS{I40}.

\subsection{MLOps Platform Related Challenges}

MLOps platforms are designed for end-to-end ML lifecycle management, as a means of abstracting the complexity of implementing MLOps as discussed in \ref{sec:sol_complexity}. This section highlights challenges and limitations related to creating and using such platforms covering platform provider and practitioner perspectives.

\subsubsection{Platform customization} 

For companies looking to build and deploy ML models, the main goal is to generate business value from data. Thus, for most of the practitioners, the underlying MLOps platform is only instrumental and their aim is to minimize the effort on implementing and managing the platform. Cloud-based managed MLOps platforms like Sagemaker are advantageous to meet this requirement. However, using fully managed MLOps platforms means the practitioners would lose fine-grained control over the technology stack and the services supported by the platform may not meet specialized requirements of many businesses \customcitePS{AwsMLOps2022}. This drives practitioners to implement and manage their own platforms which add additional complexity and cost. Practitioner expectation in this scenario is easy customization of managed platforms by integrating open-source, proprietary or homegrown tools. To meet this requirement managed MLOps platform providers need to design and implement their platforms following a modular architecture, integration and customization support to enable practitioners to integrate open-source tools. As an example, AWS highlights the design of Amazon SageMaker which provides practitioners the flexibility to integrate their own workflows with open-source APIs supporting TensorFlow, PyTorch, Kubeflow, etc. \customcitePS{AwsMLOps2022}.

\subsubsection{Scalability}
Services provided by MLOps platforms need to maintain performance requirements as the amount of service requests fluctuates and MLOps platforms should facilitate practitioners to create scalable workflows.  In such scenarios, the lack of scalability of the underlying platform design becomes a significant challenge. As an example, the use of monolithic architecture would considerably impact the scalability of a system. Thus, for MLOps platforms Microservices Architecture (MSA) demonstrated high scalability with independently scalable but interconnected services \customcitePS{I18}. It is important to highlight, scalability of MSA based MLOps platforms can be achieved by using dynamic pipeline deployment approaches (discussed in section \ref{sec:sol_pipelines}) to ensure the efficiency of MLOps pipelines to enable large-scale learning and continuous deployment of ML models.

\subsubsection{Artifact Management}
Compared to traditional software, types and amount of artifacts related to ML-enable software system is considerably higher. This includes software codes, ML models and related hyper parameters, data sets, and other metadata such as provenance data used for improving the trust of the ML-enabled system \customcitePS{MLOpsAWS}. This is further complicated by the highly explorative and iterative nature of the ML model building process where the experiments include different datasets, data preparation techniques, algorithms, hyper parameters, etc. Thus, versioning, traceability, and reproducibility become challenging \customcitePS{S204}. Lack of support for conjugated management of these artifacts is a challenge faced by practitioners when using MLOps platforms \customcitePS{I22}. To overcome this, MLOps platforms need to be built with version control, storage, and governance (access control, authentication, and authorization) support for data, model and associated meta-information \customcitePS{I22}. 

\subsubsection{Distributed computing environments}
Edge computing is introduced as a means of improving data privacy by processing data closer to the sources. This has created paradigms such as TinyML (i.e., ML models running on very resource constrained heterogeneous edge devices) and Federated Learning (i.e., a method of collaboratively training a shared model without transmitting data to a central location). These aspects affect certain ML lifecycle activities compared to ML model deployment in cloud. For example, in distributed edge computing scenarios, the model selection depends on the edge devices it is deployed on due to the heterogeneous nature of the devices, and model retraining needs to be carried out in decentralized manner using distributed streaming data. Thus, existing Cloud-based managed platforms do not meet the requirements for TinyML and Federated ML scenarios at the edge. Thus, platforms need to be designed and developed (TinyMLOps platform \customcitePS{I74}) considering such special requirements arising in distributed computing environments \customcitePS{I53}, \customcitePS{I66}.

\subsubsection{Migration to MLOps Platforms} \label{sec:migrationDecision}
As MLOps is a new and upcoming practice, many companies are migrating already existing projects to MLOps platforms to enable efficient production of the ML components of their software systems. Moreover, in ML-enabled software systems ML components are a small part of larger traditional software systems where CI/CD is achieved through DevOps practices and tools. So these companies utilize DevOps tools used for traditional software development for initial management of ML projects. Thus, when migrating MLOps platform requires additional information (artifacts and their links, workflow for creating pipelines, underlying resource requirements) needs to be extracted, which is a time consuming task \customcitePS{A6}. It is important for practitioners to conduct the migration process strategically by carefully documenting ML components and artifacts available in the current project, understanding the current architecture of the project to determine the best suited platform and utilizing tools supporting CI/CD support for software systems containing both traditional software components and ML components \customcitePS{A6}.

\begin{tcolorbox}
Practitioners face significant challenges across socio-technical shifts and barriers, ML pipeline complexity and fragmentation, scalability issues, distributed computing environments, the need for context-specific adaptations, ensuring compliance, fairness, and security of ML systems, and decision dilemmas about migrating to MLOps platforms and balancing automation with human involvement.
Proposed solutions to these challenges reflect a shift towards human-centric, systematic, and transparent communication strategies, emphasizing domain-specific architectures, enhanced governance, systematic artifact management, and structured human-in-the-loop interventions. 
 \end{tcolorbox}

\section{Future Research Directions}
Based on the MLOps practices, adoption challenges and proposed solutions, we identify the future research directions to facilitate MLOps adoption. 


\vspace{1.5mm} \noindent \textbf{Domain specific and expertise driven frameworks to facilitate cultural changes.}
Challenges such as cultural shifts, integration of responsible and secure AI practices, MLOps fragmentation, and transitioning from traditional DevOps technologies highlight the importance of sharing practitioner experience and knowledge to drive standardized MLOps adoption. To this end, future research can contribute to developing frameworks consisting of case studies 
to facilitate cultural transitions in a context-aware manner, and address best practices, solutions and tools employed to meet domain specific requirements. These frameworks need to be rooted in knowledge and experience sharing and designed to evolve as MLOps adoption increases. Studies such as \customcitePS{I24} and \customcitePS{A6} attempt to organize practitioner experience, but do not provide appropriate frameworks that can sufficiently capture the said aspects.

\vspace{1.5mm} \noindent \textbf{Domain specific MLOps reference architectures.} 
MLOps architectures, tools, and technologies developed for different domains, e.g., manufacturing domain, health domain 
capture unique requirements for ML system productionization.
While academic literature such as \customcitePS{S262}, \customcitePS{I16}, \customcitePS{S627}, and \customcitePS{W127} makes contributions towards this, 
the industry-driven and practitioner-oriented nature of MLOps demands reference architectures that combine practitioner input with platform and tool provider expertise.
Future work on reference architectures should integrate important aspects like security, ethics, transparency, compliance and governance, etc. to enable responsible and secure AI throughout the ML life cycle.

\vspace{1.5mm} \noindent \textbf{Evaluation frameworks for MLOps platforms and tools.} 
The current literature lacks comprehensive frameworks that enable evaluation of MLOps platforms and tools, hindering standardization and effective navigation of practitioners in this space.
Academic works such as \customcitePS{I8}, \customcitePS{I92} and \customcitePS{I22} evaluate existing MLOps platforms based on criteria like secure ML productionalization support, pipeline setup complexity, and artifact management support. 
Industry white papers and case studies such as \customcitePS{DataBricksAuto}, \customcitePS{AwsSecure}, \customcitePS{AmazonAI}, and \customcitePS{AwsMLOps2022} highlight the platform capabilities in automation, secure pipeline creation, responsible AI support, collaboration and architecture. 
However, this knowledge is scattered and does not provide a structured and comprehensive method for practitioners to evaluate which platforms and tools suit their needs the most.

\vspace{1.5mm} \noindent \textbf{Automation impact analysis frameworks.} 
Although the proposed best practice for automation dilemma is to automate tedious tasks and incorporate human feedback where possible \customcitePS{DataBricksAuto}, existing literature does not provide a structured approach to do this. Given that the automation decisions depend on contextual factors such as industry domain, business objectives, regulatory requirements, organization culture, availability of tools, etc., impact analysis framework needs to be created to analyze the effect of automation in a context-aware manner. Such analysis framework would help practitioners assess benefits and drawbacks using quantifiable metrics representing different criteria such as efficiency, scalability, cost, cascading risks, etc., thus enabling them to make informed decisions to avoid under or over automation.

\vspace{1.5mm} \noindent \textbf{Adoption of modular architectures such as Microservice Architecture.} 
Adoption of Microservice Architecture for MLOps pipelines and platforms facilitates their efficient and scalable deployment across distributed edge and cloud environments to support novel paradigms such as TinyML and Federated ML. 
Academic research on this demonstrates the potential of adopting microservice architecture \customcitePS{I18}, \customcitePS{S510}. 
However, the success of this approach depends on using strategic decomposition techniques for determining microservice boundaries without introducing unnecessary complexities. 
Moreover, management of such modular and distributed architectures requires development and integration of dynamic deployment, scaling, orchestration, and load balancing techniques along with data and artifact management to ensure their secure availability across distributed microservices. 

\vspace{1.5mm} \noindent \textbf{Efficient management techniques for MLOps pipelines and platforms.} 
Efficient pipeline component management techniques play a vital role in navigating the complexities 
of modular architectures, large scale pipelines, and continual learning and delivery of ML models in Big Data environments.
Future research should focus on online and dynamic resource provisioning, dynamic scaling, and fault tolerance enabled by distributed and scalable monitoring of pipelines. 
Moreover, 
these techniques need to consider the resource heterogeneity, and limited and distributed resource availability at the edge. 


\vspace{1.5mm} \noindent \textbf{Secure MLOps pipelines and platforms.} 
Practitioners can combine managed platforms with open-source, vendor supplied, or in-house tools to create pipelines. 
Security and trust of these components ultimately affect the security of the ML models produced. 
Security breaches 
can compromise data privacy, model integrity, overall system availability, etc. 
Proactive and reactive mechanisms need to be developed to monitor these pipelines and platforms and mitigate security breaches. 
As MLOps pipelines are created by integrating tools from various vendors, software supply chain vulnerabilities also become a considerable security concern. 
Intelligent approaches are needed to ensure secure configuration using technologies such as Infrastructure as Code (IaC) and Policy as Code (PaC). Moreover, with the rise of Edge-AI applications, pipelines and platforms need to be secured within distributed computing environments with limited resources. 

\vspace{1.5mm} \noindent \textbf{Open source tool integration support.} 
To meet different business requirements, customization of MLOps platforms is required where practitioners attempt to integrate open-source tools with managed MLOps platforms \customcitePS{AwsMLOps2022}. 
In this case, interoperability is achieved through standardization of interfaces, model and data formats, protocols, etc. Moreover, integration of open-source tools can cause security risks, requiring analysis before integration and continuous audits to detect and mitigate any risks after integration \customcitePS{AwsMLOps2022}, \customcitePS{I8}.

\vspace{1.5mm} \noindent \textbf{MLOps tools for responsible AI.}
As highlighted in both academic and gray literature such as \customcitePS{AmazonAI}, \customcitePS{I8}, \customcitePS{S204}, responsible AI tools (i.e., bias detection and mitigation, provenance extraction, compliance checking) need improvement. Future work on responsible AI tools requires input from experts across different domains
having various levels of security, fairness, regulatory, governance requirements that span across the ML life cycle. 

\vspace{1.5mm} \noindent \textbf{Conjugated artifact management solutions.} 
With the increased types and number of artifacts generated during MLOps productionization, 
conjugated artifact management solutions are required to track the artifacts while maintaining relationships among them \customcitePS{I22}. Future solutions should also focus on governance along with efficient and scalable storage of artifacts across distributed environments.

\section{Related Literature Reviews}\label{RelatedWork}

We examine existing systematic review studies related to MLOps and qualitatively compare them with our work based on the type of considered primary studies, the research questions covered, the number of primary studies included, and the time period for study selection. We identified eight related studies: three Multivocal Literature Reviews (MLRs), which collected data from both academic and gray literature, and five Systematic Literature Reviews (SLRs), which relied solely on academic literature. Table \ref{tab:ComparisonRelatedWork} presents a comparison of these reviews with our work, highlighting our contributions.

\begin{table}[h]
\scriptsize
\caption{Comparison of Related Work}
\label{tab:ComparisonRelatedWork}
\resizebox{0.85\textwidth}{!}{
\begin{tabular}{p{0.05\textwidth}p{0.05\textwidth}p{0.1\textwidth}p{0.1\textwidth}p{0.1\textwidth}p{0.1\textwidth}p{0.1\textwidth}p{0.08\textwidth}p{0.1\textwidth}} 
\toprule
Study & Is multivocal & MLOps Conceptualization & MLOps Activities & MLOps Roles \& Responsibilities & Practices \& Tools & Challenges & No. Primary Studies & Timeframe \\ 
\midrule
\cite{kreuzberger2023machine} & No & $\Circle$ & $\CIRCLE$ & $\CIRCLE$ &  & \Circle & 27 A & Up to May 2021 \\
\hline
\cite{lima2022mlops} & No & $\Circle$ &  & $\Circle$ & $\Circle$ & $\Circle$ & 30 A & Up to Jul, 2021 \\ 
\hline
\cite{kolltveit2022operationalizing} & No &  &  &  & $\Circle$ &  & 24 A & 2017 - 2021 \\ 
\hline
\cite{faubel2023review} & No & $\CIRCLE$ &$\Circle$ &  & $\Circle$ &  & 69 A & Jan, 2012 - May, 2022 \\ 
\hline
\cite{mboweni2022systematic} & No & $\CIRCLE$ & $\Circle$ & $\Circle$ &  &  & 18 A & 2015 - 2022 \\ 
\hline
\cite{steidl2023pipeline} & Yes & $\Circle$ & $\CIRCLE$ &  &  & $\Circle$ & 79 A, 72 G & 2010 - 2021 \\ 
\hline
\cite{john2021towards} & Yes &  & $\Circle$ &  & $\Circle$ &  & 8 A, 15 G & Jan, 2015 - Mar 31, 2021 \\ 
\hline
\cite{lwakatare2020data} & Yes &  &  &  & $\Circle$ &  & 1 A, 8 G & Gray lit.: Up to Apr, 2020, Ac. lit.: Up to Feb, 2019 \\ 
\hline
Our work & Yes & $\CIRCLE$ & $\CIRCLE$ & $\CIRCLE$ & $\CIRCLE$ & $\CIRCLE$ & 150 A, 48 G & 2013 - Sep, 2023 \\ 
\bottomrule
\multicolumn{9}{l}{$\CIRCLE$: Full coverage, $\Circle$: Partial coverage} \\
\multicolumn{9}{l}{A: Academic literature study, G: gray literature study} \\
\end{tabular}
}
\end{table}

Kreuzberger et al. \cite{kreuzberger2023machine} reported a mixed-method research, where they conducted an interview with practitioners in addition to an SLR. 
Their work targeted one RQ: ``what is MLOps'', which they answered with an overview of MLOps activities, roles and responsibilities and used these insights to create a unifying definition for MLOps. 
While they also discussed several open challenges related to adopting MLOps, it is not the main objective of the study.

Lima et al. \cite{lima2022mlops} reported an SLR, 
where their RQs aimed to identify activities related to ML model deployment, MLOps related roles and responsibilities, operationalization tools for ML models, and challenges in ML model deployment. However, this study lacks comprehensive data synthesis to provide indepth analysis of roles and responsibilities. Additionally, their discussion on MLOps related technologies is limited to MLOps platforms extracted from academic literature and their discussion on challenges does not elaborate on the challenges related to MLOps adoption.

Kolltveit et al. \cite{kolltveit2022operationalizing} reported an SLR on the operationalization of ML systems, which covers all steps following model training and evaluation. 
Specifically, they answered four RQs focused on tooling and infrastructure related aspects of MLOps such as; how are ML models operationalized, what are the main challenges in operationalizing ML models, what tools and software infrastructure are used to operationalize ML models, what are the feature gaps in the tooling and infrastructure used to operationalize ML models.
However, since the focus is only on tooling and infrastructure for operationalizing ML models, this work does not provide a comprehensive view of MLOps best practices considering the entire ML lifecycle.

Faubel et al. \cite{faubel2023review} reported an  SLR on the technical overview of MLOps. 
Specifically, the study presented MLOps definitions and listed the most common activities, and listed tools for automating activities in MLOps. 
However, this review does not consider roles and task distributions in MLOps, thus lacking emphasis on socio-technical aspects in MLOps. 

Mboweni et al. \cite{mboweni2022systematic} reported an SLR on the state-of-the-art DevOps practices in ML-enabled systems. 
Specifically, this work aimed to report how MLOps is understood and differs from DevOps. 
The reported content provides a higher-level overview of MLOps definitions, presents key roles of MLOps and associated tasks, and describes MLOps activities by constructing an MLOps architecture covering ML lifecycle.

Steidl et al. \cite{steidl2023pipeline} reported a mixed-method investigation (i.e., SLR and a semi-structured review study) focusing specifically on continuous development pipelines of AI models, which is one aspect of MLOps practices. 
This work focused on analyzing definitions of various terms related to continuous development pipelines for AI such as DevOps for AI, CI/CD for AI, MLOPs, etc., tasks handled by these pipelines, and the challenges related to implementation, adaption and usage of pipelines for continuous development of AI.

John et al. \cite{john2021towards} delivered an MLR on the adoption of MLOps in practice. More specifically, the authors derived a framework for MLOps that shows the activities that take part in the continuous development of ML models and maps the practices of companies to MLOps maturity model to show how practices evolve through stages.
However, they provided a high-level picture rather than elaborating on details regarding these questions.

Lwakatare et al. \cite{lwakatare2020data} presented an MLR on the RQ of ``how to enable modern software development process with continuous delivery for AI-enabled systems''. The authors considered one academic and eight gray literature items. 
The study discusses tools and techniques used in the continuous delivery of ML systems. 
Important details are available concerning the development stages of ML-enabled systems. 

Our analysis of the secondary studies presented in Table \ref{tab:ComparisonRelatedWork} demonstrates that no existing work contributes to building a holistic and timely depiction of MLOps landscape by answering all three research questions. Also, the majority of the related work relies solely on academic literature \cite{kreuzberger2023machine, lima2022mlops, kolltveit2022operationalizing, faubel2023review, mboweni2022systematic}. Given that MLOps is a rapidly evolving, practice-oriented field, insights from gray literature sources such as industry white papers and company websites offer valuable and timely perspectives on how MLOps is perceived by the community and adopted in real-life settings. Furthermore, the rapid improvements in MLOps call for a systematic review of the most recent literature to provide a timely view of the MLOps landscape. However, coverage by existing reviews extends only until 2022.

Considering these, our MLR aims to provide a comprehensive conceptualization of MLOps by integrating insights from a broader range of primary sources, from 150 academic studies with more practical knowledge from 48 gray literature items published up to September 2023. 
Our study extends prior research by linking MLOps conceptualization to productionalization challenges, identifying emerging activities, roles, practices, challenges, and solutions. 
Furthermore, we discuss future directions, emphasizing the need for domain-specific frameworks, secure pipelines, automation impact analysis, and responsible AI tools, aiming to enhance MLOps efficiency.

\subsection{Evolution of MLOps}
We share common RQs with related work. Hence, we compare our findings with those reported in prior studies to illustrate how MLOps has progressed over time. 

\textit{Conceptualization:} 
The conceptualization of MLOps in earlier systematic studies aligns closely with our RQ1.1 results (e.g., \cite{kreuzberger2023machine}, \cite{mboweni2022systematic}). This indicates that MLOps definitions in the literature have remained largely consistent from 2022 (the latest timeframe in prior studies) through September 2023 (our study's timeframe). Also, earlier studies explicitly defined MLOps to clarify its conceptual boundaries (e.g., \customcitePS{FIRST-GL-13}), whereas recent papers directly use the term without definitions, assuming community familiarity (e.g., \customcitePS{S90}, \customcitePS{I55}, \customcitePS{S46}). This shift highlights the growing maturity and widespread acceptance of MLOps as a recognized research area.

\textit{Activities:} While our findings align with previous works regarding core MLOps activities, we identify newer focal areas such as security \customcitePS{I8}, expanded quality assurance which includes data and model aspects \customcitePS{S276}, compliance and governance processes \customcitePS{AwsSecure}, and enhanced collaboration and communication management \customcitePS{I109}. The increased prominence of these aspects in recent literature signals a maturation in the understanding and structuring of MLOps practices (Section \ref{RQ1.2-Activities}).

\textit{Roles:} Recent literature emphasizes new roles focusing on quality assurance, compliance, and security concerns, such as data stewards, data quality boards, and compliance teams \customcitePS{AwsSecure}, \customcitePS{DataBricksMLOps}. Although an earlier study identified the MLOps engineer role \cite{kreuzberger2023machine}, our findings explicitly differentiate its responsibilities from those of ML engineers, positioning it within clearly defined workflows and clarifying its interactions with other emerging roles (Section \ref{RQ1.3-RolesAndResponsibilities}).

\textit{Practices and techniques for implementing MLOps:} Prior systematic studies identified practices like workflow automation, artifact tracking, collaboration, and continuous monitoring. Our findings extend these set of practices by highlighting recent developments, including the use of domain-specific reference architectures for specific business contexts \customcitePS{S262}, pipeline optimization strategies to enhance scalability and performance \customcitePS{S57}, containerization and service-oriented architectures \customcitePS{S627}, and systematization of regulated ML tasks to ensure compliance with emerging governance requirements \customcitePS{I40} (Section \ref{RQ2}).

\textit{MLOps adoption challenges and solutions:} Challenges such as ML pipeline complexity, socio-technical barriers, scalability and performance issues, distributed computing environments, and adapting MLOps to specific contexts have been reported in related work. 
Beyond these, our study highlights more recent and nuanced issues such as compliance \customcitePS{AwsLens}, fairness \customcitePS{AmazonAI}, and security of ML systems \customcitePS{I8}, customization of ML platforms \customcitePS{AwsMLOps2022}, pipeline fragmentation due to tool diversity (section \ref{sec:Fragmentation}), and decision-making challenges regarding automation and migration (sections \ref{sec:automationDecision}, \ref{sec:migrationDecision}).
We observe that solutions to these challenges have correspondingly evolved, increasingly incorporating human-centric, transparent, and plan-focused approaches for cultural and cross-silo adaptations \customcitePS{I9}, domain-specific MLOps architectures, frameworks supporting compliance, fairness, and security, as well as human-in-the-loop interventions \customcitePS{DataBricksAuto}. (Section \ref{RQ3}). 

\section{Conclusion} \label{conclusion}

We have systematically selected and analyzed 150 relevant academic and 48 gray literature studies to provide a comprehensive overview of the current MLOps landscape by addressing three main research questions: 1) Conceptualization of MLOps, 2) What are the state-of-the-art best practices and techniques for implementing MLOps?, 3) What are the challenges practitioners face when adopting MLOps, and what solutions have been proposed to address them?
Our findings emphasize emerging aspects of MLOps, i.e., compliance, governance, security, and quality, challenges and practices such as pipeline complexity, domain specific needs, automation trade-offs, and scalability in distributed environments. With these, MLOps practices, roles, and activities evolved toward identifying new roles, collaboration and communication oriented practices, and frameworks that address new emerging and complex aspects of MLOps such as customization and complex infrastructures.
We hope this study serves as a valuable resource for researchers and practitioners, guiding future advancements in MLOps adoption and innovation.

\section*{References}
\renewcommand\bibsection{}

\begin{thebibliography}{35}


\ifx \showCODEN    \undefined \def \showCODEN     #1{\unskip}     \fi
\ifx \showDOI      \undefined \def \showDOI       #1{#1}\fi
\ifx \showISBNx    \undefined \def \showISBNx     #1{\unskip}     \fi
\ifx \showISBNxiii \undefined \def \showISBNxiii  #1{\unskip}     \fi
\ifx \showISSN     \undefined \def \showISSN      #1{\unskip}     \fi
\ifx \showLCCN     \undefined \def \showLCCN      #1{\unskip}     \fi
\ifx \shownote     \undefined \def \shownote      #1{#1}          \fi
\ifx \showarticletitle \undefined \def \showarticletitle #1{#1}   \fi
\ifx \showURL      \undefined \def \showURL       {\relax}        \fi
\providecommand\bibfield[2]{#2}
\providecommand\bibinfo[2]{#2}
\providecommand\natexlab[1]{#1}
\providecommand\showeprint[2][]{arXiv:#2}

\bibitem[MLO(2021)]%
        {MLOpsMicrosoft1}
 \bibinfo{year}{2021}\natexlab{}.
\newblock \showarticletitle{MLOps with Azure Machine Learning}.
\newblock  (\bibinfo{year}{2021}).
\newblock
\newblock
\shownote{[White Paper]}.


\bibitem[ibm(2022)]%
        {ibmindex2022}
 \bibinfo{year}{2022}\natexlab{}.
\newblock \bibinfo{title}{{IBM Global AI Adoption Index 2022}}.
\newblock
\newblock
\urldef\tempurl%
\url{https://www.ibm.com/watson/resources/ai-adoption}
\showURL{%
\tempurl}

\bibitem[Washizaki(2024)]{washizaki2024guide}
Hironori Washizaki.
\newblock \emph{Guide to the Software Engineering Body of Knowledge (SWEBOK Guide), Version 4.0}.
\newblock IEEE Computer Society, Waseda University, Japan, 2024.
Available at: \url{http://www.swebok.org}.



\bibitem[onl(2024)]%
        {onlineAppendix}
 \bibinfo{year}{2024}\natexlab{}.
\newblock \bibinfo{title}{{Online appendix of A Multivocal Literature Review on MLOps: Overview, Solutions and Challenges}}.
\newblock
\newblock
\urldef\tempurl%
\url{https://github.com/beken/MLOpsMLR.git/}
\showURL{%
\tempurl}


\bibitem[Barbara~Kitchenham(2007)]%
        {kitchenhamSLR2007}
\bibfield{author}{\bibinfo{person}{Stuart~Charters Barbara~Kitchenham}.} \bibinfo{year}{2007}\natexlab{}.
\newblock \bibinfo{title}{Guidelines for performing systematic literature reviews in software engineering}.
\newblock
\newblock


\bibitem[Baumann et~al\mbox{.}(2022)]%
        {baumann2022dynamic}
\bibfield{author}{\bibinfo{person}{Nils Baumann}, \bibinfo{person}{Evgeny Kusmenko}, \bibinfo{person}{Jonas Ritz}, \bibinfo{person}{Bernhard Rumpe}, {and} \bibinfo{person}{Moritz~Benedikt Weber}.} \bibinfo{year}{2022}\natexlab{}.
\newblock \showarticletitle{Dynamic data management for continuous retraining}. In \bibinfo{booktitle}{\emph{Proceedings of the 25th International Conference on Model Driven Engineering Languages and Systems: Companion Proceedings}}. \bibinfo{pages}{359--366}.
\newblock


\bibitem[Beck(2000)]%
        {beck2000extreme}
\bibfield{author}{\bibinfo{person}{Kent Beck}.} \bibinfo{year}{2000}\natexlab{}.
\newblock \bibinfo{booktitle}{\emph{Extreme programming explained: embrace change}}.
\newblock \bibinfo{publisher}{addison-wesley professional}.
\newblock


\bibitem[Bourque and Fairley(2004)]%
        {bourque2004swebok}
\bibfield{author}{\bibinfo{person}{Pierre Bourque} {and} \bibinfo{person}{RJNICS Fairley}.} \bibinfo{year}{2004}\natexlab{}.
\newblock \showarticletitle{Swebok}.
\newblock \bibinfo{journal}{\emph{Nd: IEEE Computer society}} (\bibinfo{year}{2004}).
\newblock


\bibitem[Braun and Clarke(2006)]%
        {thematicAnalysisBraun}
\bibfield{author}{\bibinfo{person}{Virginia Braun} {and} \bibinfo{person}{Victoria Clarke}.} \bibinfo{year}{2006}\natexlab{}.
\newblock \showarticletitle{Using thematic analysis in psychology}.
\newblock \bibinfo{journal}{\emph{Qualitative research in psychology}} \bibinfo{volume}{3}, \bibinfo{number}{2} (\bibinfo{year}{2006}), \bibinfo{pages}{77}.
\newblock


\bibitem[Cano et~al\mbox{.}(2020)]%
        {cano2020taxonomy}
\bibfield{author}{\bibinfo{person}{Patricia~Ortegon Cano}, \bibinfo{person}{Ayrton~Mondragon Mejia}, \bibinfo{person}{Silvana De~Gyves~Avila}, \bibinfo{person}{Gloria Eva~Zagal Dominguez}, \bibinfo{person}{Ismael~Solis Moreno}, {and} \bibinfo{person}{Arianne~Navarro Lepe}.} \bibinfo{year}{2020}\natexlab{}.
\newblock \showarticletitle{A Taxonomy on Continuous Integration and Deployment Tools and Frameworks}. In \bibinfo{booktitle}{\emph{International Conference on Software Process Improvement}}. Springer, \bibinfo{pages}{323--336}.
\newblock


\bibitem[Croft et~al\mbox{.}(2022)]%
        {croft2022}
\bibfield{author}{\bibinfo{person}{Roland Croft}, \bibinfo{person}{Yongzheng Xie}, {and} \bibinfo{person}{Muhammad~Ali Babar}.} \bibinfo{year}{2022}\natexlab{}.
\newblock \showarticletitle{Data preparation for software vulnerability prediction: A systematic literature review}.
\newblock \bibinfo{journal}{\emph{IEEE Transactions on Software Engineering}} \bibinfo{volume}{49}, \bibinfo{number}{3} (\bibinfo{year}{2022}), \bibinfo{pages}{1044--1063}.
\newblock


\bibitem[Data and Team(2023)]%
        {IbmMLOps1}
\bibfield{author}{\bibinfo{person}{IBM Data} {and} \bibinfo{person}{AI Team}.} \bibinfo{year}{2023}\natexlab{}.
\newblock \bibinfo{title}{MLOps and the evolution of data science}.
\newblock
\newblock
\newblock
\shownote{IBM, [blog]}.


\bibitem[Dissanayake et~al\mbox{.}(2022)]%
        {dissanayake2022}
\bibfield{author}{\bibinfo{person}{Nesara Dissanayake}, \bibinfo{person}{Asangi Jayatilaka}, \bibinfo{person}{Mansooreh Zahedi}, {and} \bibinfo{person}{M~Ali Babar}.} \bibinfo{year}{2022}\natexlab{}.
\newblock \showarticletitle{Software security patch management-A systematic literature review of challenges, approaches, tools and practices}.
\newblock \bibinfo{journal}{\emph{Information and Software Technology}}  \bibinfo{volume}{144} (\bibinfo{year}{2022}), \bibinfo{pages}{106771}.
\newblock


\bibitem[Duvall et~al\mbox{.}(2007)]%
        {duvall2007continuous}
\bibfield{author}{\bibinfo{person}{Paul~M Duvall}, \bibinfo{person}{Steve Matyas}, {and} \bibinfo{person}{Andrew Glover}.} \bibinfo{year}{2007}\natexlab{}.
\newblock \bibinfo{booktitle}{\emph{Continuous integration: improving software quality and reducing risk}}.
\newblock \bibinfo{publisher}{Pearson Education}.
\newblock


\bibitem[Ereth(2018)]%
        {ereth2018dataops}
\bibfield{author}{\bibinfo{person}{Julian Ereth}.} \bibinfo{year}{2018}\natexlab{}.
\newblock \showarticletitle{DataOps-Towards a Definition.}
\newblock \bibinfo{journal}{\emph{LWDA}}  \bibinfo{volume}{2191} (\bibinfo{year}{2018}), \bibinfo{pages}{104--112}.
\newblock


\bibitem[Faubel and Schmid(2023)]%
        {faubel2023review}
\bibfield{author}{\bibinfo{person}{Leonhard Faubel} {and} \bibinfo{person}{Klaus Schmid}.} \bibinfo{year}{2023}\natexlab{}.
\newblock \showarticletitle{Review protocol: a systematic literature review of MLOps}.
\newblock  (\bibinfo{year}{2023}).
\newblock


\bibitem[Fowler et~al\mbox{.}(2001)]%
        {fowler2001agile}
\bibfield{author}{\bibinfo{person}{Martin Fowler}, \bibinfo{person}{Jim Highsmith}, {et~al\mbox{.}}} \bibinfo{year}{2001}\natexlab{}.
\newblock \showarticletitle{The agile manifesto}.
\newblock \bibinfo{journal}{\emph{Software development}} \bibinfo{volume}{9}, \bibinfo{number}{8} (\bibinfo{year}{2001}), \bibinfo{pages}{28--35}.
\newblock


\bibitem[Garousi et~al\mbox{.}(2019)]%
        {garousi2019guidelines}
\bibfield{author}{\bibinfo{person}{Vahid Garousi}, \bibinfo{person}{Michael Felderer}, {and} \bibinfo{person}{Mika~V M{\"a}ntyl{\"a}}.} \bibinfo{year}{2019}\natexlab{}.
\newblock \showarticletitle{Guidelines for including grey literature and conducting multivocal literature reviews in software engineering}.
\newblock \bibinfo{journal}{\emph{Information and software technology}}  \bibinfo{volume}{106} (\bibinfo{year}{2019}), \bibinfo{pages}{101--121}.
\newblock


\bibitem[John et~al\mbox{.}(2021)]%
        {john2021towards}
\bibfield{author}{\bibinfo{person}{Meenu~Mary John}, \bibinfo{person}{Helena~Holmstr{\"o}m Olsson}, {and} \bibinfo{person}{Jan Bosch}.} \bibinfo{year}{2021}\natexlab{}.
\newblock \showarticletitle{Towards mlops: A framework and maturity model}. In \bibinfo{booktitle}{\emph{2021 47th Euromicro Conference on Software Engineering and Advanced Applications (SEAA)}}. IEEE, \bibinfo{pages}{1--8}.
\newblock


\bibitem[Karamitsos et~al\mbox{.}(2020)]%
        {karamitsos2020applying}
\bibfield{author}{\bibinfo{person}{Ioannis Karamitsos}, \bibinfo{person}{Saeed Albarhami}, {and} \bibinfo{person}{Charalampos Apostolopoulos}.} \bibinfo{year}{2020}\natexlab{}.
\newblock \showarticletitle{Applying DevOps practices of continuous automation for machine learning}.
\newblock \bibinfo{journal}{\emph{Information}} \bibinfo{volume}{11}, \bibinfo{number}{7} (\bibinfo{year}{2020}), \bibinfo{pages}{363}.
\newblock


\bibitem[Steidl et~al\mbox{.}(2024)]{steidl2024past}
Monika Steidl, Rudolf Ramler, and Michael Felderer.
\newblock \emph{The Past, Present, and Future of Research on the Continuous Development of AI}.
\newblock In \emph{Proceedings of the 2024 50th Euromicro Conference on Software Engineering and Advanced Applications (SEAA)}, pages 300--308. IEEE, 2024.



\bibitem[Kolltveit and Li(2022)]%
        {kolltveit2022operationalizing}
\bibfield{author}{\bibinfo{person}{Ask~Berstad Kolltveit} {and} \bibinfo{person}{Jingyue Li}.} \bibinfo{year}{2022}\natexlab{}.
\newblock \showarticletitle{Operationalizing machine learning models: a systematic literature review}. In \bibinfo{booktitle}{\emph{Proceedings of the 1st Workshop on Software Engineering for Responsible AI}}. \bibinfo{pages}{1--8}.
\newblock


\bibitem[Kreuzberger et~al\mbox{.}(2023)]%
        {kreuzberger2023machine}
\bibfield{author}{\bibinfo{person}{Dominik Kreuzberger}, \bibinfo{person}{Niklas K{\"u}hl}, {and} \bibinfo{person}{Sebastian Hirschl}.} \bibinfo{year}{2023}\natexlab{}.
\newblock \showarticletitle{Machine learning operations (mlops): Overview, definition, and architecture}.
\newblock \bibinfo{journal}{\emph{IEEE Access}} (\bibinfo{year}{2023}).
\newblock


\bibitem[Lima et~al\mbox{.}(2022)]%
        {lima2022mlops}
\bibfield{author}{\bibinfo{person}{Anderson Lima}, \bibinfo{person}{Luciano Monteiro}, {and} \bibinfo{person}{Ana~Paula Furtado}.} \bibinfo{year}{2022}\natexlab{}.
\newblock \showarticletitle{MLOps: Practices, Maturity Models, Roles, Tools, and Challenges-A Systematic Literature Review.}
\newblock \bibinfo{journal}{\emph{ICEIS (1)}} (\bibinfo{year}{2022}), \bibinfo{pages}{308--320}.
\newblock


\bibitem[Luley et~al\mbox{.}(2023)]%
        {luley2023concept}
\bibfield{author}{\bibinfo{person}{Paul-Philipp Luley}, \bibinfo{person}{Jan~M Deriu}, \bibinfo{person}{Peng Yan}, \bibinfo{person}{Gerrit~A Schatte}, {and} \bibinfo{person}{Thilo Stadelmann}.} \bibinfo{year}{2023}\natexlab{}.
\newblock \showarticletitle{From concept to implementation: The data-centric development process for AI in industry}. In \bibinfo{booktitle}{\emph{2023 10th IEEE Swiss Conference on Data Science (SDS)}}. IEEE, \bibinfo{pages}{73--76}.
\newblock


\bibitem[Lwakatare et~al\mbox{.}(2020)]%
        {lwakatare2020data}
\bibfield{author}{\bibinfo{person}{Lucy~Ellen Lwakatare}, \bibinfo{person}{Ivica Crnkovic}, \bibinfo{person}{Ellinor R{\aa}nge}, {and} \bibinfo{person}{Jan Bosch}.} \bibinfo{year}{2020}\natexlab{}.
\newblock \showarticletitle{From a data science driven process to a continuous delivery process for machine learning systems}. In \bibinfo{booktitle}{\emph{Product-Focused Software Process Improvement: 21st International Conference, PROFES 2020, Turin, Italy, November 25--27, 2020, Proceedings 21}}. Springer, \bibinfo{pages}{185--201}.
\newblock


\bibitem[Mboweni et~al\mbox{.}(2022)]%
        {mboweni2022systematic}
\bibfield{author}{\bibinfo{person}{Tsakani Mboweni}, \bibinfo{person}{Themba Masombuka}, {and} \bibinfo{person}{Cyrille Dongmo}.} \bibinfo{year}{2022}\natexlab{}.
\newblock \showarticletitle{A systematic review of machine learning devops}. In \bibinfo{booktitle}{\emph{2022 international conference on electrical, computer and energy technologies (ICECET)}}. IEEE, \bibinfo{pages}{1--6}.
\newblock


\bibitem[Neghawi et~al\mbox{.}(2023)]%
        {neghawi2023linking}
\bibfield{author}{\bibinfo{person}{Elie Neghawi}, \bibinfo{person}{Zerui Wang}, \bibinfo{person}{Jun Huang}, {and} \bibinfo{person}{Yan Liu}.} \bibinfo{year}{2023}\natexlab{}.
\newblock \showarticletitle{Linking Team-level and Organization-level Governance in Machine Learning Operations through Explainable AI and Responsible AI Connector}. In \bibinfo{booktitle}{\emph{2023 IEEE 47th Annual Computers, Software, and Applications Conference (COMPSAC)}}. IEEE, \bibinfo{pages}{1223--1230}.
\newblock


\bibitem[Salama et~al\mbox{.}(2021)]%
        {practitionersKhalid}
\bibfield{author}{\bibinfo{person}{Khalid Salama}, \bibinfo{person}{Jarek Kazmierczak}, {and} \bibinfo{person}{Donna Schut}.} \bibinfo{year}{2021}\natexlab{}.
\newblock \showarticletitle{Practitioners guide to MLOps: A framework for continuous delivery and automation of machine learning}.
\newblock  (\bibinfo{year}{2021}).
\newblock
\newblock
\shownote{[White Paper]}.


\bibitem[Sculley et~al\mbox{.}(2015)]%
        {sculley2015hidden}
\bibfield{author}{\bibinfo{person}{David Sculley}, \bibinfo{person}{Gary Holt}, \bibinfo{person}{Daniel Golovin}, \bibinfo{person}{Eugene Davydov}, \bibinfo{person}{Todd Phillips}, \bibinfo{person}{Dietmar Ebner}, \bibinfo{person}{Vinay Chaudhary}, \bibinfo{person}{Michael Young}, \bibinfo{person}{Jean-Francois Crespo}, {and} \bibinfo{person}{Dan Dennison}.} \bibinfo{year}{2015}\natexlab{}.
\newblock \showarticletitle{Hidden technical debt in machine learning systems}.
\newblock \bibinfo{journal}{\emph{Advances in neural information processing systems}}  \bibinfo{volume}{28} (\bibinfo{year}{2015}).
\newblock


\bibitem[Shahin et~al\mbox{.}(2017)]%
        {shahin2017continuous}
\bibfield{author}{\bibinfo{person}{Mojtaba Shahin}, \bibinfo{person}{Muhammad~Ali Babar}, {and} \bibinfo{person}{Liming Zhu}.} \bibinfo{year}{2017}\natexlab{}.
\newblock \showarticletitle{Continuous integration, delivery and deployment: a systematic review on approaches, tools, challenges and practices}.
\newblock \bibinfo{journal}{\emph{IEEE access}}  \bibinfo{volume}{5} (\bibinfo{year}{2017}), \bibinfo{pages}{3909--3943}.
\newblock


\bibitem[Spieker and Gotlieb(2019)]%
        {spieker2019towards}
\bibfield{author}{\bibinfo{person}{Helge Spieker} {and} \bibinfo{person}{Arnaud Gotlieb}.} \bibinfo{year}{2019}\natexlab{}.
\newblock \showarticletitle{Towards testing of deep learning systems with training set reduction}.
\newblock \bibinfo{journal}{\emph{arXiv preprint arXiv:1901.04169}} (\bibinfo{year}{2019}).
\newblock


\bibitem[St{\aa}hl and Bosch(2014)]%
        {staahl2014modeling}
\bibfield{author}{\bibinfo{person}{Daniel St{\aa}hl} {and} \bibinfo{person}{Jan Bosch}.} \bibinfo{year}{2014}\natexlab{}.
\newblock \showarticletitle{Modeling continuous integration practice differences in industry software development}.
\newblock \bibinfo{journal}{\emph{Journal of Systems and Software}}  \bibinfo{volume}{87} (\bibinfo{year}{2014}), \bibinfo{pages}{48--59}.
\newblock


\bibitem[Steidl et~al\mbox{.}(2023)]%
        {steidl2023pipeline}
\bibfield{author}{\bibinfo{person}{Monika Steidl}, \bibinfo{person}{Michael Felderer}, {and} \bibinfo{person}{Rudolf Ramler}.} \bibinfo{year}{2023}\natexlab{}.
\newblock \showarticletitle{The pipeline for the continuous development of artificial intelligence models—Current state of research and practice}.
\newblock \bibinfo{journal}{\emph{Journal of Systems and Software}}  \bibinfo{volume}{199} (\bibinfo{year}{2023}), \bibinfo{pages}{111615}.
\newblock


\bibitem[Vartak(2022)]%
        {WhatIsMLOpsDataOps}
\bibfield{author}{\bibinfo{person}{Manasi Vartak}.} \bibinfo{year}{2022}\natexlab{}.
\newblock \bibinfo{title}{{What is MLOps? DataOps? And Why do They Matter?}}
\newblock
\newblock
\urldef\tempurl%
\url{https://devops.com/what-is-mlops-dataops-and-why-do-they-matter/}
\showURL{%
\tempurl}
\newblock
\shownote{[Online; accessed Mar 2024]}.


\bibitem[Windheuser et~al\mbox{.}(2020)]%
        {MLOpsAWS1}
\bibfield{author}{\bibinfo{person}{Christoph Windheuser}, \bibinfo{person}{Danilo Sato}, \bibinfo{person}{Alex Sadovsky}, \bibinfo{person}{David Cooperberg}, \bibinfo{person}{Greg Willis}, \bibinfo{person}{Josh Poduska}, \bibinfo{person}{Jim Falgout}, {and} \bibinfo{person}{Dylan Tong}.} \bibinfo{year}{2020}\natexlab{}.
\newblock \showarticletitle{MLOps: Continuous Delivery for Machine Learning on AWS}.
\newblock  (\bibinfo{year}{2020}).
\newblock
\newblock
\shownote{[White Paper]}.


\bibitem[Wohlin(2014)]%
        {wohlin2014guidelines}
\bibfield{author}{\bibinfo{person}{Claes Wohlin}.} \bibinfo{year}{2014}\natexlab{}.
\newblock \showarticletitle{Guidelines for snowballing in systematic literature studies and a replication in software engineering}. In \bibinfo{booktitle}{\emph{Proceedings of the 18th international conference on evaluation and assessment in software engineering}}. \bibinfo{pages}{1--10}.
\newblock


\end{thebibliography}

\begin{thebibliography}{100}

\bibitem[PS\arabic{PScounter}]{FIRST-1}
Sasu M{\"a}kinen, Henrik Skogstr{\"o}m, Eero Laaksonen, and Tommi Mikkonen.
\newblock Who needs mlops: What data scientists seek to accomplish and how can mlops help?
\newblock In {\em 2021 IEEE/ACM 1st Workshop on AI Engineering-Software Engineering for AI (WAIN)}, pages 109--112. IEEE, 2021.

\stepcounter{PScounter}

\bibitem[PS\arabic{PScounter}]{FIRST-2}
Damian~A Tamburri.
\newblock Sustainable mlops: Trends and challenges.
\newblock In {\em 2020 22nd International Symposium on Symbolic and Numeric Algorithms for Scientific Computing (SYNASC)}, pages 17--23. IEEE, 2020.

\stepcounter{PScounter} 

\bibitem[PS\arabic{PScounter}]{FIRST-3}
Yue Zhou, Yue Yu, and Bo~Ding.
\newblock Towards mlops: A case study of ml pipeline platform.
\newblock In {\em 2020 International Conference on Artificial Intelligence and Computer Engineering (ICAICE)}, pages 494--500. IEEE, 2020.

\stepcounter{PScounter} 

\bibitem[PS\arabic{PScounter}]{FIRST-4}
Aquilas~Tchanjou Njomou, Alexandra Johanne~Bifona Africa, Bram Adams, and Marios Fokaefs.
\newblock Msr4ml: Reconstructing artifact traceability in machine learning repositories.
\newblock In {\em 2021 IEEE International Conference on Software Analysis, Evolution and Reengineering (SANER)}, pages 536--540. IEEE, 2021.

\stepcounter{PScounter} 

\bibitem[PS\arabic{PScounter}]{FIRST-5}
Ziqi Guo, Tingwen Bao, Wenlong Wu, Chao Jin, and Jay Lee.
\newblock Iai devops: A systematic framework for prognostic model lifecycle management.
\newblock In {\em 2019 Prognostics and System Health Management Conference (PHM-Qingdao)}, pages 1--6. IEEE, 2019.

\stepcounter{PScounter} 

\bibitem[PS\arabic{PScounter}]{FIRST-6}
Rupesh~Raj Karn, Prabhakar Kudva, and Ibrahim Abe~M Elfadel.
\newblock Dynamic autoselection and autotuning of machine learning models for cloud network analytics.
\newblock {\em IEEE Transactions on Parallel and Distributed Systems}, 30(5):1052--1064, 2018.

\stepcounter{PScounter} 

\bibitem[PS\arabic{PScounter}]{FIRST-7}
Lucas~Cardoso Silva, Fernando~Rezende Zagatti, Bruno~Silva Sette, Lucas~Nildaimon dos Santos~Silva, Daniel Lucr{\'e}dio, Diego~Furtado Silva, and Helena de~Medeiros~Caseli.
\newblock Benchmarking machine learning solutions in production.
\newblock In {\em 2020 19th IEEE International Conference on Machine Learning and Applications (ICMLA)}, pages 626--633. IEEE, 2020.

\stepcounter{PScounter} 

\bibitem[PS\arabic{PScounter}]{FIRST-8}
Tam{\'a}s Kar{\'a}csony, Anna~Mira Loesch-Biffar, Christian Vollmar, Soheyl Noachtar, and Jo{\~a}o Paulo~Silva Cunha.
\newblock Deepepil: Towards an epileptologist-friendly ai enabled seizure classification cloud system based on deep learning analysis of 3d videos.
\newblock In {\em 2021 IEEE EMBS International Conference on Biomedical and Health Informatics (BHI)}, pages 1--5. IEEE, 2021.

\stepcounter{PScounter} 

\bibitem[PS\arabic{PScounter}]{FIRST-9}
Chandrasekar Vuppalapati, Anitha Ilapakurti, Karthik Chillara, Sharat Kedari, and Vanaja Mamidi.
\newblock Automating tiny ml intelligent sensors devops using microsoft azure.
\newblock In {\em 2020 ieee international conference on big data (big data)}, pages 2375--2384. IEEE, 2020.

\stepcounter{PScounter} 

\bibitem[PS\arabic{PScounter}]{FIRST-10}
Tuomas Granlund, Aleksi Kopponen, Vlad Stirbu, Lalli Myllyaho, and Tommi Mikkonen.
\newblock Mlops challenges in multi-organization setup: Experiences from two real-world cases.
\newblock In {\em 2021 IEEE/ACM 1st Workshop on AI Engineering-Software Engineering for AI (WAIN)}, pages 82--88. IEEE, 2021.

\stepcounter{PScounter} 

\bibitem[PS\arabic{PScounter}]{FIRST-11}
Lucy~Ellen Lwakatare, Ivica Crnkovic, and Jan Bosch.
\newblock Devops for ai--challenges in development of ai-enabled applications.
\newblock In {\em 2020 International Conference on Software, Telecommunications and Computer Networks (SoftCOM)}, pages 1--6. IEEE, 2020.

\stepcounter{PScounter} 

\bibitem[PS\arabic{PScounter}]{FIRST-12}
Benjamin Benni, Mireille Blay-Fornarino, S{\'e}bastien Mosser, Fr{\'e}d{\'e}ric Pr{\'e}cisio, and G{\"u}nther Jungbluth.
\newblock When devops meets meta-learning: A portfolio to rule them all.
\newblock In {\em 2019 ACM/IEEE 22nd International Conference on Model Driven Engineering Languages and Systems Companion (MODELS-C)}, pages 605--612. IEEE, 2019.

\stepcounter{PScounter} 

\bibitem[PS\arabic{PScounter}]{FIRST-13}
Ilias Gerostathopoulos, Stefan Kugele, Christoph Segler, Tomas Bures, and Alois Knoll.
\newblock Automated trainability evaluation for smart software functions.
\newblock In {\em 2019 34th IEEE/ACM International Conference on Automated Software Engineering (ASE)}, pages 998--1001. IEEE, 2019.

\stepcounter{PScounter} 

\bibitem[PS\arabic{PScounter}]{FIRST-14}
{\'A}lvaro~L{\'o}pez Garc{\'\i}a, Jesus~Marco De~Lucas, Marica Antonacci, Wolfgang Zu~Castell, Mario David, Marcus Hardt, Lara~Lloret Iglesias, Germ{\'a}n Molt{\'o}, Marcin Plociennik, Viet Tran, et~al.
\newblock A cloud-based framework for machine learning workloads and applications.
\newblock {\em IEEE access}, 8:18681--18692, 2020.

\stepcounter{PScounter} 

\bibitem[PS\arabic{PScounter}]{FIRST-15}
Amine Barrak, Ellis~E Eghan, and Bram Adams.
\newblock On the co-evolution of ml pipelines and source code-empirical study of dvc projects.
\newblock In {\em 2021 IEEE International Conference on Software Analysis, Evolution and Reengineering (SANER)}, pages 422--433. IEEE, 2021.

\stepcounter{PScounter} 

\bibitem[PS\arabic{PScounter}]{FIRST-16}
Fuyuki Ishikawa and Yutaka Matsuno.
\newblock Evidence-driven requirements engineering for uncertainty of machine learning-based systems.
\newblock In {\em 2020 IEEE 28th International Requirements Engineering Conference (RE)}, pages 346--351. IEEE, 2020.

\stepcounter{PScounter} 

\bibitem[PS\arabic{PScounter}]{FIRST-17}
Markus Borg, Ronald Jabangwe, Simon {\AA}berg, Arvid Ekblom, Ludwig Hedlund, and August Lidfeldt.
\newblock Test automation with grad-cam heatmaps-a future pipe segment in mlops for vision ai?
\newblock In {\em 2021 IEEE International Conference on Software Testing, Verification and Validation Workshops (ICSTW)}, pages 175--181. IEEE, 2021.

\stepcounter{PScounter} 

\bibitem[PS\arabic{PScounter}]{FIRST-18}
Eero Kauhanen, Jukka~K Nurminen, Tommi Mikkonen, and Matvei Pashkovskiy.
\newblock Regression test selection tool for python in continuous integration process.
\newblock In {\em 2021 IEEE International Conference on Software Analysis, Evolution and Reengineering (SANER)}, pages 618--621. IEEE, 2021.

\stepcounter{PScounter} 

\bibitem[PS\arabic{PScounter}]{FIRST-19}
Bojan Karla{\v{s}}, Matteo Interlandi, Cedric Renggli, Wentao Wu, Ce~Zhang, Deepak Mukunthu Iyappan~Babu, Jordan Edwards, Chris Lauren, Andy Xu, and Markus Weimer.
\newblock Building continuous integration services for machine learning.
\newblock In {\em Proceedings of the 26th ACM SIGKDD International Conference on Knowledge Discovery \& Data Mining}, pages 2407--2415, 2020.

\stepcounter{PScounter} 

\bibitem[PS\arabic{PScounter}]{FIRST-20}
Andrew Chen, Andy Chow, Aaron Davidson, Arjun DCunha, Ali Ghodsi, Sue~Ann Hong, Andy Konwinski, Clemens Mewald, Siddharth Murching, Tomas Nykodym, et~al.
\newblock Developments in mlflow: A system to accelerate the machine learning lifecycle.
\newblock In {\em Proceedings of the fourth international workshop on data management for end-to-end machine learning}, pages 1--4, 2020.

\stepcounter{PScounter} 

\bibitem[PS\arabic{PScounter}]{FIRST-21}
Cedric Renggli, Frances~Ann Hubis, Bojan Karla{\v{s}}, Kevin Schawinski, Wentao Wu, and Ce~Zhang.
\newblock Ease. ml/ci and ease. ml/meter in action: Towards data management for statistical generalization.
\newblock {\em Proceedings of the VLDB Endowment}, 12(12):1962--1965, 2019.

\stepcounter{PScounter} 

\bibitem[PS\arabic{PScounter}]{FIRST-22}
Huaizheng Zhang, Yuanming Li, Yizheng Huang, Yonggang Wen, Jianxiong Yin, and Kyle Guan.
\newblock Mlmodelci: An automatic cloud platform for efficient mlaas.
\newblock In {\em Proceedings of the 28th ACM International Conference on Multimedia}, pages 4453--4456, 2020.

\stepcounter{PScounter} 

\bibitem[PS\arabic{PScounter}]{FIRST-23}
Maurits van~der Goes.
\newblock Scaling enterprise recommender systems for decentralization.
\newblock In {\em Fifteenth ACM Conference on Recommender Systems}, pages 592--594, 2021.

\stepcounter{PScounter} 

\bibitem[PS\arabic{PScounter}]{FIRST-24}
Philipp Ruf, Manav Madan, Christoph Reich, and Djaffar Ould-Abdeslam.
\newblock Demystifying mlops and presenting a recipe for the selection of open-source tools.
\newblock {\em Applied Sciences}, 11(19):8861, 2021.

\stepcounter{PScounter} 

\bibitem[PS\arabic{PScounter}]{FIRST-25}
Emmanuel Raj, David Buffoni, Magnus Westerlund, and Kimmo Ahola.
\newblock Edge mlops: An automation framework for aiot applications.
\newblock In {\em 2021 IEEE International Conference on Cloud Engineering (IC2E)}, pages 191--200. IEEE, 2021.

\stepcounter{PScounter} 

\bibitem[PS\arabic{PScounter}]{FIRST-26}
Silverio Mart{\'\i}nez-Fern{\'a}ndez, Xavier Franch, Andreas Jedlitschka, Marc Oriol, and Adam Trendowicz.
\newblock Developing and operating artificial intelligence models in trustworthy autonomous systems.
\newblock In {\em International Conference on Research Challenges in Information Science}, pages 221--229. Springer, 2021.

\stepcounter{PScounter} 

\bibitem[PS\arabic{PScounter}]{FIRST-27}
Romeo Kienzler and Ivan Nesic.
\newblock Claimed, a visual and scalable component library for trusted ai.
\newblock {\em arXiv preprint arXiv:2103.03281}, 2021.

\stepcounter{PScounter} 

\bibitem[PS\arabic{PScounter}]{FIRST-28}
Alexandra Posoldova.
\newblock Machine learning pipelines: From research to production.
\newblock {\em IEEE Potentials}, 39(6):38--42, 2020.

\stepcounter{PScounter} 

\bibitem[PS\arabic{PScounter}]{FIRST-29}
Yan Liu, Zhijing Ling, Boyu Huo, Boqian Wang, Tianen Chen, and Esma Mouine.
\newblock Building a platform for machine learning operations from open source frameworks.
\newblock {\em IFAC-PapersOnLine}, 53(5):704--709, 2020.

\stepcounter{PScounter} 

\bibitem[PS\arabic{PScounter}]{FIRST-30}
Mirella Sangiovanni, Gerard Schouten, and Willem-Jan van~den Heuvel.
\newblock An iot beehive network for monitoring urban biodiversity: vision, method, and architecture.
\newblock In {\em Symposium and Summer School on Service-Oriented Computing}, pages 33--42. Springer, 2020.

\stepcounter{PScounter} 

\bibitem[PS\arabic{PScounter}]{FIRST-31}
Amitabha Banerjee, Chien-Chia Chen, Chien-Chun Hung, Xiaobo Huang, Yifan Wang, and Razvan Chevesaran.
\newblock Challenges and experiences with $\{$MLOps$\}$ for performance diagnostics in $\{$Hybrid-Cloud$\}$ enterprise software deployments.
\newblock In {\em 2020 USENIX Conference on Operational Machine Learning (OpML 20)}, 2020.

\stepcounter{PScounter} 

\bibitem[PS\arabic{PScounter}]{FIRST-32}
Willem-Jan van~den Heuvel and Damian~A Tamburri.
\newblock Model-driven ml-ops for intelligent enterprise applications: vision, approaches and challenges.
\newblock In {\em International Symposium on Business Modeling and Software Design}, pages 169--181. Springer, 2020.

\stepcounter{PScounter} 

\bibitem[PS\arabic{PScounter}]{FIRST-33}
Waldemar Hummer, Vinod Muthusamy, Thomas Rausch, Parijat Dube, Kaoutar El~Maghraoui, Anupama Murthi, and Punleuk Oum.
\newblock Modelops: Cloud-based lifecycle management for reliable and trusted ai.
\newblock In {\em 2019 IEEE International Conference on Cloud Engineering (IC2E)}, pages 113--120. IEEE, 2019.

\stepcounter{PScounter} 

\bibitem[PS\arabic{PScounter}]{FIRST-34}
Junsung Lim, Hoejoo Lee, Youngmin Won, and Hunje Yeon.
\newblock $\{$MLOp$\}$ lifecycle scheme for vision-based inspection process in manufacturing.
\newblock In {\em 2019 USENIX Conference on Operational Machine Learning (OpML 19)}, pages 9--11, 2019.

\stepcounter{PScounter} 

\bibitem[PS\arabic{PScounter}]{FIRST-35}
Behrouz Derakhshan, Alireza~Rezaei Mahdiraji, Tilmann Rabl, and Volker Markl.
\newblock Continuous deployment of machine learning pipelines.
\newblock In {\em EDBT}, pages 397--408, 2019.

\stepcounter{PScounter} 

\bibitem[PS\arabic{PScounter}]{FIRST-36}
Stuart Jackson, Maha Yaqub, and Cheng-Xi Li.
\newblock The agile deployment of machine learning models in healthcare.
\newblock {\em Frontiers in big Data}, page~7, 2019.

\stepcounter{PScounter} 

\bibitem[PS\arabic{PScounter}]{FIRST-37}
Ola Spjuth, Jens Frid, and Andreas Hellander.
\newblock The machine learning life cycle and the cloud: implications for drug discovery.
\newblock {\em Expert opinion on drug discovery}, 16(9):1071--1079, 2021.

\stepcounter{PScounter} 

\bibitem[PS\arabic{PScounter}]{FIRST-38}
GeumSeong Yoon, Jungsu Han, Seunghyung Lee, and JongWon Kim.
\newblock Devops portal design for smartx ai cluster employing cloud-native machine learning workflows.
\newblock In {\em Advances in Internet, Data and Web Technologies: The 8th International Conference on Emerging Internet, Data and Web Technologies (EIDWT-2020)}, pages 532--539. Springer, 2020.

\stepcounter{PScounter} 

\bibitem[PS\arabic{PScounter}]{FIRST-39}
Krzysztof Czarnecki.
\newblock Software engineering for automated vehicles: Addressing the needs of cars that run on software and data.
\newblock In {\em 2019 IEEE/ACM 41st International Conference on Software Engineering: Companion Proceedings (ICSE-Companion)}, pages 6--8. IEEE, 2019.

\stepcounter{PScounter} 

\bibitem[PS\arabic{PScounter}]{FIRST-40}
Jacopo Tagliabue.
\newblock You do not need a bigger boat: Recommendations at reasonable scale in a (mostly) serverless and open stack.
\newblock In {\em Proceedings of the 15th ACM Conference on Recommender Systems}, pages 598--600, 2021.

\stepcounter{PScounter} 

\bibitem[PS\arabic{PScounter}]{FIRST-41}
Philipp Kohl, Oliver Schmidts, Lars Kl{\"o}ser, Henri Werth, Bodo Kraft, and Albert Z{\"u}ndorf.
\newblock Stamp 4 nlp--an agile framework for rapid quality-driven nlp applications development.
\newblock In {\em International Conference on the Quality of Information and Communications Technology}, pages 156--166. Springer, 2021.

\stepcounter{PScounter} 

\bibitem[PS\arabic{PScounter}]{FIRST-42}
Aiswarya~Raj Munappy, David~Issa Mattos, Jan Bosch, Helena~Holmstr{\"o}m Olsson, and Anas Dakkak.
\newblock From ad-hoc data analytics to dataops.
\newblock In {\em Proceedings of the International Conference on Software and System Processes}, pages 165--174, 2020.

\stepcounter{PScounter} 

\bibitem[PS\arabic{PScounter}]{FIRST-43}
Haoyu Cai, Chao Wang, and Xuehai Zhou.
\newblock Deployment and verification of machine learning tool-chain based on kubernetes distributed clusters: This paper is submitted for possible publication in the special issue on high performance distributed computing.
\newblock {\em CCF Transactions on High Performance Computing}, 3(2):157--170, 2021.

\stepcounter{PScounter} 

\bibitem[PS\arabic{PScounter}]{FIRST-44}
Chaoyu Wu, E~Haihong, and Meina Song.
\newblock An automatic artificial intelligence training platform based on kubernetes.
\newblock In {\em Proceedings of the 2020 2nd International Conference on Big Data Engineering and Technology}, pages 58--62, 2020.

\stepcounter{PScounter} 

\bibitem[PS\arabic{PScounter}]{FIRST-45}
Mlops: Methods and tools of devops for machine learning, Jul 2020.
\newblock [Online; accessed 1. Apr. 2022].

\stepcounter{PScounter} 

\bibitem[PS\arabic{PScounter}]{FIRST-46}
Youngjune Lee, Oh~Joon Kwon, Haeju Lee, Joonyoung Kim, Kangwook Lee, and Kee-Eung Kim.
\newblock Augment \& valuate: A data enhancement pipeline for data-centric ai.
\newblock {\em arXiv preprint arXiv:2112.03837}, 2021.

\stepcounter{PScounter} 

\bibitem[PS\arabic{PScounter}]{FIRST-47}
Tuomas Granlund, Vlad Stirbu, and Tommi Mikkonen.
\newblock Towards regulatory-compliant mlops: Oravizio’s journey from a machine learning experiment to a deployed certified medical product.
\newblock {\em SN computer Science}, 2(5):342, 2021.

\stepcounter{PScounter} 

\bibitem[PS\arabic{PScounter}]{FIRST-48}
Sindhu Ghanta, Sriram Subramanian, Lior Khermosh, Swaminathan Sundararaman, Harshil Shah, Yakov Goldberg, Drew Roselli, and Nisha Talagala.
\newblock Ml health monitor: taking the pulse of machine learning algorithms in production.
\newblock In {\em Applications of Machine Learning}. International Society for Optics and Photonics, 2019.

\stepcounter{PScounter} 

\bibitem[PS\arabic{PScounter}]{FIRST-SB-1}
Grigori Fursin, Herve Guillou, and Nicolas Essayan.
\newblock Codereef: an open platform for portable mlops, reusable automation actions and reproducible benchmarking.
\newblock {\em The Workshop on MLOps Systems at MLSys’20}, 2020.

\stepcounter{PScounter} 

\bibitem[PS\arabic{PScounter}]{FIRST-SB-2}
Denis Baylor, Kevin Haas, Konstantinos Katsiapis, Sammy Leong, Rose Liu, Clemens Menwald, Hui Miao, Neoklis Polyzotis, Mitchell Trott, and Martin Zinkevich.
\newblock Continuous training for production ml in the tensorflow extended (tfx) platform.
\newblock In {\em 2019 USENIX Conference on Operational Machine Learning (OpML 19)}, pages 51--53, 2019.

\stepcounter{PScounter} 

\bibitem[PS\arabic{PScounter}]{FIRST-SB-3}
Rama Akkiraju, Vibha Sinha, Anbang Xu, Jalal Mahmud, Pritam Gundecha, Zhe Liu, Xiaotong Liu, and John Schumacher.
\newblock Characterizing machine learning processes: A maturity framework.
\newblock In {\em International Conference on Business Process Management}, pages 17--31. Springer, 2020.

\stepcounter{PScounter} 

\bibitem[PS\arabic{PScounter}]{FIRST-SB-4}
R~Ciucu, FC~Adochiei, Ioana-Raluca Adochiei, F~Argatu, GC~Seri{\c{t}}an, B~Enache, S~Grigorescu, and Violeta~Vasilica Argatu.
\newblock Innovative devops for artificial intelligence.
\newblock {\em The Scientific Bulletin of Electrical Engineering Faculty}, 19(1):58--63, 2019.

\stepcounter{PScounter} 

\bibitem[PS\arabic{PScounter}]{FIRST-SB-5}
Emmanuel Raj, David Buffoni, Magnus Westerlund, and Kimmo Ahola.
\newblock Edge mlops: An automation framework for aiot applications.
\newblock In {\em IEEE International Conference on Cloud Engineering (IC2E)}, pages 191--200. IEEE, 2021.

\stepcounter{PScounter} 

\bibitem[PS\arabic{PScounter}]{FIRST-SB-6}
Leonel Aguilar, David Dao, Shaoduo Gan, Nezihe~Merve Gurel, Nora Hollenstein, Jiawei Jiang, Bojan Karlas, Thomas Lemmin, Tian Li, Yang Li, et~al.
\newblock Ease. ml: A lifecycle management system for mldev and mlops.
\newblock {\em Proc. of Innovative Data Systems Research}, 2021.

\stepcounter{PScounter} 

\bibitem[PS\arabic{PScounter}]{FIRST-SB-7}
Rohith Sothilingam, Eric Yu, and Arik Senderovich.
\newblock Towards higher maturity for machine learning: A conceptual modelling approach.
\newblock {\em The iJournal: Student Journal of the Faculty of Information}, 5(1):80--97, 2019.

\stepcounter{PScounter} 

\bibitem[PS\arabic{PScounter}]{FIRST-SB-8}
Emmanuel Raj, Magnus Westerlund, and Leonardo Espinosa-Leal.
\newblock Reliable fleet analytics for edge iot solutions.
\newblock {\em arXiv preprint arXiv:2101.04414}, 2021.

\stepcounter{PScounter} 

\bibitem[PS\arabic{PScounter}]{FIRST-SB-9}
Willem-Jan van~den Heuvel and Damian~A Tamburri.
\newblock Model-driven ml-ops for intelligent enterprise applications: vision, approaches and challenges.
\newblock In {\em Business Modeling and Software Design: 10th International Symposium, BMSD 2020, Berlin, Germany, July 6-8, 2020, Proceedings 10}, pages 169--181. Springer, 2020.

\stepcounter{PScounter} 

\bibitem[PS\arabic{PScounter}]{FIRST-SB-10}
Cedric Renggli, Bojan Karla{\v{s}}, Bolin Ding, Feng Liu, Kevin Schawinski, Wentao Wu, and Ce~Zhang.
\newblock Continuous integration of machine learning models with ease. ml/ci: Towards a rigorous yet practical treatment.
\newblock {\em Proceedings of Machine Learning and Systems}, 1:322--333, 2019.

\stepcounter{PScounter} 

\bibitem[PS\arabic{PScounter}]{FIRST-SB-11}
Cedric Renggli, Luka Rimanic, Nezihe~Merve G{\"u}rel, Bojan Karla{\v{s}}, Wentao Wu, and Ce~Zhang.
\newblock A data quality-driven view of mlops.
\newblock {\em IEEE Data Engineering Bulletin}, 2021.

\stepcounter{PScounter} 

\bibitem[PS\arabic{PScounter}]{FIRST-SB-12}
Fuyuki Ishikawa and Yutaka Matsuno.
\newblock Continuous argument engineering: Tackling uncertainty in machine learning based systems.
\newblock In {\em Computer Safety, Reliability, and Security: SAFECOMP 2018 Workshops, ASSURE, DECSoS, SASSUR, STRIVE, and WAISE, V{\"a}ster{\aa}s, Sweden, September 18, 2018, Proceedings 37}, pages 14--21. Springer, 2018.

\stepcounter{PScounter} 

\bibitem[PS\arabic{PScounter}]{FIRST-SB-13}
Christian Weber, Pascal Hirmer, Peter Reimann, and Holger Schwarz.
\newblock A new process model for the comprehensive management of machine learning models.
\newblock In {\em ICEIS (1)}, pages 415--422, 2019.

\stepcounter{PScounter} 

\bibitem[PS\arabic{PScounter}]{FIRST-SB-14}
Denis Baylor, Eric Breck, Heng-Tze Cheng, Noah Fiedel, Chuan~Yu Foo, Zakaria Haque, Salem Haykal, Mustafa Ispir, Vihan Jain, Levent Koc, et~al.
\newblock Tfx: A tensorflow-based production-scale machine learning platform.
\newblock In {\em Proceedings of the 23rd ACM SIGKDD International Conference on Knowledge Discovery and Data Mining}, pages 1387--1395, 2017.

\stepcounter{PScounter} 

\bibitem[PS\arabic{PScounter}]{FIRST-SB-15}
Tim Kraska, Ameet Talwalkar, John~C Duchi, Rean Griffith, Michael~J Franklin, and Michael~I Jordan.
\newblock Mlbase: A distributed machine-learning system.
\newblock In {\em Cidr}, volume~1, pages 2--1, 2013.

\stepcounter{PScounter} 

\bibitem[PS\arabic{PScounter}]{FIRST-SB-16}
Micah~J Smith, J{\"u}rgen Cito, Kelvin Lu, and Kalyan Veeramachaneni.
\newblock Enabling collaborative data science development with the ballet framework.
\newblock {\em Proceedings of the ACM on Human-Computer Interaction}, 5(CSCW2):1--39, 2021.

\stepcounter{PScounter} 

\bibitem[PS\arabic{PScounter}]{FIRST-SB-17}
Can Kaymakci, Simon Wenninger, and Alexander Sauer.
\newblock A holistic framework for ai systems in industrial applications.
\newblock In {\em International Conference on Wirtschaftsinformatik}, pages 78--93. Springer, 2021.

\stepcounter{PScounter} 

\bibitem[PS\arabic{PScounter}]{FIRST-SB-18}
Pulkit Agrawal, Rajat Arya, Aanchal Bindal, Sandeep Bhatia, Anupriya Gagneja, Joseph Godlewski, Yucheng Low, Timothy Muss, Mudit~Manu Paliwal, Sethu Raman, et~al.
\newblock Data platform for machine learning.
\newblock In {\em Proceedings of the 2019 international conference on management of data}, pages 1803--1816, 2019.

\stepcounter{PScounter} 

\bibitem[PS\arabic{PScounter}]{FIRST-GL-1}
Sridhar Alla and Suman~Kalyan Adari.
\newblock {\em Beginning MLOps with MLFlow}.
\newblock Springer, 2021.

\stepcounter{PScounter} 

\bibitem[PS\arabic{PScounter}]{FIRST-GL-3}
{\em MLOps for IT Teams: How to Transform the Machine Learning Lifecycle}.
\newblock DataRobot, 2021.

\stepcounter{PScounter} 

\bibitem[PS\arabic{PScounter}]{FIRST-GL-4}
{\em Practical MLOps - How to get ready for production models}.
\newblock Valohai, 2020.

\stepcounter{PScounter} 

\bibitem[PS\arabic{PScounter}]{FIRST-GL-5}
Niko Vuokko, Jukka Remes, Alexander Finn, Harry Souris, Hossein Yousefi, Joanna Purosto, and Pauliina Alanen.
\newblock Reliable and scalable ai with mlops, 2021.
\newblock SILO AI, [ebook].

\stepcounter{PScounter} 

\bibitem[PS\arabic{PScounter}]{FIRST-GL-6}
Zechu Li, Xiao-Yang Liu, Jiahao Zheng, Zhaoran Wang, Anwar Walid, and Jian Guo.
\newblock Finrl-podracer: High performance and scalable deep reinforcement learning for quantitative finance.
\newblock {\em arXiv preprint arXiv:2111.05188}, 2021.

\stepcounter{PScounter} 

\bibitem[PS\arabic{PScounter}]{FIRST-GL-7}
Vlad Stirbu, Tuomas Granlund, Jere Hel{\'e}n, and Tommi Mikkonen.
\newblock Extending soup to ml models when designing certified medical systems.
\newblock In {\em 2021 IEEE/ACM 3rd International Workshop on Software Engineering for Healthcare (SEH)}, pages 32--35. IEEE, 2021.

\stepcounter{PScounter} 

\bibitem[PS\arabic{PScounter}]{FIRST-GL-8}
Markus Borg.
\newblock Agility in software 2.0--notebook interfaces and mlops with buttresses and rebars.
\newblock In {\em International Conference on Lean and Agile Software Development}, pages 3--16. Springer, 2022.

\stepcounter{PScounter} 

\bibitem[PS\arabic{PScounter}]{FIRST-GL-9}
Frances~Ann Hubis, Wentao Wu, and Ce~Zhang.
\newblock Quantitative overfitting management for human-in-the-loop ml application development with ease. ml/meter.
\newblock {\em arXiv preprint arXiv:1906.00299}, 2019.

\stepcounter{PScounter} 

\bibitem[PS\arabic{PScounter}]{FIRST-GL-10}
Yizheng Huang, Huaizheng Zhang, Yonggang Wen, Peng Sun, and Nguyen Binh~Duong TA.
\newblock Modelci-e: Enabling continual learning in deep learning serving systems.
\newblock {\em arXiv preprint arXiv:2106.03122}, 2021.

\stepcounter{PScounter} 

\bibitem[PS\arabic{PScounter}]{FIRST-GL-11}
Dennis Muiruri, Lucy~Ellen Lwakatare, Jukka~K Nurminen, and Tommi Mikkonen.
\newblock Practices and infrastructures for machine learning systems: An interview study in finnish organizations.
\newblock {\em Computer}, 55(6):18--29, 2022.

\stepcounter{PScounter} 

\bibitem[PS\arabic{PScounter}]{FIRST-GL-12}
Nikhil Muralidhar, Sathappah Muthiah, Patrick Butler, Manish Jain, Yu~Yu, Katy Burne, Weipeng Li, David Jones, Prakash Arunachalam, Hays'~Skip' McCormick, et~al.
\newblock Using antipatterns to avoid mlops mistakes.
\newblock {\em arXiv preprint arXiv:2107.00079}, 2021.

\stepcounter{PScounter} 

\bibitem[PS\arabic{PScounter}]{FIRST-GL-13}
Georgios Symeonidis, Evangelos Nerantzis, Apostolos Kazakis, and George~A Papakostas.
\newblock Mlops-definitions, tools and challenges.
\newblock In {\em 2022 IEEE 12th Annual Computing and Communication Workshop and Conference (CCWC)}, pages 0453--0460. IEEE, 2022.

\stepcounter{PScounter} 

\bibitem[PS\arabic{PScounter}]{FIRST-GL-14}
Razvan Ciobanu, Alexandru Purdila, Laurentiu Piciu, and Andrei Damian.
\newblock Solis--the mlops journey from data acquisition to actionable insights.
\newblock {\em arXiv preprint arXiv:2112.11925}, 2021.

\stepcounter{PScounter} 

\bibitem[PS\arabic{PScounter}]{FIRST-GL-15}
Xiao-Yang Liu, Zechu Li, Zhuoran Yang, Jiahao Zheng, Zhaoran Wang, Anwar Walid, Jian Guo, and Michael~I Jordan.
\newblock Elegantrl-podracer: Scalable and elastic library for cloud-native deep reinforcement learning.
\newblock {\em arXiv preprint arXiv:2112.05923}, 2021.

\stepcounter{PScounter} 

\bibitem[PS\arabic{PScounter}]{FIRST-GL-16}
Peizheng Li, Jonathan Thomas, Xiaoyang Wang, Ahmed Khalil, Abdelrahim Ahmad, Rui Inacio, Shipra Kapoor, Arjun Parekh, Angela Doufexi, Arman Shojaeifard, et~al.
\newblock Rlops: Development life-cycle of reinforcement learning aided open ran.
\newblock {\em arXiv preprint arXiv:2111.06978}, 2021.

\stepcounter{PScounter} 

\bibitem[PS\arabic{PScounter}]{FIRST-GL-17}
Chulhong Min, Akhil Mathur, Utku~G{\"u}nay Acer, Alessandro Montanari, and Fahim Kawsar.
\newblock Sensix++: Bringing mlops and multi-tenant model serving to sensory edge devices.
\newblock {\em ACM Transactions on Embedded Computing Systems}, 22(6):1--27, 2023.

\stepcounter{PScounter} 

\bibitem[PS\arabic{PScounter}]{FIRST-GL-19}
Grigori Fursin.
\newblock The collective knowledge project: Making ml models more portable and reproducible with open apis, reusable best practices and mlops.
\newblock {\em arXiv preprint arXiv:2006.07161}, 2020.

\stepcounter{PScounter} 

\bibitem[PS\arabic{PScounter}]{FIRST-GL-22}
Fan Yun, Toshiki Shibahara, Yuichi Ohsita, Daiki Chiba, Mitsuaki Akiyama, and Masayuki Murata.
\newblock Understanding machine learning model updates based on changes in feature attributions.
\newblock 2020.

\stepcounter{PScounter} 

\bibitem[PS\arabic{PScounter}]{MLOpsAWS}
Christoph Windheuser, Danilo Sato, Alex Sadovsky, David Cooperberg, Greg Willis, Josh Poduska, Jim Falgout, and Dylan Tong.
\newblock Mlops: Continuous delivery for machine learning on aws.
\newblock 2020.
\newblock [White Paper].

\stepcounter{PScounter} 

\bibitem[PS\arabic{PScounter}]{FIRST-GL-26}
Sumeet Agrawal and Anant Mittal.
\newblock Mlops: 5 steps to operationalize machine learning models.
\newblock 2020.
\newblock [White Paper].

\stepcounter{PScounter} 

\bibitem[PS\arabic{PScounter}]{MLOpsMicrosoft}
Mlops with azure machine learning.
\newblock 2021.
\newblock [White Paper].

\stepcounter{PScounter} 

\bibitem[PS\arabic{PScounter}]{FIRST-GL-28}
Rohit Panikkar, Tamim Saleh, Maxime Szybowski, and Rob Whiteman.
\newblock Operationalizing machine learning in processes.
\newblock {\em McKinsey\&Company, September}, 2021.

\stepcounter{PScounter} 

\bibitem[PS\arabic{PScounter}]{FIRST-GL-29}
Khalid Salama, Jarek Kazmierczak, and Donna Schut.
\newblock Practitioners guide to mlops: A framework for continuous delivery and automation of machine learning.
\newblock 2021.
\newblock [White Paper].

\stepcounter{PScounter} 

\bibitem[PS\arabic{PScounter}]{FIRST-GL-23}
Building versus buying an ml management platform, 2020.
\newblock Algorithmia Inc, [white paper].

\stepcounter{PScounter} 

\bibitem[PS\arabic{PScounter}]{FIRST-GL-33}
The complete guide to machine learning operations, 2020.
\newblock run.ai, [white paper].

\stepcounter{PScounter} 

\bibitem[PS\arabic{PScounter}]{FIRST-WA-2}
Stephen Watts.
\newblock {What Is MLOps? Machine Learning Operations Explained}, Jul 2020.
\newblock [Online; accessed 1. Apr. 2022].

\stepcounter{PScounter} 

\bibitem[PS\arabic{PScounter}]{FIRST-WA-3}
Harshit Tyagi.
\newblock {What is MLOps? Machine Learning Operations Explained}, Mar 2021.
\newblock [Online; accessed 1. Apr. 2022].

\stepcounter{PScounter} 

\bibitem[PS\arabic{PScounter}]{FIRST-WA-4}
Alessandro Gaggia.
\newblock Mlops essentials: four pillars for machine learning operations on aws, Oct 2021.
\newblock [Online; accessed 1. Apr. 2022].

\stepcounter{PScounter} 

\bibitem[PS\arabic{PScounter}]{FIRST-WA-5}
Erin Wolpert.
\newblock An overview of machine learning operations - building mature machine learning production systems through mlops, Nov 2021.
\newblock [Online; accessed 1. Apr. 2022].

\stepcounter{PScounter} 

\bibitem[PS\arabic{PScounter}]{FIRST-WA-6}
Ville~Tuulos Outerbounds.
\newblock Mlops vs. devops: Why data makes it different, Oct 2021.
\newblock [Online; accessed 1. Apr. 2022].

\stepcounter{PScounter} 

\bibitem[PS\arabic{PScounter}]{FIRST-WA-9}
Navdeep~Singh Gill.
\newblock {MLOps Platform - Productionizing Machine Learning Models}, June 2023.
\newblock [Online; accessed Nov 2023].

\stepcounter{PScounter} 

\bibitem[PS\arabic{PScounter}]{FIRST-WA-11}
Mlops: Methods and tools of devops for machine learning, Jul 2020.
\newblock [Online; accessed 1. Apr. 2022].

\stepcounter{PScounter} 

\bibitem[PS\arabic{PScounter}]{FIRST-WA-15}
Seldon~Technologies Ltd.
\newblock Mlops software: Open source vs enterprise.
\newblock \url{https://www.seldon.io/mlops-software-comparison}, 2022.
\newblock Accessed: 2023.

\stepcounter{PScounter} 

\bibitem[PS\arabic{PScounter}]{FIRST-WA-17}
Danilo Sato, Arif Wider, and Christoph Windheuser.
\newblock Continuous delivery for machine learning.
\newblock \url{https://martinfowler.com/articles/cd4ml.html}, 2019.
\newblock Accessed: 2023.

\stepcounter{PScounter} 

\bibitem[PS\arabic{PScounter}]{pillarsofMLOpsDominick}
Dominick Rocco.
\newblock The 4 pillars of mlops: How to deploy ml models to production, May 2021.
\newblock [Online; accessed 1. Apr. 2022].

\stepcounter{PScounter} 

\bibitem[PS\arabic{PScounter}]{GoogleCloud}
Mlops: Continuous delivery and automation pipelines in machine learning, Jan 2020.
\newblock [Online; accessed 1. Apr. 2022].

\stepcounter{PScounter} 

\bibitem[PS\arabic{PScounter}]{I9}
Beatriz~MA Matsui and Denise~H Goya.
\newblock Mlops: A guide to its adoption in the context of responsible ai.
\newblock In {\em Proceedings of the 1st Workshop on Software Engineering for Responsible AI}, pages 45--49, 2022.

\stepcounter{PScounter} 

\bibitem[PS\arabic{PScounter}]{S58}
Rohith Sothilingam, Vik Pant, and Eric Yu.
\newblock Using i* to analyze collaboration challenges in mlops project teams.
\newblock 2022.

\stepcounter{PScounter} 

\bibitem[PS\arabic{PScounter}]{I8}
Xinrui Zhang and Jason Jaskolka.
\newblock Conceptualizing the secure machine learning operations (secmlops) paradigm.
\newblock In {\em 2022 IEEE 22nd International Conference on Software Quality, Reliability and Security (QRS)}, pages 127--138. IEEE, 2022.

\stepcounter{PScounter} 

\bibitem[PS\arabic{PScounter}]{AwsSecure}
David Ping, Stefan Natu, Qingwei Li, Saeed Aghabozori, Vadim Dabravolski, Simon Zamarin, Ibrahim Gabr, and Amir Imani.
\newblock Build a secure enterprise machine learning platform on aws, 2021.
\newblock Amazon Web Services. [AWS Technical Guide].

\stepcounter{PScounter} 

\bibitem[PS\arabic{PScounter}]{AwsMLOps2021}
Bruno Klein.
\newblock Aws prescriptive guidance - planning for successful mlops, 2021.
\newblock Amazon Web Services.

\stepcounter{PScounter} 

\bibitem[PS\arabic{PScounter}]{AwsMLOps2022}
Natalie~Heard Vin~Sharma, Rob~Ferguson.
\newblock Mlops: Emerging trends in data, code, and infrastructure., 2022.
\newblock Amazon Web Services.

\stepcounter{PScounter} 

\bibitem[PS\arabic{PScounter}]{Canonical}
Canonical Ubuntu.
\newblock A guide to mlops, 2023.

\stepcounter{PScounter} 

\bibitem[PS\arabic{PScounter}]{DataBricksAuto}
Databricks.
\newblock How to automate your machine learning pipeline, 2022.
\newblock [ebook].

\stepcounter{PScounter} 

\bibitem[PS\arabic{PScounter}]{AwsLens}
Amazon~Web Services.
\newblock Machine learning lens - aws well-architected framework, 2023.

\stepcounter{PScounter} 

\bibitem[PS\arabic{PScounter}]{AwsApn}
Sergio Zavota, Ajish Palakadan, Salman Taherian, and Mandeep Sehmi.
\newblock Machine learning lens - aws well-architected framework, 2023.
\newblock Amazon Web Services.

\stepcounter{PScounter} 

\bibitem[PS\arabic{PScounter}]{G2}
What is mlops?, 2023.
\newblock Databricks, [website].

\stepcounter{PScounter} 

\bibitem[PS\arabic{PScounter}]{DataBricksMLOps}
Joseph Bradley, Rafi Kurlansik, Matt Thomson, and Niall Turbitt.
\newblock The big book of mlops, 2023.
\newblock Databricks, [ebook].

\stepcounter{PScounter} 

\bibitem[PS\arabic{PScounter}]{IbmMLOps}
IBM Data and AI~Team.
\newblock Mlops and the evolution of data science, 2023.
\newblock IBM, [blog].

\stepcounter{PScounter} 

\bibitem[PS\arabic{PScounter}]{G3}
Machine learning operations (mlops), 2023.

\stepcounter{PScounter} 

\bibitem[PS\arabic{PScounter}]{I40}
Elie Neghawi, Zerui Wang, Jun Huang, and Yan Liu.
\newblock Linking team-level and organization-level governance in machine learning operations through explainable ai and responsible ai connector.
\newblock In {\em 2023 IEEE 47th Annual Computers, Software, and Applications Conference (COMPSAC)}, pages 1223--1230. IEEE, 2023.

\stepcounter{PScounter} 

\bibitem[PS\arabic{PScounter}]{I48}
Janvi Prasad, Arushi Jain, and Ushus~Elizabeth Zachariah.
\newblock Comparative evaluation of machine learning development lifecycle tools.
\newblock In {\em 2022 International Conference on Recent Trends in Microelectronics, Automation, Computing and Communications Systems (ICMACC)}, pages 1--6. IEEE, 2022.

\stepcounter{PScounter} 

\bibitem[PS\arabic{PScounter}]{shivakumar2020web}
Shailesh~Kumar Shivakumar.
\newblock Web performance monitoring and infrastructure planning.
\newblock In {\em Modern Web Performance Optimization}, pages 175--212. Springer, 2020.

\stepcounter{PScounter} 

\bibitem[PS\arabic{PScounter}]{I18}
Seol Roh, Ki-Moon Jeong, Hye-Young Cho, and Eui-Nam Huh.
\newblock An efficient microservices architecture for mlops.
\newblock In {\em 2023 Fourteenth International Conference on Ubiquitous and Future Networks (ICUFN)}, pages 652--654. IEEE, 2023.

\stepcounter{PScounter} 

\bibitem[PS\arabic{PScounter}]{I24}
Qinghua Lu, Liming Zhu, Xiwei Xu, Jon Whittle, David Douglas, and Conrad Sanderson.
\newblock Software engineering for responsible ai: An empirical study and operationalised patterns.
\newblock In {\em Proceedings of the 44th International Conference on Software Engineering: Software Engineering in Practice}, pages 241--242, 2022.

\stepcounter{PScounter} 

\bibitem[PS\arabic{PScounter}]{I22}
Gorka Z{\'a}rate, Ra{\'u}l Mi{\~n}{\'o}n, Josu D{\'\i}az-de Arcaya, and Ana~I Torre-Bastida.
\newblock K2e: Building mlops environments for governing data and models catalogues while tracking versions.
\newblock In {\em 2022 IEEE 19th International Conference on Software Architecture Companion (ICSA-C)}, pages 206--209. IEEE, 2022.

\stepcounter{PScounter} 

\bibitem[PS\arabic{PScounter}]{I74}
Sam Leroux, Pieter Simoens, Meelis Lootus, Kartik Thakore, and Akshay Sharma.
\newblock Tinymlops: Operational challenges for widespread edge ai adoption.
\newblock In {\em 2022 IEEE International Parallel and Distributed Processing Symposium Workshops (IPDPSW)}, pages 1003--1010. IEEE, 2022.

\stepcounter{PScounter} 

\bibitem[PS\arabic{PScounter}]{I53}
Mariam Barry, Albert Bifet, and Jean-Luc Billy.
\newblock Streamai: Dealing with challenges of continual learning systems for serving ai in production.
\newblock In {\em 2023 IEEE/ACM 45th International Conference on Software Engineering: Software Engineering in Practice (ICSE-SEIP)}, pages 134--137. IEEE, 2023.

\stepcounter{PScounter} 

\bibitem[PS\arabic{PScounter}]{I66}
Mariam Barry, Jacob Montiel, Albert Bifet, Sameer Wadkar, Nikolay Manchev, Max Halford, Raja Chiky, Saad~EL Jaouhari, Katherine~B Shakman, Joudi Al~Fehaily, et~al.
\newblock Streammlops: Operationalizing online learning for big data streaming \& real-time applications.
\newblock In {\em 2023 IEEE 39th International Conference on Data Engineering (ICDE)}, pages 3508--3521. IEEE, 2023.

\stepcounter{PScounter} 

\bibitem[PS\arabic{PScounter}]{I47}
Ayush Shridhar and Deepak Nadig.
\newblock Heuristic-based resource allocation for cloud-native machine learning workloads.
\newblock In {\em 2022 IEEE International Conference on Advanced Networks and Telecommunications Systems (ANTS)}, pages 415--418. IEEE, 2022.

\stepcounter{PScounter} 

\bibitem[PS\arabic{PScounter}]{I16}
Philipp Ruf, Christoph Reich, and Djaffar Ould-Abdeslam.
\newblock Aspects of module placement in machine learning operations for cyber physical systems.
\newblock In {\em 2022 11th Mediterranean Conference on Embedded Computing (MECO)}, pages 1--6. IEEE, 2022.

\stepcounter{PScounter} 

\bibitem[PS\arabic{PScounter}]{I43}
Ilias Syrigos, Nikolaos Angelopoulos, and Thanasis Korakis.
\newblock Optimization of execution for machine learning applications in the computing continuum.
\newblock In {\em 2022 IEEE Conference on Standards for Communications and Networking (CSCN)}, pages 118--123. IEEE, 2022.

\stepcounter{PScounter} 

\bibitem[PS\arabic{PScounter}]{I92}
Alexandre Carqueja, Bruno Cabral, Jo{\~a}o~Paulo Fernandes, and Nuno Louren{\c{c}}o.
\newblock On the democratization of machine learning pipelines.
\newblock In {\em 2022 IEEE Symposium Series on Computational Intelligence (SSCI)}, pages 455--462. IEEE, 2022.

\stepcounter{PScounter} 

\bibitem[PS\arabic{PScounter}]{S565}
Fotis Psallidas, Megan~Eileen Leszczynski, Mohammad~Hossein Namaki, Avrilia Floratou, Ashvin Agrawal, Konstantinos Karanasos, Subru Krishnan, Pavle Subotic, Markus Weimer, Yinghui Wu, et~al.
\newblock Demonstration of geyser: Provenance extraction and applications over data science scripts.
\newblock In {\em Companion of the 2023 International Conference on Management of Data}, pages 123--126, 2023.

\stepcounter{PScounter} 

\bibitem[PS\arabic{PScounter}]{A36}
Dezhan Tu, Yeye He, Weiwei Cui, Song Ge, Haidong Zhang, Shi Han, Dongmei Zhang, and Surajit Chaudhuri.
\newblock Auto-validate by-history: Auto-program data quality constraints to validate recurring data pipelines.
\newblock In {\em Proceedings of the 29th ACM SIGKDD Conference on Knowledge Discovery and Data Mining}, pages 4991--5003, 2023.

\stepcounter{PScounter} 

\bibitem[PS\arabic{PScounter}]{S627}
Alec Gunny, Dylan Rankin, Philip Harris, Erik Katsavounidis, Ethan Marx, Muhammed Saleem, Michael Coughlin, and William Benoit.
\newblock A software ecosystem for deploying deep learning in gravitational wave physics.
\newblock In {\em Proceedings of the 12th Workshop on AI and Scientific Computing at Scale using Flexible Computing Infrastructures}, pages 9--17, 2022.

\stepcounter{PScounter} 

\bibitem[PS\arabic{PScounter}]{W127}
Steve Harris, Tim Bonnici, Thomas Keen, Watjana Lilaonitkul, Mark~J White, and Nel Swanepoel.
\newblock Clinical deployment environments: Five pillars of translational machine learning for health.
\newblock {\em Frontiers in Digital Health}, 4:939292, 2022.

\stepcounter{PScounter} 

\bibitem[PS\arabic{PScounter}]{S262}
Tim Raffin, Tobias Reichenstein, Jonas Werner, Alexander K{\"u}hl, and J{\"o}rg Franke.
\newblock A reference architecture for the operationalization of machine learning models in manufacturing.
\newblock {\em Procedia CIRP}, 115:130--135, 2022.

\stepcounter{PScounter} 

\bibitem[PS\arabic{PScounter}]{I36}
Rustem Dautov, Erik~Johannes Husom, and Fotis Gonidis.
\newblock Towards mlops in mobile development with a plug-in architecture for data analytics.
\newblock In {\em 2022 6th International Conference on Computer, Software and Modeling (ICCSM)}, pages 22--27. IEEE, 2022.

\stepcounter{PScounter} 

\bibitem[PS\arabic{PScounter}]{S57}
Fabian Stieler and Bernhard Bauer.
\newblock 18th international conference on evaluation of novel approaches to software engineering (enase 2023).
\newblock 04 2023.

\stepcounter{PScounter} 

\bibitem[PS\arabic{PScounter}]{I171}
Xiaoyu Zhang, Jorge~Piazentin Ono, Huan Song, Liang Gou, Kwan-Liu Ma, and Liu Ren.
\newblock Sliceteller: A data slice-driven approach for machine learning model validation.
\newblock {\em IEEE Transactions on Visualization and Computer Graphics}, 29(1):842--852, 2022.

\stepcounter{PScounter} 

\bibitem[PS\arabic{PScounter}]{S660}
Ruibo Chen and Wenjun Wu.
\newblock Parallelizing automatic model management system for aiops on microservice platforms.
\newblock In {\em European Conference on Parallel Processing}, pages 376--387. Springer, 2021.

\stepcounter{PScounter} 

\bibitem[PS\arabic{PScounter}]{I19}
John Wickstr{\"o}m, Magnus Westerlund, and Emmanuel Raj.
\newblock Decentralizing machine learning operations using web3 for iot platforms.
\newblock In {\em 2022 IEEE International Conference on Cloud Computing Technology and Science (CloudCom)}, pages 238--245. IEEE, 2022.

\stepcounter{PScounter} 

\bibitem[PS\arabic{PScounter}]{I31}
Nikos Psaromanolakis, Vasileios Theodorou, Dimitris Laskaratos, Ioannis Kalogeropoulos, Maria-Eleftheria Vlontzou, Eleni Zarogianni, and Georgios Samaras.
\newblock Mlops meets edge computing: an edge platform with embedded intelligence towards 6g systems.
\newblock In {\em 2023 Joint European Conference on Networks and Communications \& 6G Summit (EuCNC/6G Summit)}, pages 496--501. IEEE, 2023.

\stepcounter{PScounter} 

\bibitem[PS\arabic{PScounter}]{I41}
Alvaro Luis, Miguel~A Patricio, Antonio Berlanga, and Jos{\'e}~M Molina.
\newblock Seamless transition from machine learning on the cloud to industrial edge devices with thinger. io.
\newblock {\em IEEE Internet of Things Journal}, 2023.

\stepcounter{PScounter} 

\bibitem[PS\arabic{PScounter}]{I64}
Guillermo~Chinarro {\'A}lvarez, C{\'e}sar Alejandro~Achig Ram{\'\i}rez, Javier Andi{\'o}n, Juan~C Due{\~n}as, et~al.
\newblock Toward an integrated and supported machine learning process.
\newblock In {\em 2023 7th International Young Engineers Forum (YEF-ECE)}, pages 37--42. IEEE, 2023.

\stepcounter{PScounter} 

\bibitem[PS\arabic{PScounter}]{S66}
Sachchidanand Singh, Naveen Singh, and Vinay Singh.
\newblock Comparative analysis of open standards for machine learning model deployments.
\newblock In {\em ICT Systems and Sustainability: Proceedings of ICT4SD 2021, Volume 1}, pages 499--507. Springer, 2022.

\stepcounter{PScounter} 

\bibitem[PS\arabic{PScounter}]{I55}
Mattia Antonini, Miguel Pincheira, Massimo Vecchio, and Fabio Antonelli.
\newblock Tiny-mlops: A framework for orchestrating ml applications at the far edge of iot systems.
\newblock In {\em 2022 IEEE international conference on evolving and adaptive intelligent systems (EAIS)}, pages 1--8. IEEE, 2022.

\stepcounter{PScounter} 

\bibitem[PS\arabic{PScounter}]{S46}
Hemadri Jayalath and Lakshmish Ramaswamy.
\newblock Enhancing performance of operationalized machine learning models by analyzing user feedback.
\newblock In {\em Proceedings of the 2022 4th International Conference on Image, Video and Signal Processing}, pages 197--203, 2022.

\stepcounter{PScounter} 

\bibitem[PS\arabic{PScounter}]{I170}
Laurent Bou{\'e}, Pratap Kunireddy, and Pavle Suboti{\'c}.
\newblock Automatically resolving data source dependency hell in large scale data science projects.
\newblock In {\em 2023 IEEE/ACM 2nd International Conference on AI Engineering--Software Engineering for AI (CAIN)}, pages 1--6. IEEE, 2023.

\stepcounter{PScounter} 

\bibitem[PS\arabic{PScounter}]{I1}
DR~Niranjan et~al.
\newblock Jenkins pipelines: A novel approach to machine learning operations (mlops).
\newblock In {\em 2022 International Conference on Edge Computing and Applications (ICECAA)}, pages 1292--1297. IEEE, 2022.

\stepcounter{PScounter} 

\bibitem[PS\arabic{PScounter}]{RedHatOpenShiftAI}
Mlops: Machine learning operations with red hat openshift.
\newblock 2023.
\newblock [White Paper].

\stepcounter{PScounter} 

\bibitem[PS\arabic{PScounter}]{RedHatDataMLOpsIOT}
Data, mlops, and iot for the next-generation insurance industry.
\newblock 2021.
\newblock [Report].

\stepcounter{PScounter} 

\bibitem[PS\arabic{PScounter}]{RedHatResources}
Top 5 ways developers and data scientists can collaborate.
\newblock 2022.
\newblock [Blog].

\stepcounter{PScounter} 

\bibitem[PS\arabic{PScounter}]{I109}
Nadia Nahar, Shurui Zhou, Grace Lewis, and Christian K{\"a}stner.
\newblock Collaboration challenges in building ml-enabled systems: Communication, documentation, engineering, and process.
\newblock In {\em Proceedings of the 44th international conference on software engineering}, pages 413--425, 2022.

\stepcounter{PScounter} 

\bibitem[PS\arabic{PScounter}]{AmazonAI}
Amazon ai fairness and explainability whitepaper.
\newblock 2023.
\newblock [White Paper].

\stepcounter{PScounter} 

\bibitem[PS\arabic{PScounter}]{Airflow}
Operationalizing machine learning.
\newblock [Online; accessed 1. Apr. 2022].

\stepcounter{PScounter} 

\bibitem[PS\arabic{PScounter}]{NVIDIA2023}
William Benton.
\newblock {Demystifying Enterprise MLOps}, Mar 2023.
\newblock [Online; accessed Nov 2023].

\stepcounter{PScounter} 

\bibitem[PS\arabic{PScounter}]{I96}
Souma~Kanti Paul, Sadia Riaz, and Suchismita Das.
\newblock A conceptual architecture for ai in supply chain risk management.
\newblock In {\em TENCON 2022-2022 IEEE Region 10 Conference (TENCON)}, pages 1--5. IEEE, 2022.

\stepcounter{PScounter} 

\bibitem[PS\arabic{PScounter}]{S243}
Ruibo Chen, Yanjun Pu, Bowen Shi, and Wenjun Wu.
\newblock An automatic model management system and its implementation for aiops on microservice platforms.
\newblock {\em The Journal of Supercomputing}, 79(10):11410--11426, 2023.

\stepcounter{PScounter} 

\bibitem[PS\arabic{PScounter}]{S33}
Juan Pineda-Jaramillo and Francesco Viti.
\newblock Mlops in freight rail operations.
\newblock {\em Engineering Applications of Artificial Intelligence}, 123:106222, 2023.

\stepcounter{PScounter} 

\bibitem[PS\arabic{PScounter}]{S68}
Sebastian Schelter, Stefan Grafberger, Shubha Guha, Bojan Karlas, and Ce~Zhang.
\newblock Proactively screening machine learning pipelines with arguseyes.
\newblock In {\em Companion of the 2023 International Conference on Management of Data}, pages 91--94, 2023.

\stepcounter{PScounter} 

\bibitem[PS\arabic{PScounter}]{S21}
Tobias M{\"u}ller, Milena Zahn, and Florian Matthes.
\newblock On the adoption of federated machine learning: Roles, activities and process life cycle.
\newblock In {\em ICEIS (1)}, pages 525--531, 2023.

\stepcounter{PScounter} 

\bibitem[PS\arabic{PScounter}]{S10}
Rakshith Subramanya, Seppo Sierla, and Valeriy Vyatkin.
\newblock From devops to mlops: Overview and application to electricity market forecasting.
\newblock {\em Applied Sciences}, 12(19):9851, 2022.

\stepcounter{PScounter} 

\bibitem[PS\arabic{PScounter}]{S43}
Vlad Stirbu, Tuomas Granlund, and Tommi Mikkonen.
\newblock Continuous design control for machine learning in certified medical systems.
\newblock {\em Software Quality Journal}, 31(2):307--333, 2023.

\stepcounter{PScounter} 

\bibitem[PS\arabic{PScounter}]{I89}
Harsha Moraliyage, Dilantha Haputhanthri, Chamod Samarajeewa, Nishan Mills, Daswin De~Silva, Milos Manic, and Andrew Jennings.
\newblock Automated machine learning in critical energy infrastructure for net zero carbon emissions.
\newblock In {\em 2023 IEEE 32nd International Symposium on Industrial Electronics (ISIE)}, pages 1--7. IEEE, 2023.

\stepcounter{PScounter} 

\bibitem[PS\arabic{PScounter}]{I45}
Ra{\'u}l Mi{\~n}{\'o}n, Josu D{\'\i}az-de Arcaya, Ana~I Torre-Bastida, Gorka Zarate, and Aitor Moreno-Fernandez-de Leceta.
\newblock Mlpacker: A unified software tool for packaging and deploying atomic and distributed analytic pipelines.
\newblock In {\em 2022 7th International Conference on Smart and Sustainable Technologies (SpliTech)}, pages 1--6. IEEE, 2022.

\stepcounter{PScounter} 

\bibitem[PS\arabic{PScounter}]{S204}
Marius Schlegel and Kai-Uwe Sattler.
\newblock Mlflow2prov: extracting provenance from machine learning experiments.
\newblock In {\em Proceedings of the Seventh Workshop on Data Management for End-to-End Machine Learning}, pages 1--4, 2023.

\stepcounter{PScounter} 

\bibitem[PS\arabic{PScounter}]{IBM2023}
Dominik Kreuzberger, Julianne~Forgo Kreuzberger, Mihai Criveti, Przemek Czuba, and Yvette Machowski.
\newblock What is mlops?
\newblock [Online; accessed 1. Nov. 2023].

\stepcounter{PScounter} 

\bibitem[PS\arabic{PScounter}]{S625}
David Nigenda, Zohar Karnin, Muhammad~Bilal Zafar, Raghu Ramesha, Alan Tan, Michele Donini, and Krishnaram Kenthapadi.
\newblock Amazon sagemaker model monitor: A system for real-time insights into deployed machine learning models.
\newblock In {\em Proceedings of the 28th ACM SIGKDD Conference on Knowledge Discovery and Data Mining}, pages 3671--3681, 2022.

\stepcounter{PScounter} 

\bibitem[PS\arabic{PScounter}]{S615}
Nathalie Rauschmayr, Sami Kama, Muhyun Kim, Miyoung Choi, and Krishnaram Kenthapadi.
\newblock Profiling deep learning workloads at scale using amazon sagemaker.
\newblock In {\em Proceedings of the 28th ACM SIGKDD Conference on Knowledge Discovery and Data Mining}, pages 3801--3809, 2022.

\stepcounter{PScounter} 

\bibitem[PS\arabic{PScounter}]{S454}
Houssem Guissouma, Moritz Zink, and Eric Sax.
\newblock Continuous safety assessment of updated supervised learning models in shadow mode.
\newblock In {\em 2023 IEEE 20th International Conference on Software Architecture Companion (ICSA-C)}, pages 301--308. IEEE, 2023.

\stepcounter{PScounter} 

\bibitem[PS\arabic{PScounter}]{S632}
Josu D{\'\i}az-de Arcaya, Ana~I Torre-Bastida, Ra{\'u}l Mi{\~n}{\'o}n, and Aitor Almeida.
\newblock Orfeon: An aiops framework for the goal-driven operationalization of distributed analytical pipelines.
\newblock {\em Future Generation Computer Systems}, 140:18--35, 2023.

\stepcounter{PScounter} 

\bibitem[PS\arabic{PScounter}]{I86}
Peini Liu, Gusseppe Bravo-Rocca, Jordi Guitart, Ajay Dholakia, David Ellison, and Miroslav Hodak.
\newblock Scanflow-k8s: Agent-based framework for autonomic management and supervision of ml workflows in kubernetes clusters.
\newblock In {\em 2022 22nd IEEE International Symposium on Cluster, Cloud and Internet Computing (CCGrid)}, pages 376--385. IEEE, 2022.

\stepcounter{PScounter} 

\bibitem[PS\arabic{PScounter}]{S646}
Igor~L Markov, Hanson Wang, Nitya~S Kasturi, Shaun Singh, Mia~R Garrard, Yin Huang, Sze Wai~Celeste Yuen, Sarah Tran, Zehui Wang, Igor Glotov, et~al.
\newblock Looper: An end-to-end ml platform for product decisions.
\newblock In {\em Proceedings of the 28th ACM SIGKDD Conference on Knowledge Discovery and Data Mining}, pages 3513--3523, 2022.

\stepcounter{PScounter} 

\bibitem[PS\arabic{PScounter}]{S163}
Semo Yang, Jihwan Moon, Jinsoo Kim, Kwangkee Lee, and Kangyoon Lee.
\newblock Flscalize: Federated learning lifecycle management platform.
\newblock {\em IEEE Access}, 2023.

\stepcounter{PScounter} 

\bibitem[PS\arabic{PScounter}]{I168}
Hanqing Cao, Joshua Finer, Murad Megjhani, Daniel~C Nametz, Virginia Lorenzi, Lena Mamykina, Richard Meyers, Sarah~C Rossetti, and Soojin Park.
\newblock Machine learning model deployment using real-time physiological monitoring: Use case of detecting delayed cerebral ischemia.
\newblock In {\em 2022 IEEE Healthcare Innovations and Point of Care Technologies (HI-POCT)}, pages 42--45. IEEE, 2022.

\stepcounter{PScounter} 

\bibitem[PS\arabic{PScounter}]{I68}
Paul-Philipp Luley, Jan~M Deriu, Peng Yan, Gerrit~A Schatte, and Thilo Stadelmann.
\newblock From concept to implementation: The data-centric development process for ai in industry.
\newblock In {\em 2023 10th IEEE Swiss Conference on Data Science (SDS)}, pages 73--76. IEEE, 2023.

\stepcounter{PScounter} 

\bibitem[PS\arabic{PScounter}]{I28}
Qi~Cheng and Guodong Long.
\newblock Federated learning operations (flops): Challenges, lifecycle and approaches.
\newblock In {\em 2022 International Conference on Technologies and Applications of Artificial Intelligence (TAAI)}, pages 12--17. IEEE, 2022.

\stepcounter{PScounter} 

\bibitem[PS\arabic{PScounter}]{S521}
Shir Chorev, Philip Tannor, Dan~Ben Israel, Noam Bressler, Itay Gabbay, Nir Hutnik, Jonatan Liberman, Matan Perlmutter, Yurii Romanyshyn, and Lior Rokach.
\newblock Deepchecks: A library for testing and validating machine learning models and data.
\newblock {\em Journal of Machine Learning Research}, 23(285):1--6, 2022.

\stepcounter{PScounter} 

\bibitem[PS\arabic{PScounter}]{I32}
Keita Sakuma, Ryuta Matsuno, and Yoshio Kameda.
\newblock A method of identifying causes of prediction errors to accelerate mlops.
\newblock In {\em 2023 IEEE/ACM International Workshop on Deep Learning for Testing and Testing for Deep Learning (DeepTest)}, pages 9--16. IEEE, 2023.

\stepcounter{PScounter} 

\bibitem[PS\arabic{PScounter}]{A6}
Aquilas~Tchanjou Njomou, Marios Fokaefs, Dimitry~Fumtim Silatchom~Kamga, and Bram Adams.
\newblock On the challenges of migrating to machine learning life cycle management platforms.
\newblock In {\em Proceedings of the 32nd Annual International Conference on Computer Science and Software Engineering}, pages 42--51, 2022.

\stepcounter{PScounter} 

\bibitem[PS\arabic{PScounter}]{S510}
Bo~Wen, Yan Koyfman, Hongfei Tian, Boris Lublinsky, Raquel Norel, Carla Agurto, Dean Wampler, and Narciso Albarracin.
\newblock Accelerating automation of digital health applications via cloud native approach.
\newblock In {\em Proceedings of the Eighth International Workshop on Container Technologies and Container Clouds}, pages 1--6, 2022.

\stepcounter{PScounter} 

\bibitem[PS\arabic{PScounter}]{I5}
Gregor Cerar and Jernej Hribar.
\newblock Machine learning operations model store: Optimizing model selection for ai as a service.
\newblock In {\em 2023 International Balkan Conference on Communications and Networking (BalkanCom)}, pages 1--5. IEEE, 2023.

\stepcounter{PScounter} 

\bibitem[PS\arabic{PScounter}]{I17}
Rustem Dautov, Erik~Johannes Husom, Fotis Gonidis, Spyridon Papatzelos, and Nikolaos Malamas.
\newblock Bridging the gap between java and python in mobile software development to enable mlops.
\newblock In {\em 2022 18th International Conference on Wireless and Mobile Computing, Networking and Communications (WiMob)}, pages 363--368. IEEE, 2022.

\stepcounter{PScounter} 

\bibitem[PS\arabic{PScounter}]{I27}
Qinghua Lu, Liming Zhu, Xiwei Xu, Jon Whittle, and Zhenchang Xing.
\newblock Towards a roadmap on software engineering for responsible ai.
\newblock In {\em Proceedings of the 1st International Conference on AI Engineering: Software Engineering for AI}, pages 101--112, 2022.

\stepcounter{PScounter} 

\bibitem[PS\arabic{PScounter}]{I33}
Rizal~Broer Bahaweres, Aldi Zulfikar, Irman Hermadi, Arif~Imam Suroso, and Yandra Arkeman.
\newblock Docker and kubernetes pipeline for devops software defect prediction with mlops approach.
\newblock In {\em 2022 2nd International Seminar on Machine Learning, Optimization, and Data Science (ISMODE)}, pages 248--253. IEEE, 2022.

\stepcounter{PScounter} 

\bibitem[PS\arabic{PScounter}]{I91}
Bradley Eck, Duygu Kabakci-Zorlu, Yan Chen, France Savard, and Xiaowei Bao.
\newblock A monitoring framework for deployed machine learning models with supply chain examples.
\newblock In {\em 2022 IEEE International Conference on Big Data (Big Data)}, pages 2231--2238. IEEE, 2022.

\stepcounter{PScounter} 

\bibitem[PS\arabic{PScounter}]{I152}
Rohan Gawhade, Lokesh~Ramdev Bohara, Jesvin Mathew, and Poonam Bari.
\newblock Computerized data-preprocessing to improve data quality.
\newblock In {\em 2022 Second International Conference on Power, Control and Computing Technologies (ICPC2T)}, pages 1--6. IEEE, 2022.

\stepcounter{PScounter} 

\bibitem[PS\arabic{PScounter}]{S4}
Claudia~Patricia Ayala~Mart{\'\i}nez, Besim Bilalli, Cristina G{\'o}mez~Seoane, and Silverio~Juan Mart{\'\i}nez~Fern{\'a}ndez.
\newblock Dogo4ml: Development, operation and data governance for ml-based software systems.
\newblock In {\em Joint Proceedings of RCIS 2022 Workshops and Research Projects Track: co-located with the 16th International Conference on Research Challenges in Information Science (RCIS 2022): Barcelona, Spain, May 17-20, 2022}. CEUR-WS. org, 2022.

\stepcounter{PScounter} 

\bibitem[PS\arabic{PScounter}]{S24}
Ashish~Singh Parihar, Umesh Gupta, Utkarsh Srivastava, Vishal Yadav, and Vaibhav~Kumar Trivedi.
\newblock Automated machine learning deployment using open-source ci/cd tool.
\newblock In {\em Proceedings of Data Analytics and Management: ICDAM 2022}, pages 209--222. Springer, 2023.

\stepcounter{PScounter} 

\bibitem[PS\arabic{PScounter}]{S90}
Ra{\'u}l Mi{\~n}{\'o}n, Josu Diaz-de Arcaya, Ana~I Torre-Bastida, and Philipp Hartlieb.
\newblock Pangea: an mlops tool for automatically generating infrastructure and deploying analytic pipelines in edge, fog and cloud layers.
\newblock {\em Sensors}, 22(12):4425, 2022.

\stepcounter{PScounter} 

\bibitem[PS\arabic{PScounter}]{S156}
Dhia~Elhaq Rzig, Foyzul Hassan, Chetan Bansal, and Nachiappan Nagappan.
\newblock Characterizing the usage of ci tools in ml projects.
\newblock In {\em Proceedings of the 16th ACM/IEEE International Symposium on Empirical Software Engineering and Measurement}, pages 69--79, 2022.

\stepcounter{PScounter} 

\bibitem[PS\arabic{PScounter}]{S261}
Songzhu Mei, Cong Liu, Qinglin Wang, and Huayou Su.
\newblock Model provenance management in mlops pipeline.
\newblock In {\em Proceedings of the 2022 8th International Conference on Computing and Data Engineering}, pages 45--50, 2022.

\stepcounter{PScounter} 

\bibitem[PS\arabic{PScounter}]{S276}
Ayan Chatterjee, Bestoun~S Ahmed, Erik Hallin, and Anton Engman.
\newblock Quality assurance in mlops setting: An industrial perspective.
\newblock {\em arXiv preprint arXiv:2211.12706}, 2022.

\stepcounter{PScounter} 

\bibitem[PS\arabic{PScounter}]{I268}
Joran Leest, Ilias Gerostathopoulos, and Claudia Raibulet.
\newblock Evolvability of machine learning-based systems: An architectural design decision framework.
\newblock In {\em 2023 IEEE 20th International Conference on Software Architecture Companion (ICSA-C)}, pages 106--110. IEEE, 2023.

\stepcounter{PScounter} 

\bibitem[PS\arabic{PScounter}]{I271}
Antonio Guerriero, Roberto Pietrantuono, and Stefano Russo.
\newblock Iterative assessment and improvement of dnn operational accuracy.
\newblock In {\em 2023 IEEE/ACM 45th International Conference on Software Engineering: New Ideas and Emerging Results (ICSE-NIER)}, pages 43--48. IEEE, 2023.

\stepcounter{PScounter} 

\bibitem[PS\arabic{PScounter}]{I35}
Georgios Samaras, Vasileios Theodorou, Dimitris Laskaratos, Nikolaos Psaromanolakis, Marinela Mertiri, and Alexandros Valantasis.
\newblock Qmp: A cloud-native mlops automation platform for zero-touch service assurance in 5g systems.
\newblock In {\em 2022 IEEE International Mediterranean Conference on Communications and Networking (MeditCom)}, pages 86--89. IEEE, 2022.

\stepcounter{PScounter} 

\bibitem[PS\arabic{PScounter}]{S1}
Nils Baumann, Evgeny Kusmenko, Jonas Ritz, Bernhard Rumpe, and Moritz~Benedikt Weber.
\newblock Dynamic data management for continuous retraining.
\newblock In {\em Proceedings of the 25th International Conference on Model Driven Engineering Languages and Systems: Companion Proceedings}, pages 359--366, 2022.

\stepcounter{PScounter} 

\bibitem[PS\arabic{PScounter}]{W115}
Jing Peng, Shiliang Zheng, Yutao Li, and Zhe Shuai.
\newblock From source code to model service: A framework’s perspective.
\newblock In {\em The International Conference on Natural Computation, Fuzzy Systems and Knowledge Discovery}, pages 1355--1362. Springer, 2022.

\stepcounter{PScounter} 

\bibitem[PS\arabic{PScounter}]{S291}
Markus Borg.
\newblock Agility in software 2.0--notebook interfaces and mlops with buttresses and rebars.
\newblock In {\em International Conference on Lean and Agile Software Development}, pages 3--16. Springer, 2022.

\stepcounter{PScounter} 

\bibitem[PS\arabic{PScounter}]{S523}
Eyad Kannout, Micha{\l} Grodzki, and Marek Grzegorowski.
\newblock Considering various aspects of models’ quality in the ml pipeline-application in the logistics sector.
\newblock In {\em 2022 17th Conference on Computer Science and Intelligence Systems (FedCSIS)}, pages 403--412. IEEE, 2022.

\end{thebibliography}

\newpage

 \begin{appendices}
 \section{Primary Studies}
 \renewcommand\bibsection{}


\section{Background}\label{background}
\subsection{CI/CD practices in traditional software engineering and DevOps}\label{Background-DevOps}
Continuous integration (CI) practices within agile development approaches \cite{beck2000extreme} have widespread usage in today's software development community. CI refers to frequent integration of software artifacts (e.g., code base, dependencies) during development \cite{staahl2014modeling}. The benefits of CI include reducing release cycles by keeping software always ready for release, increasing team productivity, finding bugs as early as possible, and improving software quality \cite{spieker2019towards, staahl2014modeling}.
On the other hand, continuous deployment (CD) indicates the automatic and frequent delivery of software to the customer or production environment \cite{shahin2017continuous, karamitsos2020applying, staahl2014modeling}. 
This practice allows for the customer to receive early feedback on the functionality delivered of the software, helping to identify further improvements \cite{cano2020taxonomy}. 
CI/CD contributes to reducing risks and improving software quality \cite{duvall2007continuous}.

DevOps (Dev for development and Ops for operations) is a natural evolution of agile and CI/CD practices. It is a paradigm of continuous and collaborative software engineering \cite{karamitsos2020applying}. 
It aims to deliver software more frequently using an automated, repeatable, and continuous process flow.
Furthermore, DevOps involves practices, tools, and sociotechnical aspects that facilitate easier collaboration between development and operations teams \cite{FIRST-42}.

\subsection{MLOps} \label{background-MLOps}
CI/CD practices have been very useful in the development of ML-enabled systems, much like their success in traditional software development. 
Based on its benefits in traditional software engineering, the ML development community adopts CI / CD practices in ML development \cite{lima2022mlops}. 
The experimental and probabilistic nature of AI/ML systems requires frequent repetition of development and deployment flow. Therefore, the ML development and deployment process demands practices similar to those of CI / CD \cite{FIRST-3}. 

MLOps concept has emerged with the goal of establishing a set of practices for more efficient ML development and deployment flow by mimicking the DevOps practices \cite{FIRST-1}. 
MLOps introduces additional practices specific to ML, emphasizing the automation of development and monitoring at all stages of ML, and deployment \cite{FIRST-1, FIRST-10}.

One important practice of MLOps is monitoring. 
Monitoring helps to track data and model performance shifts to ensure the reliability and quality of system over time \cite{FIRST-7}, and enables continuous feedback loop between development, operations, and production environments. 
At this point, along with CI and CD, MLOps incorporates the continuous training (CT) concept, because the continuous delivery of an ML system is not independent from data collection, analysis, and model training phases \cite{FIRST-GL-13}.

Further, MLOps introduces new practices regarding the roles, responsibilities and collaboration between people with different backgrounds. MLOps engineer is a new role introduced with MLOps, referring to the professionals responsible for maintaining the operational stability of ML pipelines and the entire ML system \cite{AwsSecure, neghawi2023linking}. Moreover, new collaboration practices, i.e., collaboration points \cite{I109}, have been introduced to improve the efficiency of MLOps projects by enabling better collaboration between data science and software engineering teams.

Here we also mention DataOps and ModelOps which are two concepts that share similarities with MLOps. However, our work focuses on MLOps. Thus, we only highlight differences between concepts and only cover MLOps in our RQs.

DataOps is coined to refer to the automation of data analytics life-cycle \cite{ereth2018dataops, FIRST-42}. DataOps establishes a set of practices for data-centric systems mimicking DevOps practices and aims to ship data frequently in an automated \cite{WhatIsMLOpsDataOps}.

ModelOps is defined as continuous practices for AI-enabled software systems \cite{FIRST-33}. The key activities revolve around continuous model training, model metadata versioning, deployment, and management of AI pipelines. 

While both DataOps and ModelOps are related to MLOps, these three terms refer to different concepts.
DataOps centralizes data analytics, management, and the shipment of data. ModelOps centralizes building and managing various versions of models and serving them in AI systems. 
On the other hand, MLOps is primarily concerned with productionalization of ML systems and automating the whole ML pipeline from data collection through model training, deployment, and monitoring.
Further, the skills required by these three concepts vary; DataOps mainly involves data science, data analysis, and data engineering, and ModelOps involves data science. MLOps utilizes a broader range of skills such as data and ML engineering, data science, and MLOps engineering (see Section \ref{RQ1.3-RolesAndResponsibilities}).

.
 \end{appendices}

\end{document}